\documentclass[showpacs, amsmath,amssymb,aps]{revtex4}
\usepackage{tabularx}
\usepackage{comment}
\usepackage{graphicx}% Include figure files
\usepackage{float}
\usepackage{dcolumn}% Align table columns on decimal point
\usepackage{bm}% bold math
\usepackage{mathrsfs}
\usepackage{amsmath}

\begin{document}
%\nofiles

%\preprint{APS/123-QED}
\title{Possible Connection between Dark Matter and Supermassive Black Holes}
\author{D. Batic}
\email{davide.batic@ku.ac.ae}
\affiliation{%
Department of Mathematics,\\  Khalifa University of Science and Technology,\\ Main Campus, Abu Dhabi,\\ United Arab Emirates}\
\author{J. Mojahed Faraji}
\email{100045863@ku.ac.ae}
\affiliation{%
Department of Mathematics,\\  Khalifa University of Science and Technology,\\ Main Campus, Abu Dhabi,\\ United Arab Emirates}
\author{M. Nowakowski}
\email{mnowakos@uniandes.edu.co}
\affiliation{
Departamento de Fisica,\\ Universidad de los Andes, Cra.1E
No.18A-10, Bogota, Colombia
}

\date{\today}
\date{\today}

\begin{abstract}
  Dark Matter (DM) is usually studied in connection with rotational curves in the outskirts of the galaxies. However,  the role of DM might be different in the galactic bulges and centers where Supermassive Black Holes (SMBHs) dominate the gravitational interaction. Indeed, given the fact that DM is the dominant matter species in the Universe, it is natural to assume a close connection between DM and SMBHs.
Here we probe into this possibility by constructing stable objects with fuzzy mass distributions based on standard DM profiles. These astrophysical objects come out in three types: a fuzzy droplet without horizon and fuzzy Black Holes (BHs) with one or two horizons. We emphasize that all objects are solutions of Einstein equations. Their effective potentials which govern the motion of a test body, can display a reasonable similarity to the effective potential of a Schwarzschild BH at the galactic center. Therefore, some of our solutions could, in principle, replace the standard BH-picture of the galactic center and, at the same time,  have the advantage that they have been composed of the main matter ingredient of the Universe. 
\end{abstract}
%\pacs{04.70.-s,95.30.Sf,95.35.+d,98.35.Gi}
\maketitle

\section{Introduction}
Recently, \cite{Ghez1,Ghez2} were able to verify experimentally the presence of a supermassive gravitational object known as Sagittarius A$^{*}$ at the center of the Milky Way. This was achieved by measuring the trajectories of the so-called S-stars, which are  celestial objects orbiting at relativistic speeds in proximity of the central galactic region where a SMBH is thought to reside. However, the mechanisms behind the formation of such gigantic SMBHs are not yet clearly understood. Up to now, several suggestions have been brought forward. Computer simulations conducted by \cite{jap} indicated that SMBHs may arise from the collapse of extremely massive clouds of gas at the time when galaxies were forming. Other studies \cite{Kulier,Pacucci} suggested that a SMBH started as a normal sized black hole (BH) and it became supermassive either by swallowing enormous quantities of matter over time or by merging with a cluster of BHs. Another model predicts that a dense stellar cluster may undergo core collapse because of the negative heat capacity of the system pushing the velocity dispersion in the galactic central region to relativistic speeds \cite{Spitzer,Boekholt}. In that regard, it is worth mentioning that \cite{Begelman} proposed a process according to which {\it{quasi-stars}} may initially form from the collapse of large gas clouds and later implode under the action of their own gravity to give rise to seed BHs of approximately twenty solar masses. Moreover, \cite{Yoshida} showed numerically that proto-galactic DM halos may trigger rapid gas condensation leading to the formation of supermassive protostar immersed in a dense gas cloud where mass accretion allows the protostar to increase its mass up to $3.4\cdot 10^4$ solar masses. On the other hand, \cite{John} found by a radiation hydrodynamics simulation of early galaxy formation that massive black holes may form in rapidly growing pre-galactic gas clouds. More precisely, the idea brought forward there is that bright ultraviolet light emitted during star formation in young galaxies may stop a  nearby gas cloud from producing stars until it reaches a critical mass leading to gravitational collapse and BH formation. However, a scenario where a huge gas cloud lies in the proximity of a star-forming galaxy may turn to be quite rare. In addition to SMBH, ultra massive BHs (UMBH) such as ULAS J1342$+$0928 and have been recently reported in \cite{Ban,Wang}. \cite{Balberg,Poll,Feng} invoked DM collapse with self-interaction as an ingredient behind their formation while \cite{Ban} suggested that these objects may provide evidence that our Universe could have originated from a Big Bounce instead of a Big Bang. 

In addition to the attempts described above, many other authors dived into the possibility of alternative mechanisms triggering the formation of SMBHs.  For example, gravitational vacuum condensate objects also know as gravastars were proposed and discussed in \cite{MazMo,ChiRe} while \cite{RuBo,ScMi} suggested the presence of boson stars. Naked singularities were brought into the picture by \cite{Joshi,BaMa,Chowd}, burning disks appeared in \cite{Kundt}, quantum cores (Ruffini-Arg\"{u}elles-Rueda model) were invoked by \cite{Ru1,Ru2} and DM gravitationally bound clamps relying on the exponential-sphere density profile were introduced in \cite{Bosh,Sofue,Leu}. Finally, \cite{Becerra} showed by a numerical simulation that if the central SMBH is replaced by an object made of darkinos, this model does not only reproduce the same kinematics for S-stars but also explain the G2 anomaly \cite{Park}. Similar conclusions as in \cite{Becerra} has been obtained from a theoretical point of view in \cite{EPJCus} where the authors chose the Einasto DM profile motivated by similarities to the Gaussian distribution used in \cite{Piero,BHnoncomm,DavidePiero,PIEROBOSS}. In the present work, we extended the study performed by \cite{EPJCus} to the case of the Zhao DM profile \cite{Zhao} in order to not only understand the similarities and differences emerging from the latter and the Einasto profile but also to check the sensitivity of our results against the choice of different DM profiles. To this purpose, the Zhao profile is ideal because it encompasses several known DM profiles and it helps us to understand the role of DM in the central region of Milky Way. We discover that similarly as in \cite{EPJCus} fuzzy DM droplets and black holes with a horizon structure reminiscent of that encountered in a Reissner-Nordstr\"{o}om
geometry or DM droplets can be constructed if the Zhao profile is coupled with an energy-momentum tensor for an anisotropic fluid and an EOS of the de Sitter-type, i.e. $p_r=-\rho$. Such an EOS  has been widely used in the literature to model regular BHs \cite{EPJCus,d1,d2,d3,d4,d5,d6,d7,d8,Mazur,Felten,negativeP}. Furthermore, the BH solutions we found are regular at $r=0$ while the analysis of the Hawking temperature for  the Zhao inspired BH shows that the BH gets hotter as the horizon radius decreases. The temperature exhibits a maximum after which the BH cools down and its temperature vanishes at the radius of the extremal BH. Interestingly it is possible to tune the parameters in the Zaho profile so that the effective potential of the BH solution can be reasonably fitted at the minimum of the Schwarzschild effective potential modelling the central black hole in our galaxy. This procedure ensures that in both models the kinematics of the S-stars will be very similar. Differently as in \cite{EPJCus}, in the context of this model the effective potential associated to the DM droplet does not fit well at the minimum of the Schwarzschild effective potential. In the second model, we introduce a nonlocal EOS for an anisotropic fluid. More precisely, we construct a self-gravitating fuzzy DM droplet which is regular at the origin and whose effective
potential allows bound states for massive and massless particles. In the massive case the effective potential exhibits two minima and one maximum signalizing that we may have stable and unstable bound orbits. It is surprising to discover that in the case of light the effective potential admits a minimum and a maximum close to the central region. They correspond to a stable and an unstable photon sphere. Since the detection of a photon sphere is well within the capabilities of the Event Horizon Telescope  (EHT)\cite{EHT}, we computed the shadow which turned to be considerably larger than the corresponding shadow of a Schwarzschild BH.

The paper is organized as follows: In Section II, we introduce the Zhao profile and some relevant formulae needed in the sections to follow. In Section III and IV, we study DM objects assuming a de Sitter-like EOS and their feasibility in reproducing the kinematics of the S-stars. In section V, we adopt a nonlocal EOS to show that a fuzzy DM droplet consisting of an anisotropic fluid allows for stable orbits in the case of both massive and massless particles. In particular, this new solution of the Einstein field equations exhibits an inner stable photon sphere surrounded by an unstable photon sphere. We conclude this section by computing the shadow of the DM droplet and show that it might in principle be detected by the EHT. We present our conclusions in Section VI.

\section{Zaho's dark matter master profile}
The Zhao density profile is given by \cite{Zhao}
\begin{equation}\label{ro}
    \rho(r)=\frac{\rho_0}{(\frac{r}{r_0})^\gamma \big[1+(\frac{r}{r_0})^\alpha\big] ^{\frac{\beta-\gamma}{\alpha}}} 
\end{equation}
where $r_0$ and $\rho_0$ are the characteristic radius and density ,respectively. Moreover, $\alpha, \beta,$ and $\gamma$ are real parameters and whenever useful we will use the triple $(\alpha,\beta,\gamma)$ to refer to a specific model. It should be said that, while the model parameters $\beta$ and $\gamma$ are the same in \cite{Zhao} and \cite{Krav}, one should be careful with the parameter $\alpha$ since $\alpha$ in \cite{Krav} corresponds to $1/\alpha$ in \cite{Zhao}. Furthermore, the family of density profiles represented by (\ref{ro}) succeeds to include cuspy profiles as those appearing in the Navarro-Frenk-White model as well as the so-called modified isothermal profile which is usually adopted to describe the halo density distribution in studies of observed rotation curves. For an overview of the $(\alpha,\beta,\gamma)$-models covered by (\ref{ro}) we refer to Table~\ref{overview}.
\begin{table}[!ht]
\caption{Principal analytical models associated to the density profile (\ref{ro}). Note that the MIS model is an $\alpha$-model with $\alpha=2$ while the Dehnen model $(1,4,0)$ is a special case of the $(k,n)$-model with $k=n=1$ and $\gamma=0$. Legend abbreviations appearing in the table: NFW=Navarro-Frenk-White; MHP=modified Hubble profile; PS=perfect sphere; MIS=modified isothermal sphere.}
\begin{center}
\begin{tabular}{ | l | l | l | l|l|}
\hline
Model  &  $\alpha$  & $\beta$                    &$\gamma$       \\ \hline
NFW \cite{struct}   &  $1$   & $3$        &$1$     \\ \hline
Jaffe \cite{Jaf}      & $1$    & $4$         & $2$    \\ \hline
Hernquist \cite{Hern} & $1$    & $4$    & $1$ \\ \hline
Dehnen \cite{Deh}   &  $1$   & $4$       & $0\leq\gamma<3$    \\ \hline
$\beta$-model \cite{Zhao}  &  $1$ & $\beta>0$      &  $1$ \\ \hline
Burkert \cite{Burk}   &  $2$   & $2$         & $0$   \\ \hline
MHP \cite{Binn}& $2$ & $3$ &$0$    \\ \hline
Kravtsov \cite{Krav}  &  $2$   & $3$          & $0.2$ \\ \hline
PS \cite{Binn,deZ}& $2$ & $4$ &$0$    \\ \hline
MIS \cite{Pl,Sack}& $2$ & $5$ &$0$    \\ \hline
$(k,n)$-model \cite{Zhao}  & $1/n$, $n\in\mathbb{N}$ & $3+\frac{k}{n}$, $k\in\mathbb{N}$ &$0\leq\gamma<3$    \\ \hline
$\alpha$-model \cite{Zhao}  &  $\alpha>0$  & $3+\alpha$      &  $2-\alpha$ \\ \hline
\end{tabular}
\label{overview}
\end{center}
\end{table}
In what follows, we will assume that $\alpha>0$. Moreover, the density profile (\ref{rho}) is regular at $r=0$ provided that $\gamma=0$. Such a constraint is not too strong because it still allows to study the Dehnen model $(1,4,0)$, the perfect sphere, the MIS and the $(k,n)$-model with $\gamma=0$. In the next section, we will show that all these models coupled to a certain equation of state (EOS) give rise to regular black hole solutions. The characteristic density $\rho_0$ can be written in terms of the total mass $M$ by means of the formula
\begin{equation}\label{massr}
    M=4\pi \int_{0}^{\infty} r^2 \rho(r) \,dr .
\end{equation}
If we introduce the variable transformation
\begin{equation}\label{xr}
    x=\frac{(r/r_0)^\alpha}{1+(r/r_0)^\alpha}.
\end{equation}
mapping the radial interval $[0,\infty)$ to $[0,1)$, the formula for the total mass becomes
\begin{equation}\label{mpq}
    M=\frac{4 \pi r_0^3 \rho_0}{\alpha} \int_{0}^{1} x^p (1-x)^q \,dx,\quad p=\frac{3-\alpha}{\alpha}, \quad q=\frac{\beta-\alpha-3}{\alpha}.
\end{equation}
At this point, a remark is in order. The total mass of the system is finite provided that the integrand in (\ref{mpq}) is integrable at $x=0$ and $x=1$. The corresponding conditions are $p>-1$ and $q>-1$. While the first inequality is satisfied for any $\alpha>0$, the second constraint requires that $\beta>3$. However, as it can be evinced from Table~\ref{overview}, this additional limitation is not too stringent because there are still several interesting models characterized by triplets $(\alpha,\beta,0)$ fulfilling the aforementioned integrability condition such as the Dehnen's model $(1,4,0)$, the Perfect Sphere model $(2,4,0)$, the MIS model $(2,5,0)$ and the $(k,n)$-model $(1/n,3+k/n,0)$. If the integrability condition is not satisfied, as for instance in the Navarro-Frenk-White model, then one needs to introduce an appropriate cut-off distance in the interval of integration in order to obtain a finite total mass. In the present work, we will take under scrutiny the class of models characterized by
\begin{equation}\label{general_condition}
\alpha>0,\quad\beta>3,\quad \gamma=0.
\end{equation}
All the aforementioned models satisfying (\ref{general_condition}) allows to express the integral in (\ref{mpq}) in terms of the Beta function or equivalently as a ratio of Gamma functions. Hence, with the help of $6.2.1$ or $6.2.2$ in \cite{Abra} we find that 
\begin{equation}\label{M}
    M=\frac{4 \pi r_0^3 \rho_0}{\alpha}B(p+1,q+1)=\frac{4\pi r_0^3\rho_0}{\alpha}\frac{\Gamma(p+1)\Gamma(q+1)}{\Gamma(p+q+2)},\quad
    p+1=\frac{3}{\alpha},\quad
    q+1=\frac{\beta-3}{\alpha}.
\end{equation}
Using the above result to express $\rho_0$ in terms of the total mass, we can rewrite (\ref{ro}) as follows
\begin{equation}\label{rho}
\rho(r)=\frac{\alpha M}{4\pi r_0^3 B(p+1,q+1)}\left[1+\left(\frac{r}{r_0}\right)^\alpha\right]^{-\frac{\beta}{\alpha}}.
\end{equation}
In order to compute the associated mass function $m$ defined as 
\begin{equation}\label{massf}
m(r)=4\pi \int_{0}^{r} u^2 \rho(u) \,du,
\end{equation}
it is convenient to first apply the variable transformation (\ref{xr}) to the above integral. This leads to the  integral representation
\begin{equation}\label{int}
m(x)=\frac{M}{B(p+1,q+1)}\int_0^x s^p(1-s)^q~ds
\end{equation}
with parameters $p$ and $q$ defined as in (\ref{mpq}). Note that the condition $M(0)=0$ is trivially satisfied. The constraints on the parameters $\alpha$, $\beta$ and $\gamma$ introduced in (\ref{general_condition}) allow to express (\ref{int}) in terms of the incomplete Beta function by means of $6.6.1$ in \cite{Abra} and we end up with the following analytical expression for the mass function
\begin{equation}\label{dieMasse}
m(x)=\frac{M}{B(p+1,q+1)}B_x(p+1,q+1),
\end{equation}
where $p$ and $q$ have been specified in (\ref{M}).

\section{Zaho's fuzzy black holes}
\begin{figure}[!ht]\label{6in1}
\includegraphics[scale=0.50]{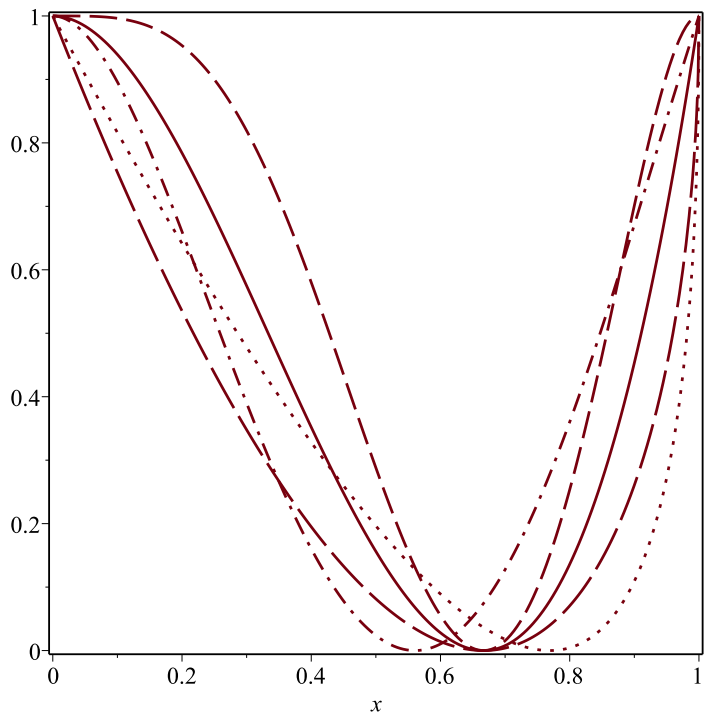}
\caption{\label{figure1}
Plot of the metric coefficient $g_{00}$ as a function of $x$ for different models with  $\mu=\mu_c$ (see Table~\ref{tableEins}) corresponding to the case of an extreme black hole. Legend: solid line Dehnen $(1,4,0)$, dotted line PS, long-dashed line MIS, dashed line $(k,n)=(1,2)$ and dash-dotted line $(k,n)=(2,1)$.}
\end{figure}

\begin{figure}[!ht]\label{6in6}
\centering
    \includegraphics[width=0.3\textwidth]{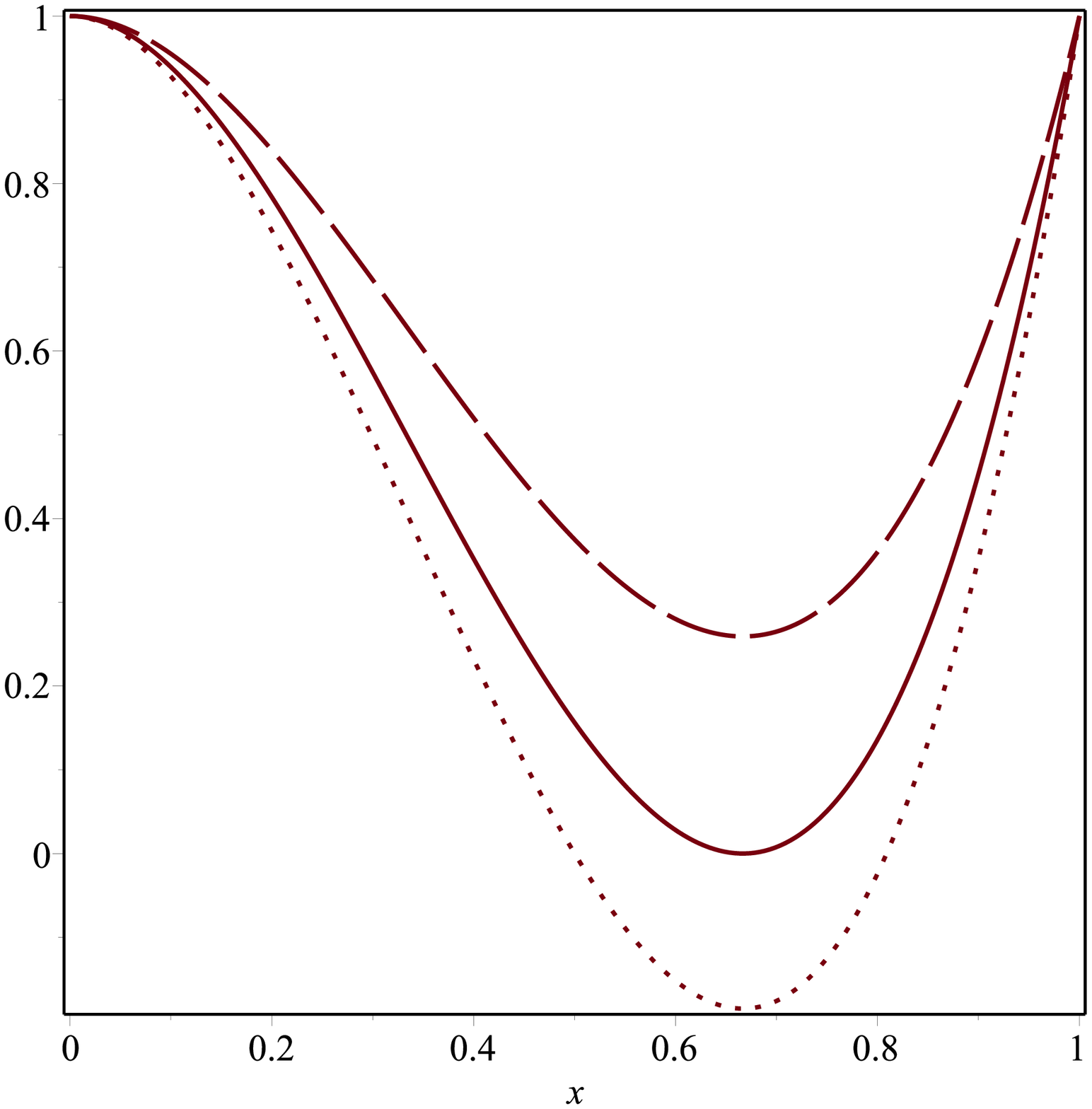}
    \includegraphics[width=0.3\textwidth]{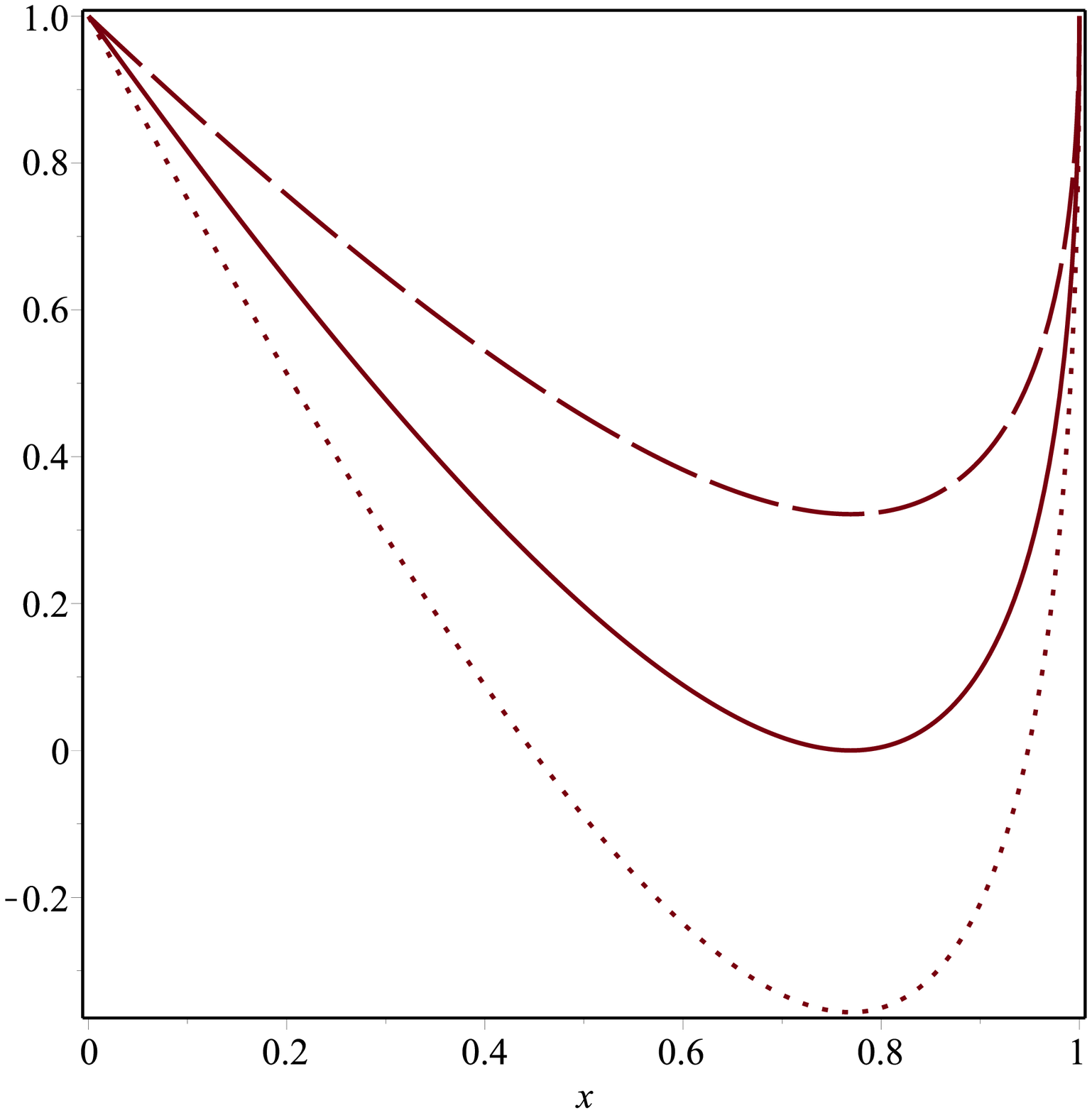}
    \includegraphics[width=0.3\textwidth]{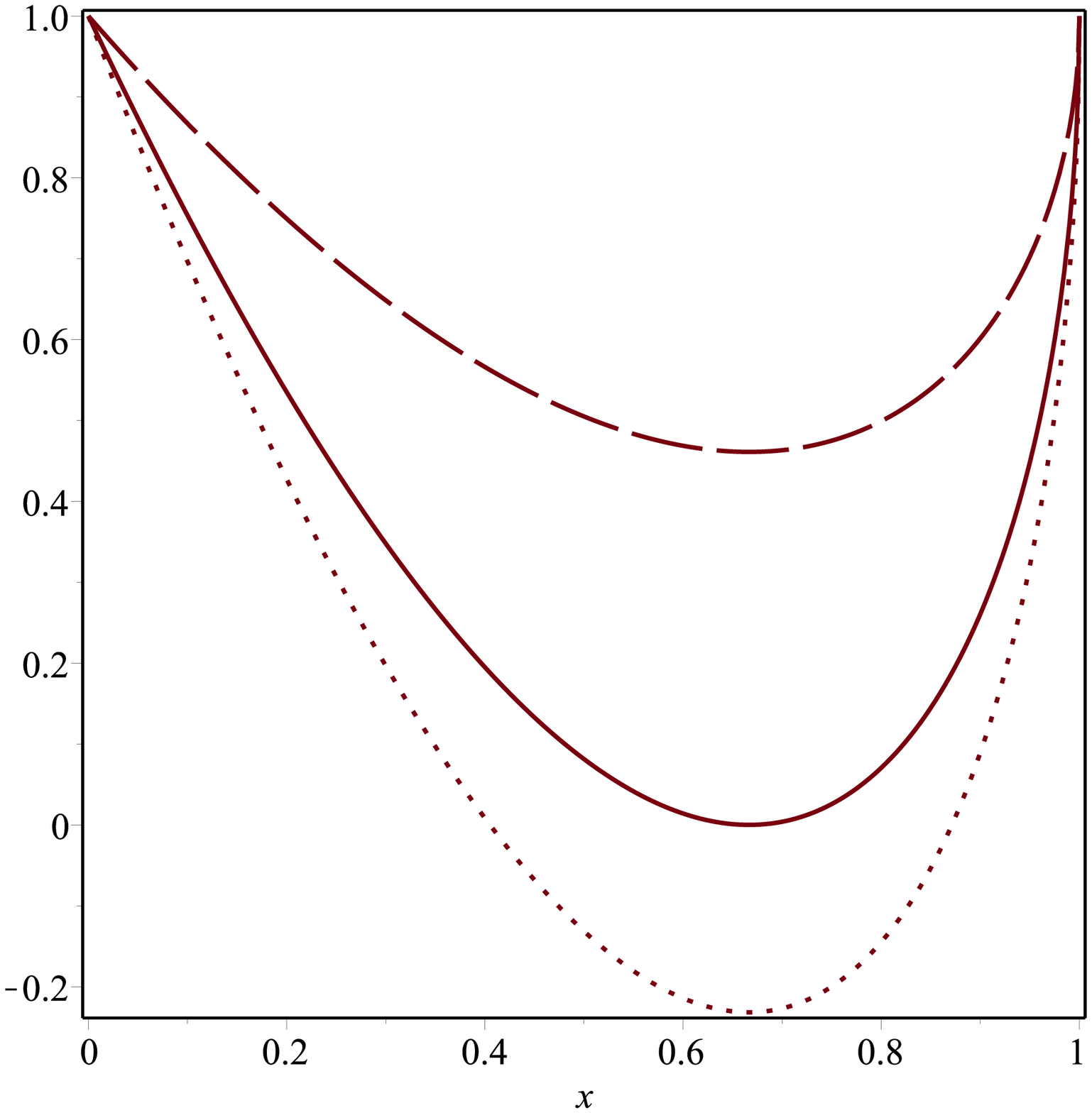}
    \includegraphics[width=0.3\textwidth]{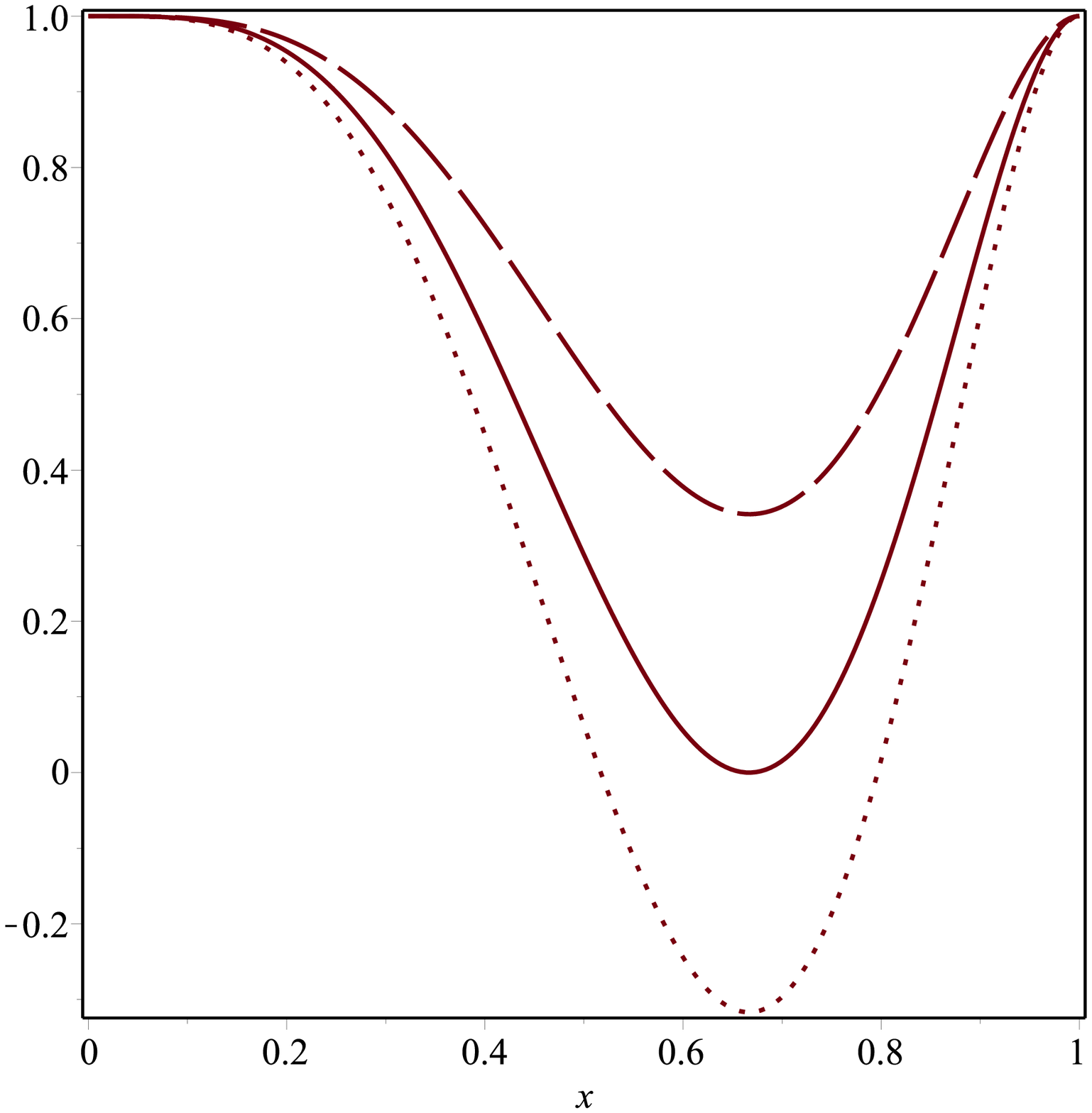}
    \includegraphics[width=0.3\textwidth]{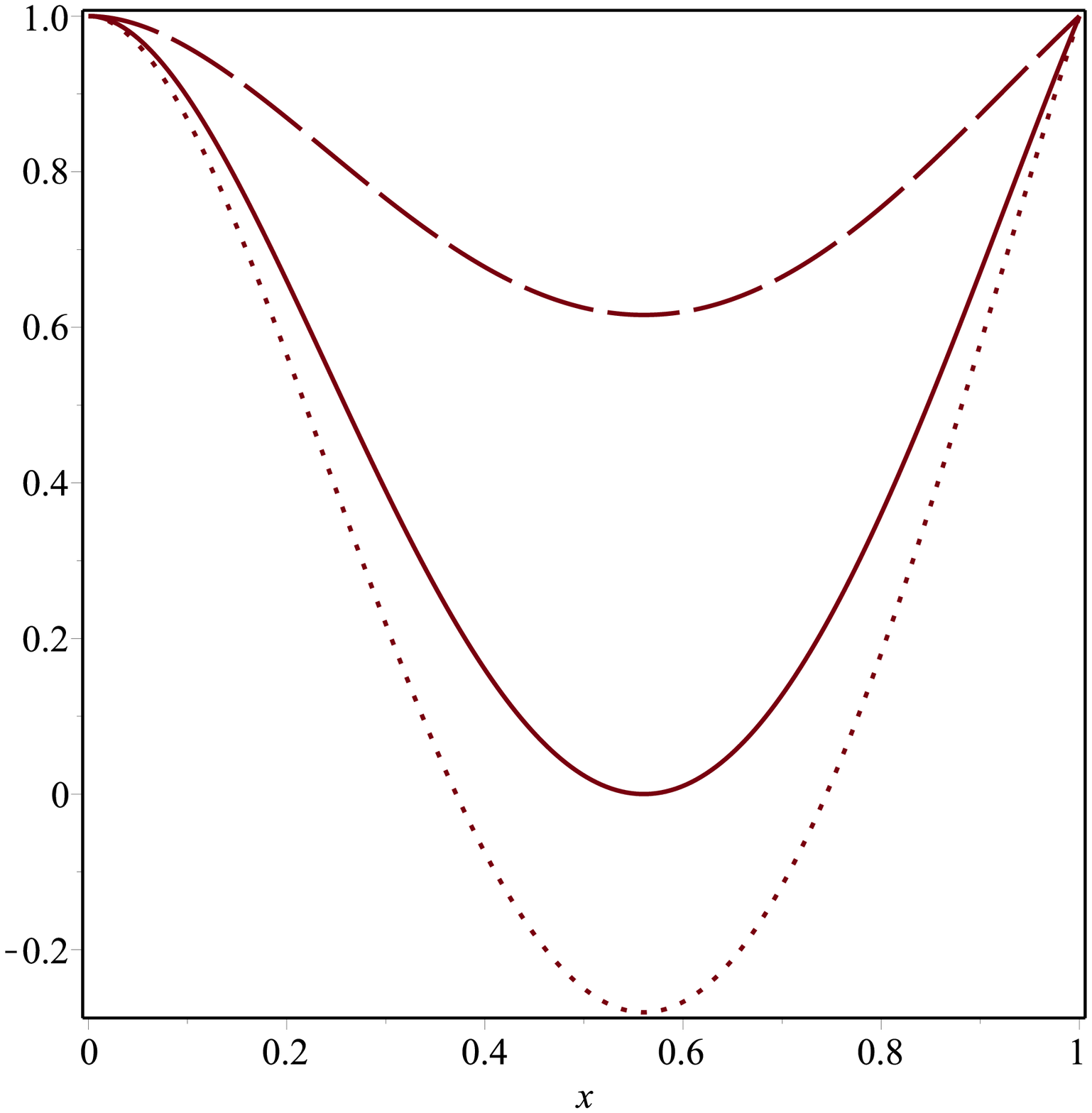}
    \includegraphics[width=0.3\textwidth]{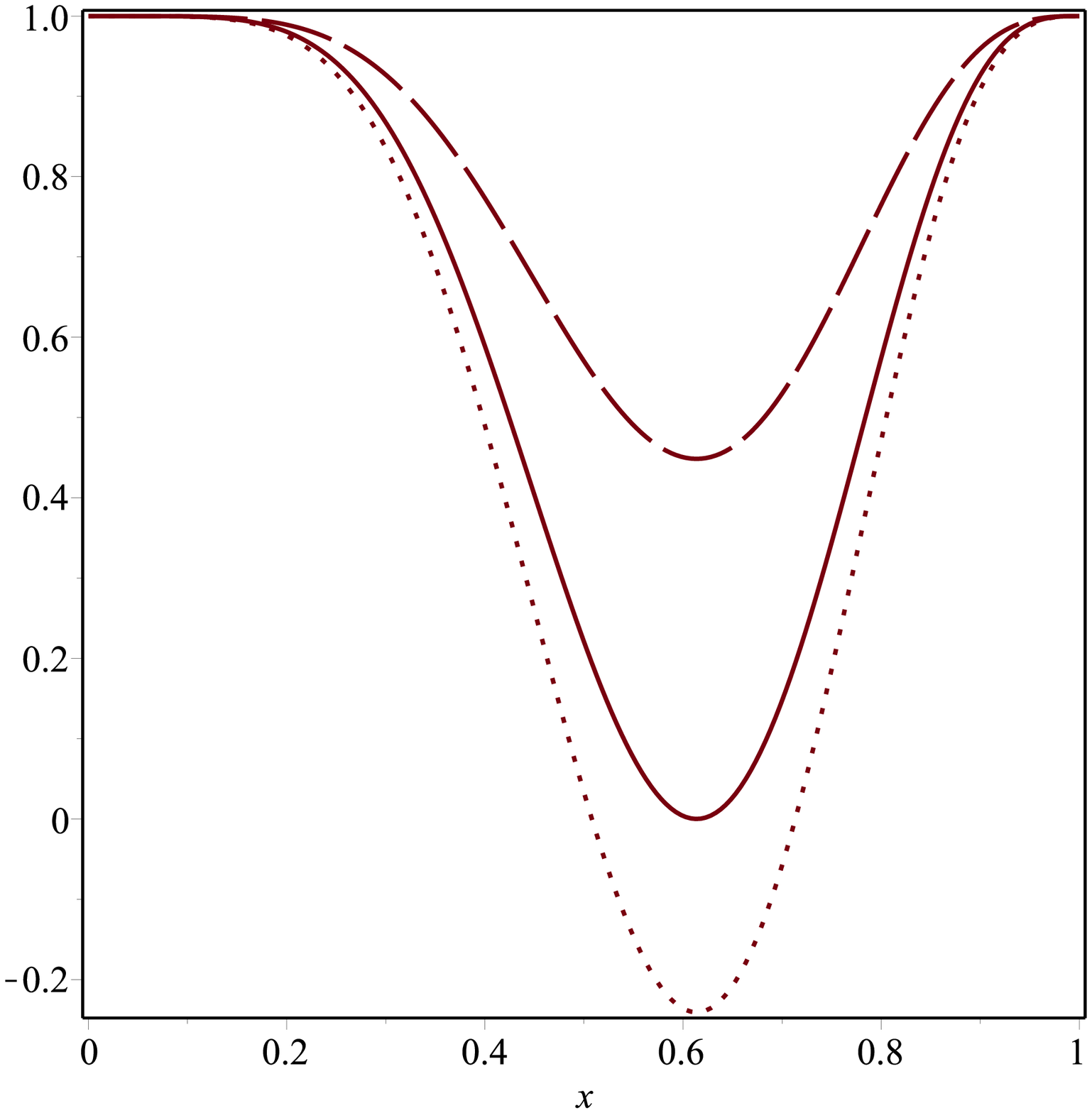}
\caption{\label{figure2}
Plots of the metric coefficient $g_{00}$ for different models and 3 different values of $\mu$. For each model, $\mu>\mu_c$ (dotted line) gives rise to two distinct horizons, $\mu=\mu_c$ (solid line) corresponds to the case of an extreme black hole  while $0<\mu<\mu_c$ (long-dashed line) produces a self-gravitating DM droplet. First row from left: Dehnen $(1,4,0)$, PS and MIS models. Second row from left: $(k,n)=(1,2)$,  $(k,n)=(2,1)$ and $(k,n)=(2,3)$ models.}
\end{figure}

This section is devoted to the construction of black hole solutions from the Zhao density profile. We will assume that the mass density function associated to the gravitational object is static, spherically symmetric and given by (\ref{dieMasse}). Moreover, the gravitational source has total mass $M$. We will focus our attention on those models characterized by triples $(\alpha,\beta,\gamma)$ satisfying the constraint (\ref{general_condition}). Furthermore, we consider the following ansatz for the metric
\begin{equation}\label{ansatz}
    ds^2=g_{00}(r)dt^2-g_{00}^{-1}(r)dr^2-r^2(d\vartheta^2+\sin{\vartheta}d\varphi^2), \quad 0\leq \vartheta \leq \pi, \quad 0 \leq \varphi < 2\pi. 
\end{equation}
In order to find the unknown metric coefficient $g_{00}$, we consider the Einstein field equations 
\begin{equation}\label{EFE}
    R_{\mu \nu}= -8 \pi \Big( T_{\mu\nu}-\frac{T}{2}g_{\mu\nu} \Big), \quad T= g^{\mu \nu} T_{\mu \nu}
\end{equation}
in the presence of a static, anisotropic fluid for which the energy-momentum tensor is given by 
\begin{equation}\label{EMT}
    T^\mu{}_\nu=\mathrm{diag}(\rho,-p_r,-p_\perp, -p_\perp), \quad p_r \neq p_\perp.
\end{equation}
Here, the density function $\rho$ is chosen according to (\ref{rho}) while $p_r$ and $p_\perp$ are the radial and tangential pressures, respectively. If we use the conservation equation $T^{\mu\nu}{}_{;\nu}=0$ with $\mu=1$, i.e.
\begin{equation}
    -\frac{d p_r}{dr}=\frac{1}{2}g^{00} \frac{d g^{00}}{d r}(p_r+\rho)+\frac{2}{r}(p_r-p_\perp)
\end{equation}
in the $(\mu,\nu)= (2,2)$ equation in (\ref{EFE}), we end up with the  Tolman-Oppenheimer-Volkoff equation
\begin{equation}\label{TOV}
    \frac{d p_r}{dr}+(\rho+p_r)\frac{m(r)+4 \pi r^3 p_r}{r[r-2m(r)]}+\frac{2}{r}(p_r-p_\perp)=0,
\end{equation}
where the mass function $m$ is represented by (\ref{massf}). Note that by means of the coordinate transformation (\ref{xr}) and under the parametric constraint (\ref{general_condition}) it is possible to express $m$ in terms of an incomplete Beta function as in (\ref{dieMasse}). Similarly as in  \cite{Piero,BHnoncomm,DavidePiero,PIEROBOSS},  we introduce an EOS of de Sitter type
\begin{equation}\label{p_r}
    p_r=-\rho=-\frac{\alpha M}{4\pi r_0^3B(p+1,q+1)}\left[1+\left(\frac{r}{r_0}\right)^\alpha\right]^{-\frac{\beta}{\alpha}},
\end{equation}
where according to the previous section we set $\gamma=0$. Note that such an EOS has been often used in connection with models of regular BHs \cite{d1,d2,d3,d4,d5,d6,d7,d8,Mazur,Felten,negativeP,EPJCus} which do not need to be mini BHs because no scale factor appears. If we impose (\ref{p_r}) in (\ref{TOV}), we find that the tangential pressure is 
\begin{equation}\label{p_TT}
p_\bot=-\rho-\frac{r}{2}\frac{d\rho}{dr}=-\frac{\alpha M}{4\pi r_0^3B(p+1,q+1)}\left[1+\left(\frac{r}{r_0}\right)^\alpha\right]^{-\frac{\beta}{\alpha}-1}\left[1+\left(1-\frac{\beta}{2}\right)\left(\frac{r}{r_0}\right)^\alpha\right].
\end{equation}
We observe that both pressures have a finite value at $r=0$ where
\begin{equation}
p_r(0)=p_\bot(0)=\frac{\alpha M}{4\pi r_0^3B(p+1,q+1)}.
\end{equation}
The radial pressure is a monotone increasing function which vanishes as $r\to\infty$. Moreover, the tangential pressure vanishes at the radius
\begin{equation}
r=r_0\left(\frac{2}{2-\beta}\right)^\frac{1}{\alpha}
\end{equation}
after which it becomes positive, exhibits a maximum at 
\begin{equation}
r_m=r_0\left(\frac{\alpha+2}{2-\beta}\right)^\frac{1}{\alpha},\quad
p_\bot(r_m)=\frac{\alpha^2 M}{8\pi r_0^3B(p+1,q+1)}\left(\frac{\beta-2}{\alpha+\beta}\right)^\frac{\alpha+\beta}{\alpha}
\end{equation}
and becomes zero as $r\to\infty$. If we solve the Einstein field equations (\ref{EFE}) with metric, energy-momentum tensor, and pressures $p_r$ and $p_\bot$ as given by (\ref{ansatz}), (\ref{EMT}), (\ref{p_r}) and (\ref{p_TT}), respectively, together with the requirement that the metric goes over into the Minkowski metric asymptotically at infinity, we end up with the line element (\ref{ansatz}) with
\begin{equation}\label{g_00r}
    g_{00}(r)= 1-\frac{2m(r)}{r},
\end{equation}
where the mass function can be obtained from (\ref{dieMasse}) by switching back to the radial variable. However, in order to study the properties of $g_{00}$, it results convenient to express it in the variable $x$, namely
\begin{equation}\label{g_00x}
    g_{00}(x)= 1-\frac{2M}{r_0 B(p+1,q+1)}\left(\frac{1-x}{x}\right)^{\frac{1}{\alpha}}B_x(p+1,q+1).
\end{equation}
We have summarized in Table~\ref{overview2} the analytic expressions for the mass function and the corresponding $g_{00}$ in the models considered in the present work.
\begin{table}[!ht]
\caption{Analytic results for the mass function and the metric coefficient $g_{00}$ expressed as functions of the variable $x$ defined in (\ref{xr}). For the abbreviations we refer to Table~\ref{overview}.}
\begin{center}
\begin{tabular}{ | l | l | l | l|l|}
\hline
Model          &  $m(x)$  & $g_{00}(x)$    \\ \hline
Dehnen (1,4,0) &  $Mx^3$     & $1-\frac{2M}{r_0}(1-x)x^2$            \\ \hline
PS             &  $\frac{2M}{\pi}\left[\frac{\pi}{4}-\sqrt{x(1-x)}+\frac{1}{2}\sin^{-1}{(2x-1)}\right]$     & $1-\frac{4M}{\pi r_0}\left[\sqrt{\frac{1-x}{x}}\left(\frac{\pi}{4}+\frac{1}{2}\sin^{-1}{(2x-1)}\right)+x-1\right]$          \\ \hline
MIS            &  $\frac{2}{3}Mx\sqrt{x}$     & $1-\frac{2M}{r_0}x\sqrt{1-x}$            \\ \hline
$(k,n)$-model $\gamma=0$  & $\frac{M}{B(3n,k)}B_x(3n,k)$ & $1-\frac{2M}{r_0 B(3n,k)}\left(\frac{1-x}{x}\right)^n B_x(3n,k)$\\ \hline
\end{tabular}
\label{overview2}
\end{center}
\end{table}
In order to study the regularity of the metric coefficient $g_{00}$, it is convenient to introduce the scaled mass $\mu=M/r_0$. Moreover, by $\mu_c$ we denote the critical value of the scaled mass such that $g_{00}$ has two coinciding roots (see Fig.~\ref{figure1}). If $\mu>\mu_c$, there are two distinct real roots and no real roots for $0<\mu<\mu_c$. The picture emerging from Table~\ref{tableEins} is that depending on the value of the mass parameter all models treated here can describe a black hole with two horizons, an extreme black hole or a self-gravitating DM droplet. This behaviour can be explicitly seen in Figure~\ref{figure2}. Numerical values of $\mu_c$ and the corresponding horizon $x_e$ are displayed in Table~\ref{tableEins}. In the extreme and non-extreme regimes, i.e. $\mu\geq\mu_c$, the behaviour of the metric coefficient $g_{00}$ as $x\to 0$ has been displayed in Table~\ref{overview3} where all expansions around the point $x=0$ are quite straightforward made exception for that one related to the $(k,n)$-model where we made use of the following result in \cite{Abra}
\begin{equation}
\int_0^x s^{3n-1}(1-s)^{3k-1}~ds=\frac{x^{3n}}{3n}{}_2 F_1(3n,1-k;1+3n;x),
\end{equation}
which allows to express the incomplete Beta function in terms of the hypergeometric function. We discover that $g_{00}$ never blows up at $x=0$. The regularity of the metric at $x=0$ can also be analyzed by means of the Kretschmann scalar. Since we verified that the latter stays finite as $x\to 0$ thus confirming the outcome of the previous analysis, we do not need to go into more detail about that. Hence, instead of having a point of infinite curvature at $x=0$, there is always a regular core which is of de Sitter type only in the Dehnen and $(k,1)$-models. This finding signalizes that the effect of coupling the DM models treated here with an anisotropic fluid characterized with an EOS of the type $p_r=-\rho$ is that of replacing the curvature singularity with a regular region. Finally note that also in the regime $0<\mu<\mu_c$ there is no naked singularity and a self-gravitating DM droplet emerges in this case. 
\begin{table}
\caption{Exact and numerical estimates for the critical values of the scaled mass $\mu_c$ along with the corresponding values of the horizon denoted by $x_e$ where $g_{00}$ exhibits two coinciding roots.}
\begin{center}
\begin{tabular}{ | l | l | l | l|}
\hline
Model             & $\mu_c$            & $x_e$        \\ \hline
Dehnen $(1,4,0)$  & 27/8               & 2/3          \\ \hline
PS                & 2.211570492        & 0.769136195  \\ \hline
MIS               & $3\sqrt{3}/4$      & 2/3          \\ \hline
$(k,n)=(1,2)$     & 729/32             & 2/3          \\ \hline
$(k,n)=(1,3)$     & 19683/128          & 2/3          \\ \hline
$(k,n)=(2,1)$     & 1.561898379        & 0.560434506  \\ \hline
$(k,n)=(2,2)$     & 7.090679869        & 0.596349742  \\ \hline
$(k,n)=(2,3)$     & 36.26612447        & 0.613667840  \\ \hline
\end{tabular}
\label{tableEins}
\end{center}
\end{table}
\begin{table}[!ht]
\caption{Behaviour of the metric coefficient $g_{00}$ in a neighbourhood of $x=0$. For the abbreviations we refer to Table~\ref{overview}.}
\begin{center}
\begin{tabular}{ | l | l | l | l|l|}
\hline
Model          &  $g_{00}(x)$    \\ \hline
Dehnen (1,4,0) &  $1-\frac{2M}{r_0}x^2+\mathcal{O}(x^3)$ \\ \hline
PS             &  $1-\frac{8M}{3\pi r_0}x+\frac{8M}{15\pi r_0}x^2+\mathcal{O}(x^3)$\\ \hline
MIS            & $1-\frac{2M}{r_0}x+\frac{M}{r_0}x^2++\mathcal{O}(x^3)$\\ \hline
$(k,n)$-model $\gamma=0$  & $1-\frac{2M}{3nr_0 B(3n,k)}\left[x^{2n}+\mathcal{O}(x^{2n+1})\right]$\\ \hline
\end{tabular}
\label{overview3}
\end{center}
\end{table}
In Table~\ref{table2}, we considered the central galactic BH in the Milky Way. More precisely, we gave numerical estimates for the scale factor $r_0$ and the corresponding extreme horizon in the case $\mu=\mu_c$. We observe that for the models represented by Dehnen $(1,4,0)$, PS and MIS the degenerate horizon never exceeds $10$ times the radius of the sun while the scale factor $r_0$ is consistently smaller than the extreme horizon. Note that in the case of the $(k,n)$ models, the extreme horizon shrinks as $n$ increases.
\begin{table}
\caption{Typical numerical values of the scale factor $r_0$ and the corresponding position of the extreme horizon for choices of $\mu_c$ as given in Table~\ref{tableEins} and a BH mass $M=4.1\cdot 10^6 M_\odot$.}
\begin{center}
\begin{tabular}{ | l | l | l | l | l|}
\hline
$\mbox{Model}$ & $\frac{r_{0,c}}{r_\odot}=\frac{G_N M}{c^2\mu_c}$   & $\frac{r_e}{r_\odot}=\frac{r_{0,c}}{r_\odot}\left(\frac{x_e}{1-x_e}\right)^{1/\alpha}$ \\ \hline
Dehnen (1,4,0) & 2.58 & 5.15  \\ \hline
PS             & 3.93 & 7.18 \\ \hline
MIS            & 6.69 & 9.47 \\ \hline
$(k,n)=(1,2)$  & 0.38 & 1.53 \\ \hline
$(k,n)=(1,3)$  & 0.06 & 0.45 \\ \hline
\end{tabular}
\label{table2}
\end{center}
\end{table}
\begin{figure}[!ht]\label{hr}
\includegraphics[scale=0.50]{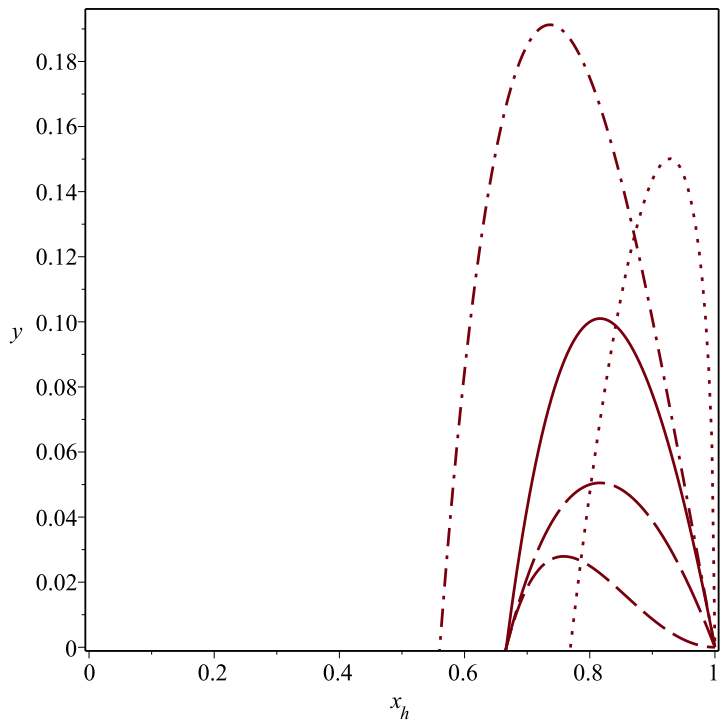}
\caption{\label{hr1}
Plot of the Hawking temperature $4\pi r_0 T_H$ versus the radius of the event horizon $x_h$. We have $T_H=0$ for $x_h=x_e$, i.e. when the event horizon coincides with the horizon of the corresponding extremal black hole. Legend: solid line Dehnen $(1,4,0)$, dotted line PS, long-dashed line MIS, dashed line $(k,n)=(1,2)$ and dash-dotted line $(k,n)=(2,1)$.}
\end{figure}
We conclude this section by considering the Hawking temperature for this new class of black holes. The black hole temperature can be computed from the formula \cite{Piero}
\begin{equation}
T_H=\frac{1}{4\pi}\left.\frac{dg_{00}}{dr}\right|_{r=r_h}=\frac{\alpha}{4\pi r_0}\left.x^{1-\frac{1}{\alpha}}(1-x)^{1+\frac{1}{\alpha}}\frac{dg_{00}}{dx}\right|_{x=x_h}.
\end{equation}
Here, $r_h$ and $x_h$ represents the position of the event horizon depending whether we use the radial variable or the transformed radial variable $x$ defined in (\ref{xr}). Moreover, the mass parameter $\mu$ has been expressed in terms of $x_h$ by means of the horizon equation $g_{00}(x_h)=0$. Fig.~\ref{hr1}, which displays the temperature $T_H$ as a function of $x_h$, indicates that a Zhao inspired black hole increases its temperature, as the horizon radius gets smaller, until $T_H$ exhibits a maximum after which $T_H$ decreases sharply and vanishes exactly at the radius of the extremal black hole, that is at $x_h=x_e$. We recall that for an extreme black hole the Hawking temperature must be zero because the metric component $g_{00}$ has a double root at $x=x_e$. Hence, differently as in Schwarzschild where $T_H$ blows up as the radius of the event horizon shrinks, we find  that the final fate of the evaporation process is a zero temperature extremal black hole whose final configuration is entirely controlled in addition to the black hole mass by the parameter $\alpha$ in the Dehnen $(1,4,0)$, PS and MIS models while it also depends on the parameter $\beta$ in the case of the $(k,n)$ model. We remind the reader that according to \cite{DavidePiero}, a final configuration with finite temperature inhibits any relevant back reaction, i.e a self-interaction of the radiated energy with its source. As a consequence our solution is stable versus back reaction and therefore, it can describe the entire black hole life until the final configuration is reached. Finally, the presence of an inner Cauchy horizon may signalize that the inner region of our  black holes is unstable, however one may follow the procedure outlined in \cite{DavidePiero} to show the stability of the Zhao inspired black hole interior.

\section{Dark Matter inspired galactic black holes}\label{DMG}
In this section we want to understand if the black hole located at the centre of our galaxy whose mass and Schwarzschild radius are  $M_{BH}=4.1\cdot 10^6~M_\odot$ and $R_{BH}=2G_N M_{BH}/c^2=17.4~R_\odot=3.92\cdot 10^{-7}$ pc, respectively \cite{Ghez1,Ghez2}, can be modelled in terms of the diffuse DM black holes derived in the previous section. This requires that we find estimates for the relevant parameters in the models. More precisely, this is accomplished by imposing first that the total mass $M$ entering in the line element (\ref{ansatz}) through the metric coefficient $g_{00}$ in (\ref{g_00x})  coincides with $M_{BH}$ followed by the condition that the mass function $m$ provides a good approximation for $M_{BH}$ when it is evaluated at the minimum $r_{min}$ of the effective potential for a massive test particle. In other words, we require that
\begin{equation}\label{condizione}
1-\frac{m(r_{min})}{M_{BH}}\leq 10^{-2}.
\end{equation}
In the analysis to follow, we focus on the PS, MIS, Dehnen $(1,4,0)$ and some examples of  $(k,n)$ models. If we replace the corresponding mass function for each of the aforementioned models into (\ref{condizione}), we can solve (\ref{condizione}) numerically and express the solution in the form 
\begin{equation}\label{cond2}
\frac{r_{min}}{r_0}\geq\lambda,
\end{equation}
where $\lambda$ is a lower bound whose numerical value depends on the particular model considered (see Table~\ref{tableCond}).  
\begin{table}
\caption{Typical numerical values of the lower bound $\lambda$ appearing in the inequality (\ref{cond2}).}
\begin{center}
\begin{tabular}{ | l | l | l | l|}
\hline
Model             & $\lambda$          \\ \hline
Dehnen $(1,4,0)$  & 298                \\ \hline
PS                & 120                \\ \hline
MIS               & 12                 \\ \hline
$(k,n)=(1,2)$     & $3.56\cdot 10^5$            \\ \hline
$(k,n)=(1,3)$     & $7.17\cdot 10^8$          \\ \hline
$(k,n)=(2,1)$     & 23                 \\ \hline
$(k,n)=(2,2)$     & 1860        \\ \hline
$(k,n)=(2,3)$     & $2.54\cdot 10^5$        \\ \hline
\end{tabular}
\label{tableCond}
\end{center}
\end{table}
Note that (\ref{cond2}) alone is not sufficient in order to find the optimal choice of the parameter $r_0$ such that the Schwarzschild effective potential and the effective potential of our diffused black hole share the same minimum and at the same time they both coincide  in a large neighbourhood of it and asymptotically away. As we will see, one first needs to identify the optimal $r_0$ and then, verify that (\ref{cond2}) is fulfilled. Moreover, it turns out that once $r_0$ is determined, the matching of the effective potentials at the minimum remains stable over a large range of the angular momentum of the test particle. In this regard, we recall that in the case of a spherically symmetric metric such as (\ref{ansatz}) with $g_{00}$ given as in (\ref{g_00r}) the radial geodesic can be cast into the form of an energy conservation equation \cite{Fliessbach}
\begin{equation} \label{C}
\frac{\dot{r}^2}{2}+V_{eff}(r)=const\equiv C,
\end{equation}
where the dot means differentiation with respect to the proper time or an affine parameter, depending whether a massive or a  massless particle is considered $V_{eff}$ denotes the effective potential associated to the geometry described by the line element (\ref{ansatz}), i.e. 
\begin{equation}\label{realmassive}
V_{eff}(r)=-\epsilon\frac{m(r)}{r}+\frac{\ell^2}{2r^2}\left(1-\frac{2m(r)}{r}\right),\quad\epsilon=\left\{
\begin{array}{cc}
1 &\mbox{if}~m_p\neq 0,\\
0 &\mbox{if}~m_p=0,
\end{array}
\right.
\end{equation}
Here, $m_p$ stands for the mass of a test particle and $\ell$ is its total angular momentum per unit mass. We also recall that the effective potential in the case of the Schwarzschild metric can be directly obtained from (\ref{realmassive}) by replacing the mass function with $M_{BH}$. Let $r_s=2M_{BH}$. If we rescale the radial variable and the angular momentum per unit mass as $y=r/r_s$ and $L=\ell/r_s$, the Schwarzschild effective potential becomes
\begin{equation}\label{VSmassive}
V_{eff,S}(y)=-\frac{\epsilon}{2y}+\frac{L^2}{2y^2}-\frac{L^2}{2y^3}.
\end{equation}

\subsection{The Dehnen model $(1,4,0)$}
\begin{figure}[!ht]\label{Dmodel}
\centering
    \includegraphics[width=0.3\textwidth]{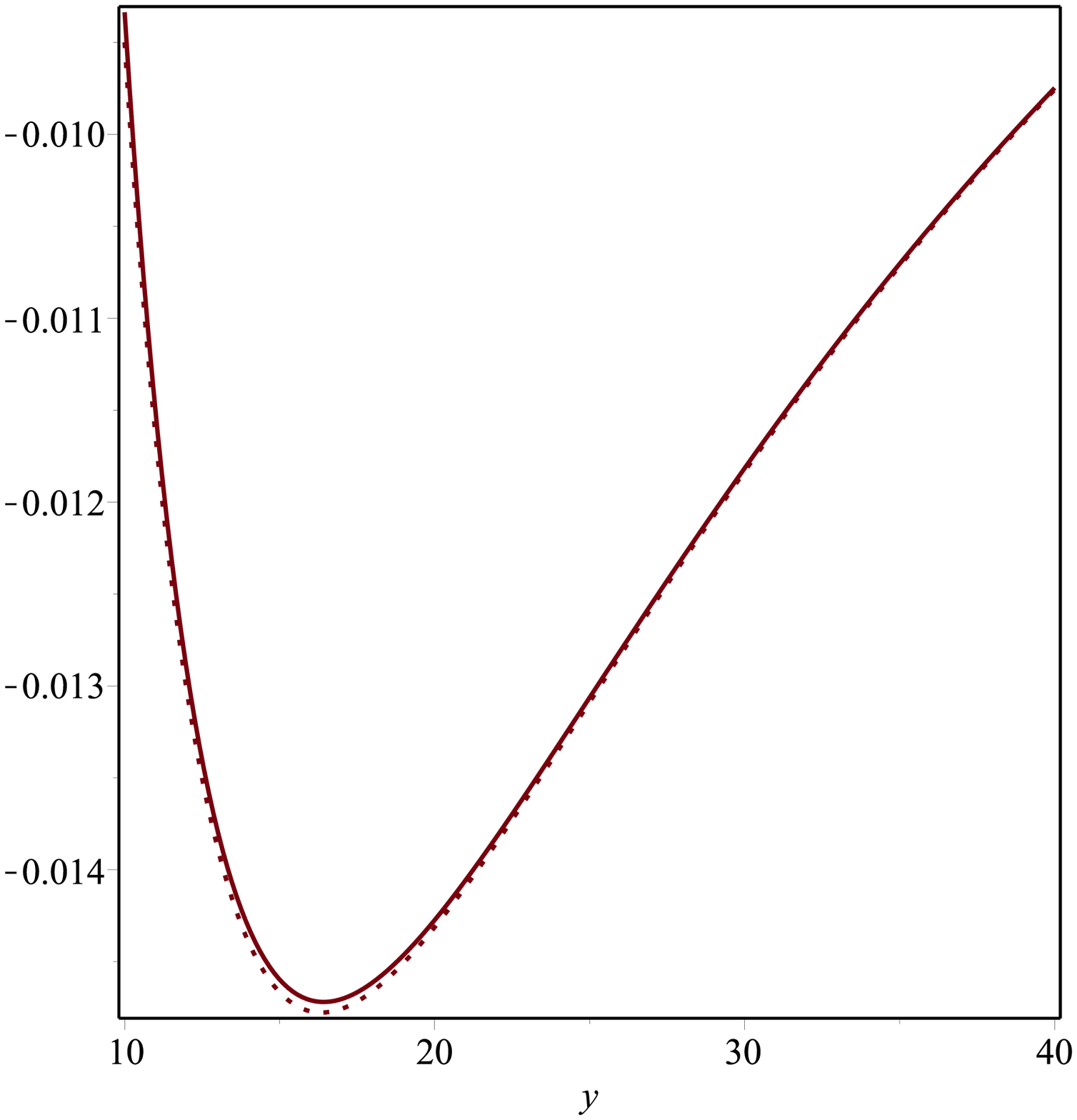}
    \includegraphics[width=0.3\textwidth]{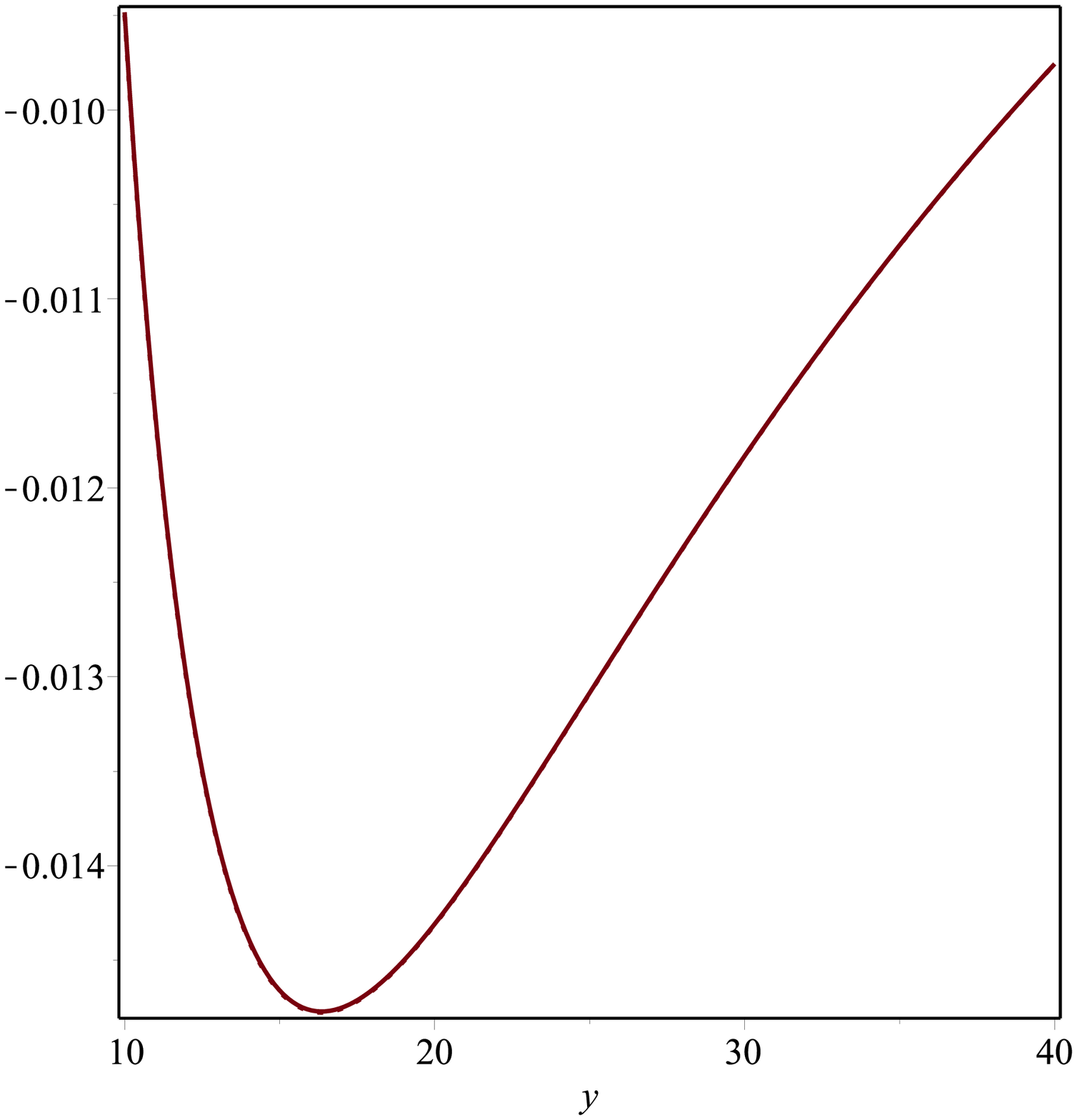}
\caption{\label{figure4}
Plots of the effective potential (\ref{VeffD}) (solid line) and the Schwarzschild effective potential (\ref{VSmassive}) (dotted line) in the massive case for $L=3$. The figure on the left refers to the case $\sigma=0.01$ while the one on the right to $\sigma=0.001$ for which the two potentials agree remarkably well in a large neighbourhood of the minimum.}
\end{figure}

\begin{figure}[!ht]\label{Dmodelphoton}
\centering
    \includegraphics[width=0.3\textwidth]{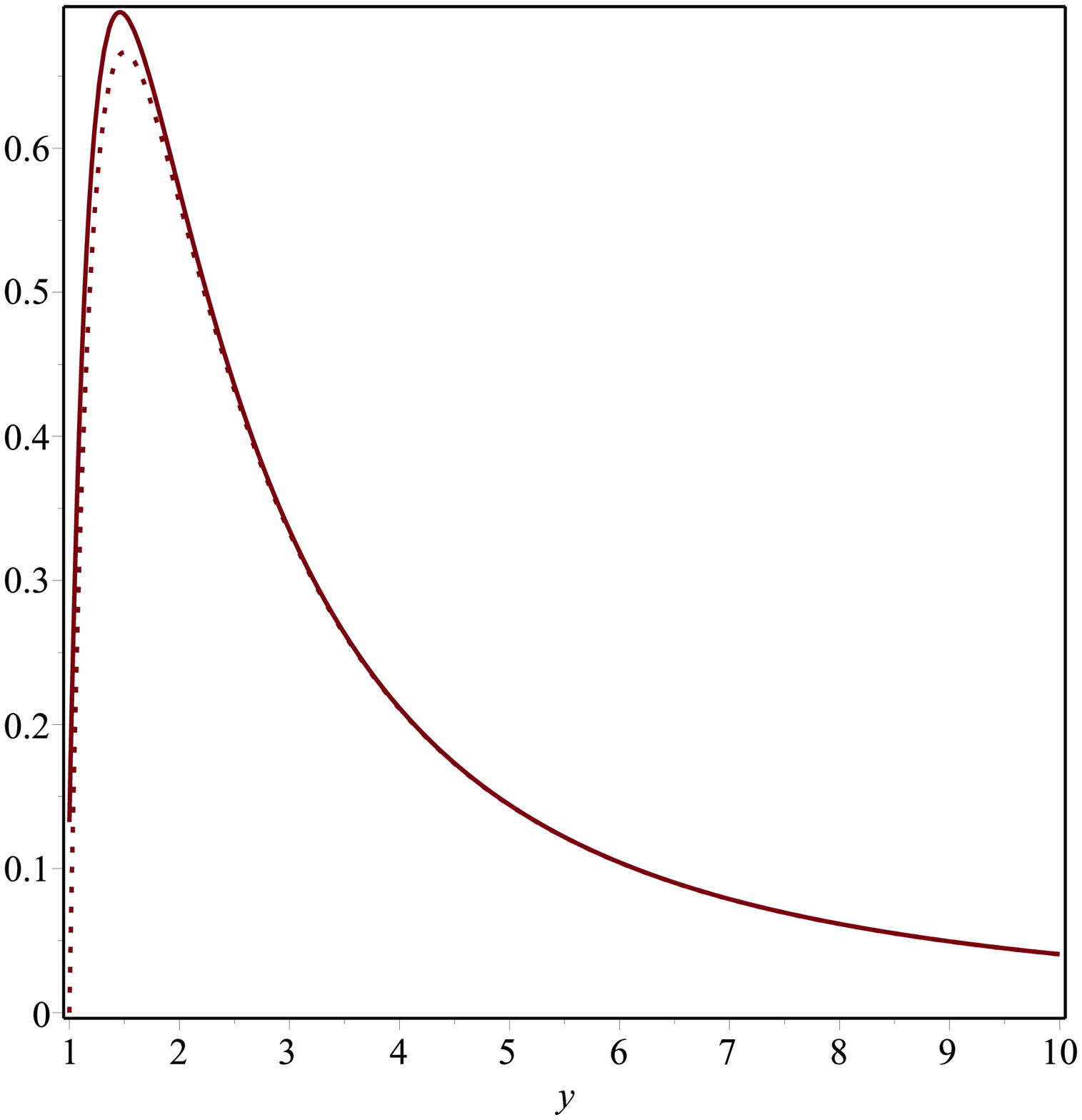}
    \includegraphics[width=0.3\textwidth]{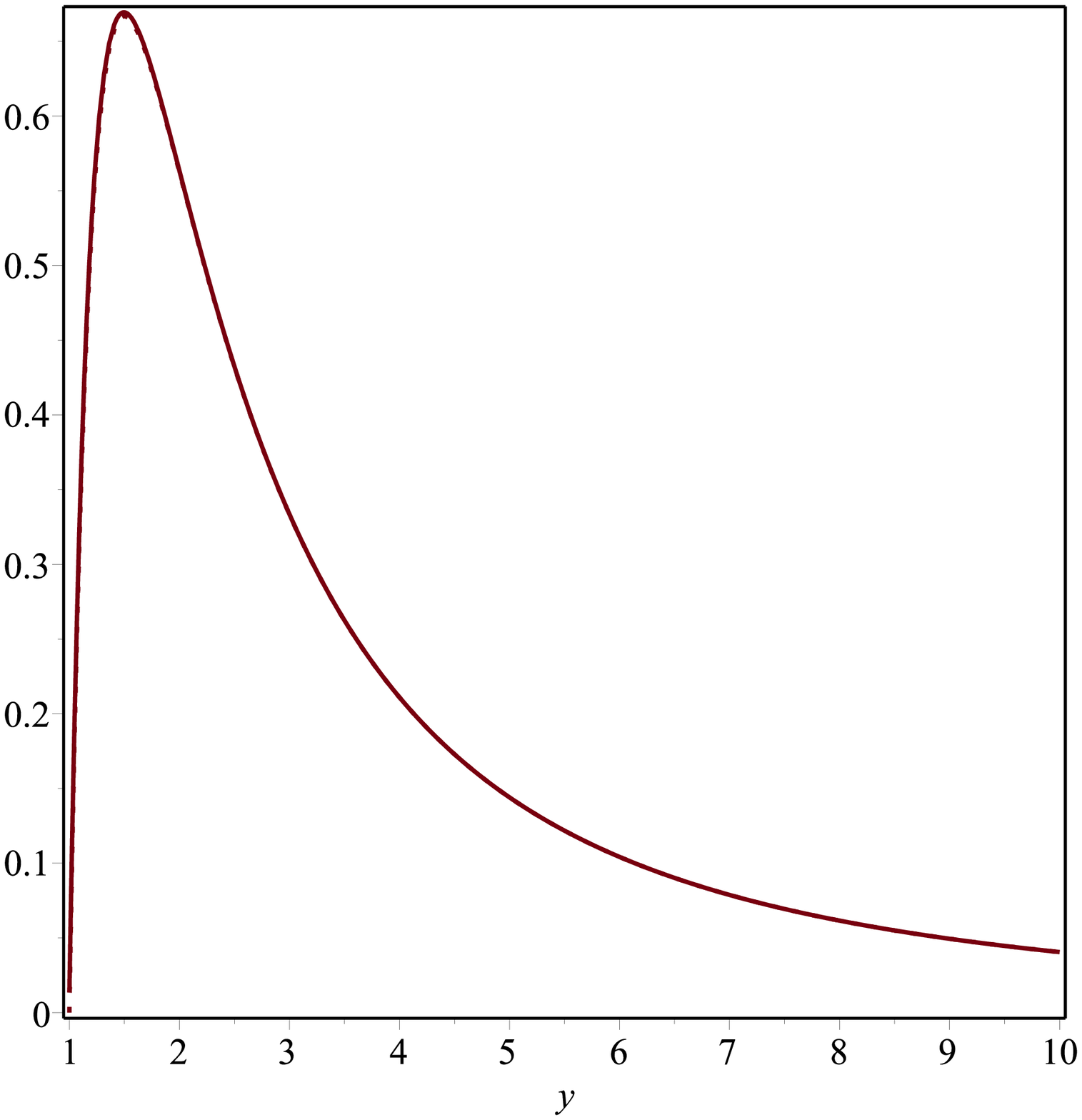}
\caption{\label{figure5}
Plots of the effective potential (\ref{VeffD}) (solid line) and the Schwarzschild effective potential (\ref{VSmassive}) (dotted line) in the massless case for $L=3$. The figure on the left refers to the case $\sigma=0.01$ while the one on the right to $\sigma=0.001$ for which both potentials shares the same photon sphere and both black holes have the same event horizon.}
\end{figure}

In this case, the density and mass function are 
\begin{equation}
\rho(r)=\frac{3M_{BH}}{4\pi r_0^3\left(1+\frac{r}{r_0}\right)^4},\quad
m(r)=M_{BH}\frac{r^3}{r_0^3\left(1+\frac{r}{r_0}\right)^3}
\end{equation}
Let $y=r/r_s$ and $L=\ell/r_s$. Moreover, assume that $r_0/r_s=\sigma$ where $\sigma$ is a free parameter to be chosen so that in the massive case both potentials $V_{eff,S}$ and $V_{eff}$ have the same minimum and they agree in a large neighbourhood of it and asymptotically away. We find that
\begin{equation}\label{VeffD}
V_{eff}(y)=V_{eff,S}(y)+\sigma\frac{\sigma^2+3\sigma y+y^2}{(\sigma+y)^3}\left(\frac{\epsilon}{2y}+\frac{L^2}{2y^3}\right).
\end{equation}
According to Table~\ref{tableEins}, there will be a black hole with two distinct horizons if $\mu>27/8$ which is equivalent to the condition $\sigma<4/27$. Moreover, (\ref{cond2}) requires that $y_{min}/\sigma>298$. As we will see here below, these constraints are easily satisfied. For $\epsilon=0$ the radius of the photon sphere is $y_\gamma=3/2-4\sigma+\mathcal{O}(\sigma^2)$ while the event horizon is located at $y_h=1-3\sigma+\mathcal{O}(\sigma^2)$. From Fig.~\ref{figure4} we observe that in the massive case with $L=3$ the choice $\sigma=0.001$ already ensures that both potentials match well both at the minimum and in a large interval containing it. Moreover, Table~\ref{tableDm} indicates that the choice of $\sigma$ is not sensitive to the angular momentum $L$ of the test particle. Finally, we observe in Fig.~\ref{figure5} that in the massless case with $L=3$ and for the choice $\sigma=0.001$ both potentials practically shares the same photon sphere and both black holes have almost the same event horizon. 
\begin{table}
\caption{Dehnen model (1,4,0): numerical values of the minima $y_{min}$ and $y_{min,s}$ in the effective potentials (\ref{VeffD}) and (\ref{VSmassive}) for $\sigma=0.001$ and different values of $L$.}
\begin{center}
\begin{tabular}{ | l | l | l | l|}
\hline
$L$             & $y_{min}$        &$y_{min,s}$          \\ \hline
2             & 6.011            &6                    \\ \hline
3             & 16.355           &16.348               \\ \hline
4             & 30.429           &30.422       \\ \hline
5             & 48.458           &48.452       \\ \hline
10            & 198.495          &198.489      \\ \hline
50            & 4998.506         &4998.500       \\ \hline
100           & 19998.506        &19998.500    \\ \hline
\end{tabular}
\label{tableDm}
\end{center}
\end{table}

\subsection{The PS model $(2,4,0)$}
Taking into account that the density and mass function are 
\begin{equation}
\rho(r)=\frac{M_{BH}}{\pi^2 r_0^3\left(1+\frac{r^2}{r_0^2}\right)^2},\quad
m(r)=\frac{2M_{BH}}{\pi}\left[\arctan{\left(\frac{r}{r_0}\right)}+\frac{r}{r_0\left(1+\frac{r^2}{r_0^2}\right)}\right]
\end{equation}
and letting $y=r/r_s$, $L=\ell/r_s$ and $r_0/r_s=\sigma$, the effective potential reads
\begin{equation}\label{VeffPS}
V_{eff}(y)=\frac{L^2}{2y^2}-\left(\frac{\epsilon}{\pi y}+\frac{L^2}{\pi y^3}\right)\left[\arctan{\left(\frac{y}{\sigma}\right)}+\frac{y}{\sigma\left(1+\frac{y^2}{\sigma^2}\right)}\right].
\end{equation}
The free parameter $\sigma$ must be picked so that in the massive case both potentials $V_{eff,S}$ and $V_{eff}$ have the same minimum and they agree in a large neighbourhood of it and asymptotically away. From Table~\ref{tableEins}, we see that we have a black hole with two distinct horizons if $\mu>\mu_c$ which is equivalent to the condition $\sigma<0.2261$. Moreover, (\ref{cond2}) requires that $y_{min}/\sigma>120$. Also in the present case it turns out that both constraints are easy to fulfill.
\begin{figure}[!ht]\label{PSmodel}
\centering
    \includegraphics[width=0.3\textwidth]{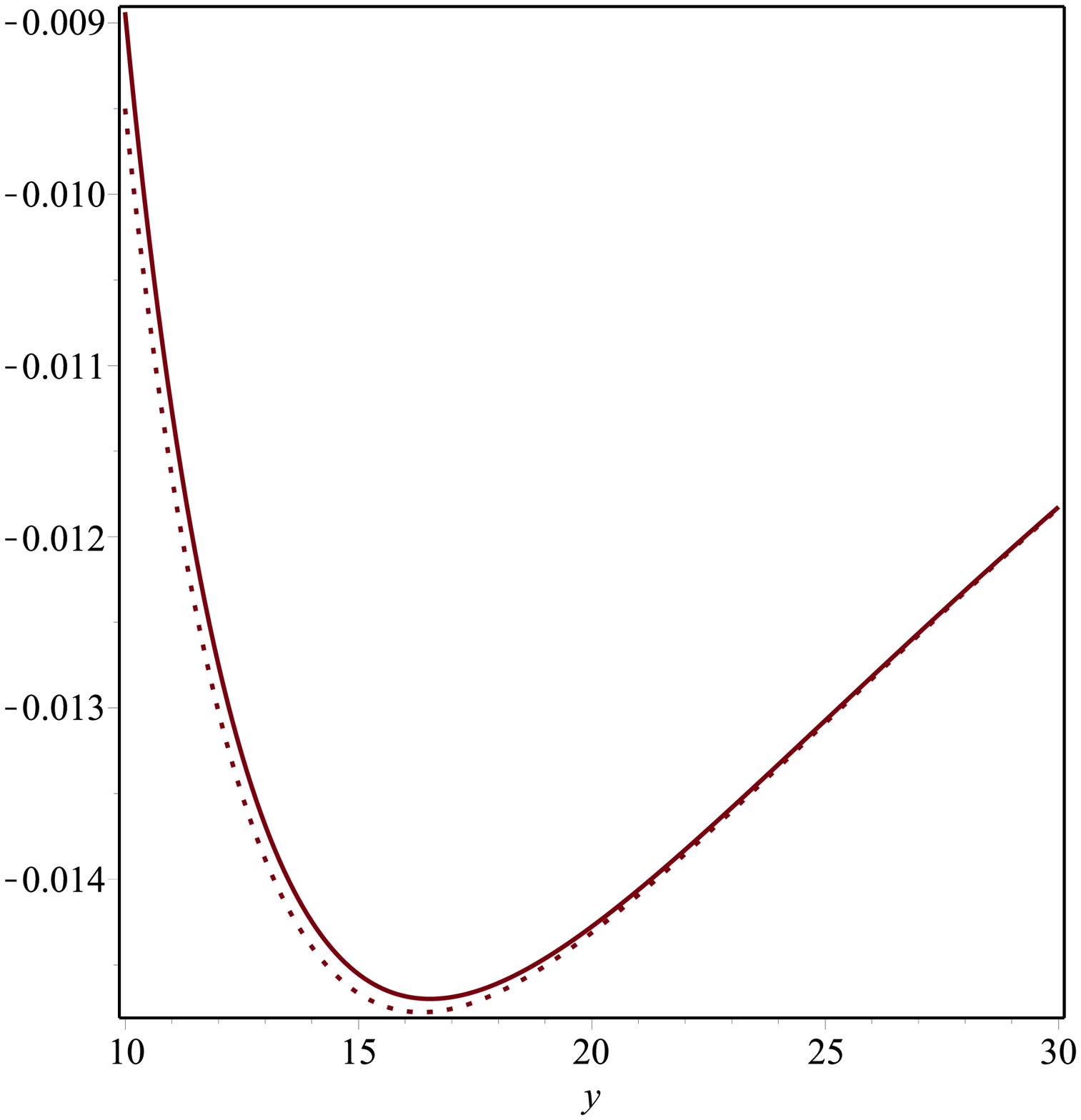}
    \includegraphics[width=0.3\textwidth]{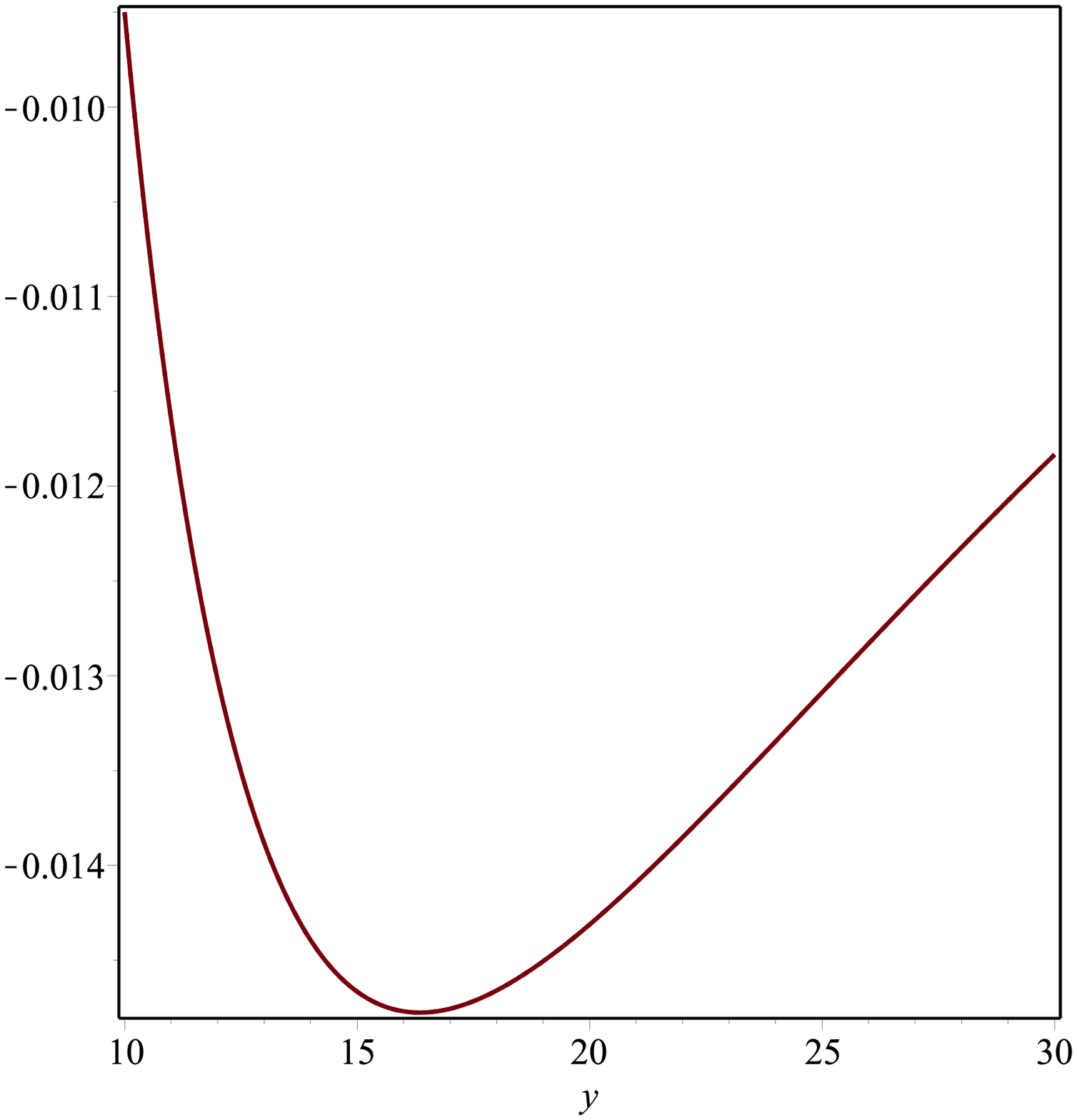}
\caption{\label{figure6}
Plots of the effective potential (\ref{VeffPS}) (solid line) and the Schwarzschild effective potential (\ref{VSmassive}) (dotted line) in the massive case for $L=3$. The figure on the left refers to the case $\sigma=3$ while the one on the right to $\sigma=0.1$ for which the two potentials agree remarkably well in a large neighbourhood of the minimum.}
\end{figure}
\begin{table}
\caption{PS model: numerical values of the minima $y_{min}$ and $y_{min,s}$ in the effective potentials (\ref{VeffPS}) and (\ref{VSmassive}) for $\sigma=0.1$ and different values of $L$ .}
\begin{center}
\begin{tabular}{ | l | l | l | l|}
\hline
$L$           & $y_{min}$        &$y_{min,s}$          \\ \hline
2             & 6.000082481      &6.000000000                    \\ \hline
3             & 16.34847665      &16.34846923               \\ \hline
4             & 30.42220709      &30.42220510       \\ \hline
5             & 48.45207956      &48.45207880       \\ \hline
10            & 198.4885781      &198.4885780      \\ \hline
50            & 4998.499550      &4998.499550       \\ \hline
100           & 19998.49989      &19998.49989     \\ \hline
\end{tabular}
\label{tablePS}
\end{center}
\end{table}
From Fig.~\ref{figure6} we observe that in the massive case with $L=3$ the choice $\sigma=0.1$ already ensures an excellent fit for both potentials. Furthermore, Table~\ref{tablePS} signalizes that the choice of $\sigma$ is not sensitive to the parameter $L$. Finally, we observe in Fig.~\ref{figure7} that in the massless case with $L=3$ and for the choice $\sigma=0.1$ both potentials have the same photon sphere and both black holes share the same event horizon. 
\begin{figure}[!ht]\label{PSmodelphoton}
\centering
    \includegraphics[width=0.3\textwidth]{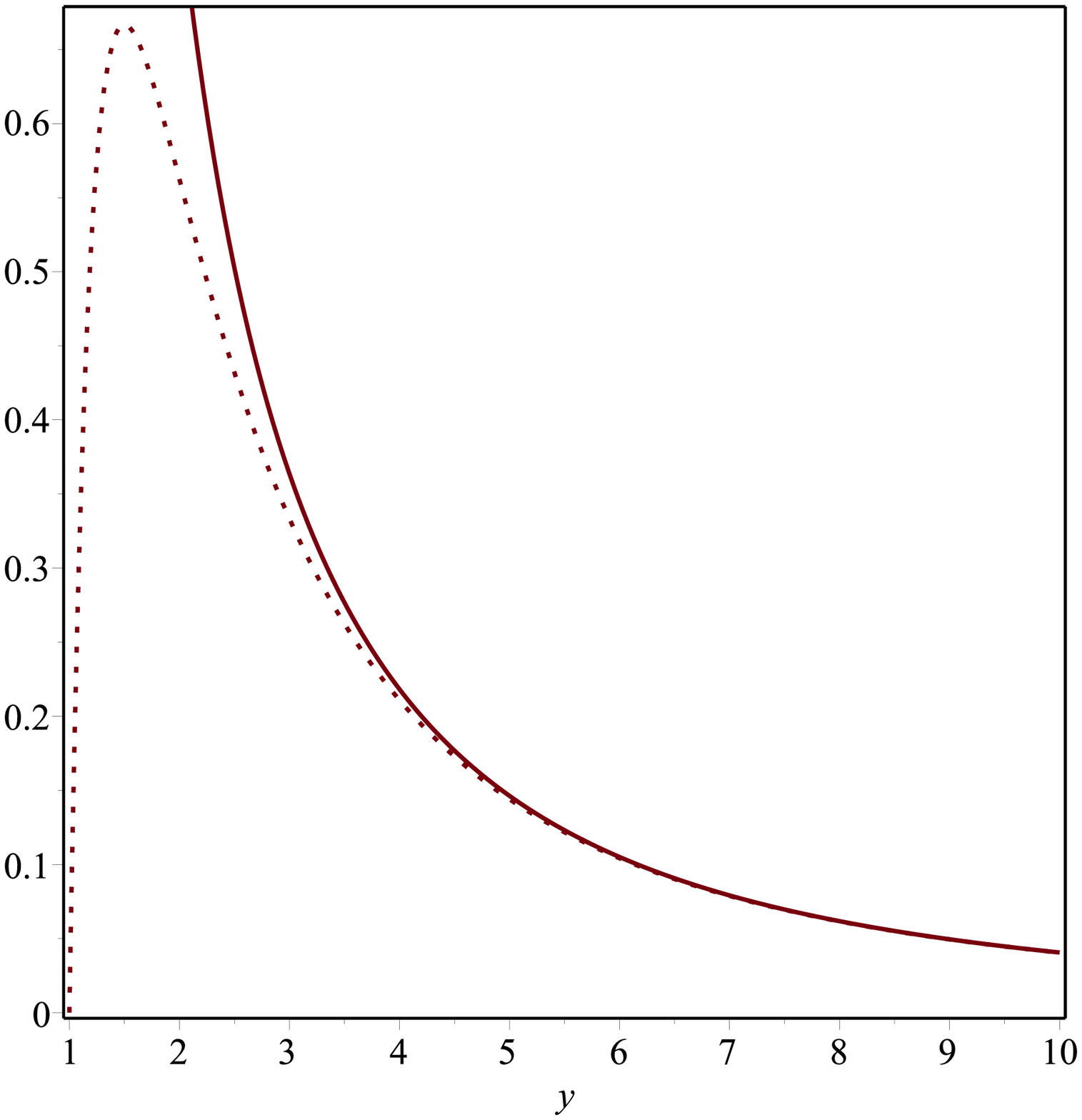}
    \includegraphics[width=0.3\textwidth]{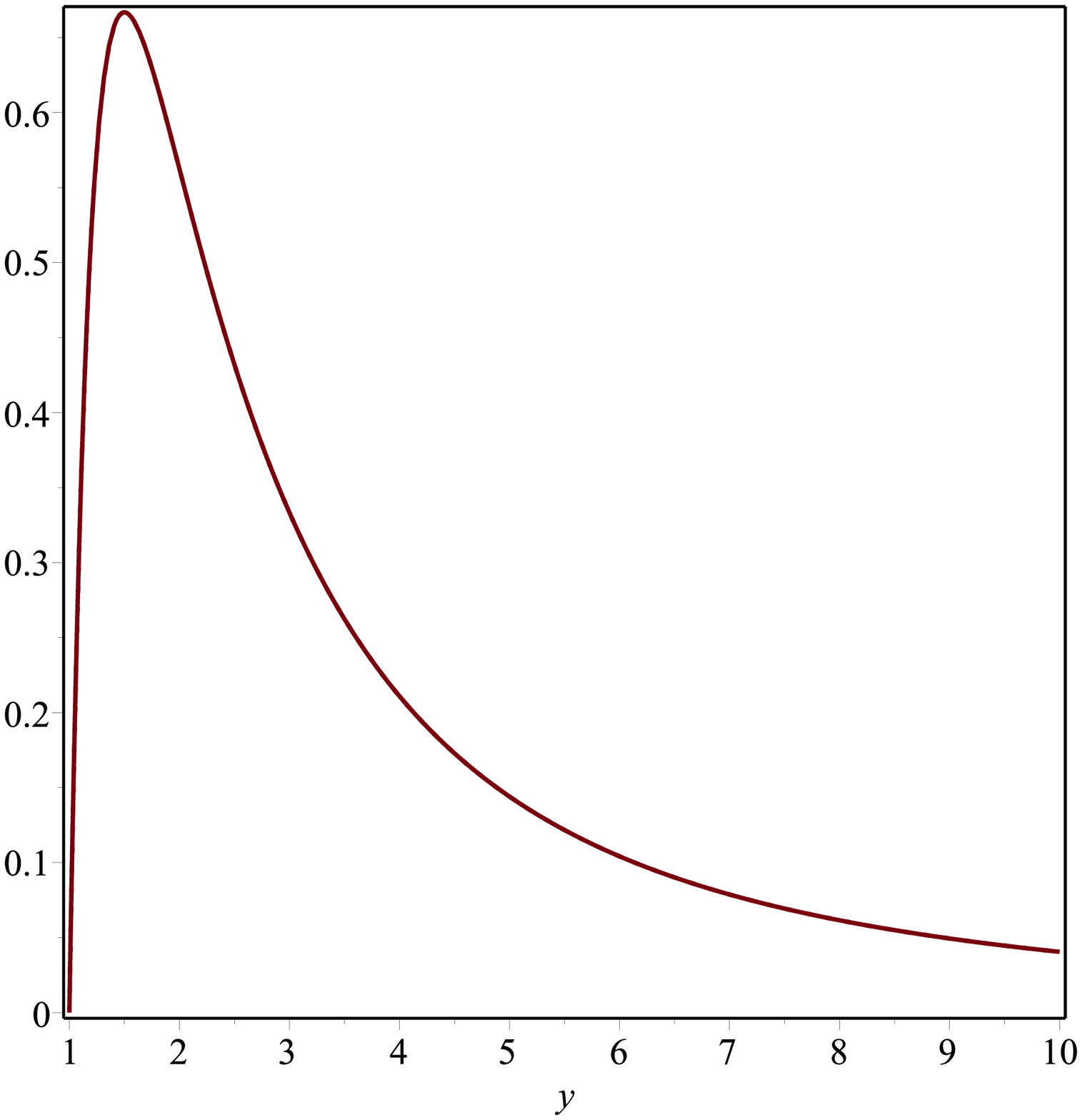}
\caption{\label{figure7}
Plots of the effective potential (\ref{VeffPS}) (solid line) and the Schwarzschild effective potential (\ref{VSmassive}) (dotted line) in the massless case for $L=3$. The figure on the left refers to the case $\sigma=3$ while the one on the right to $\sigma=0.1$.}
\end{figure}

\subsection{The MIS model $(2,5,0)$}
\begin{figure}[!ht]\label{MISmodel}
\centering
    \includegraphics[width=0.3\textwidth]{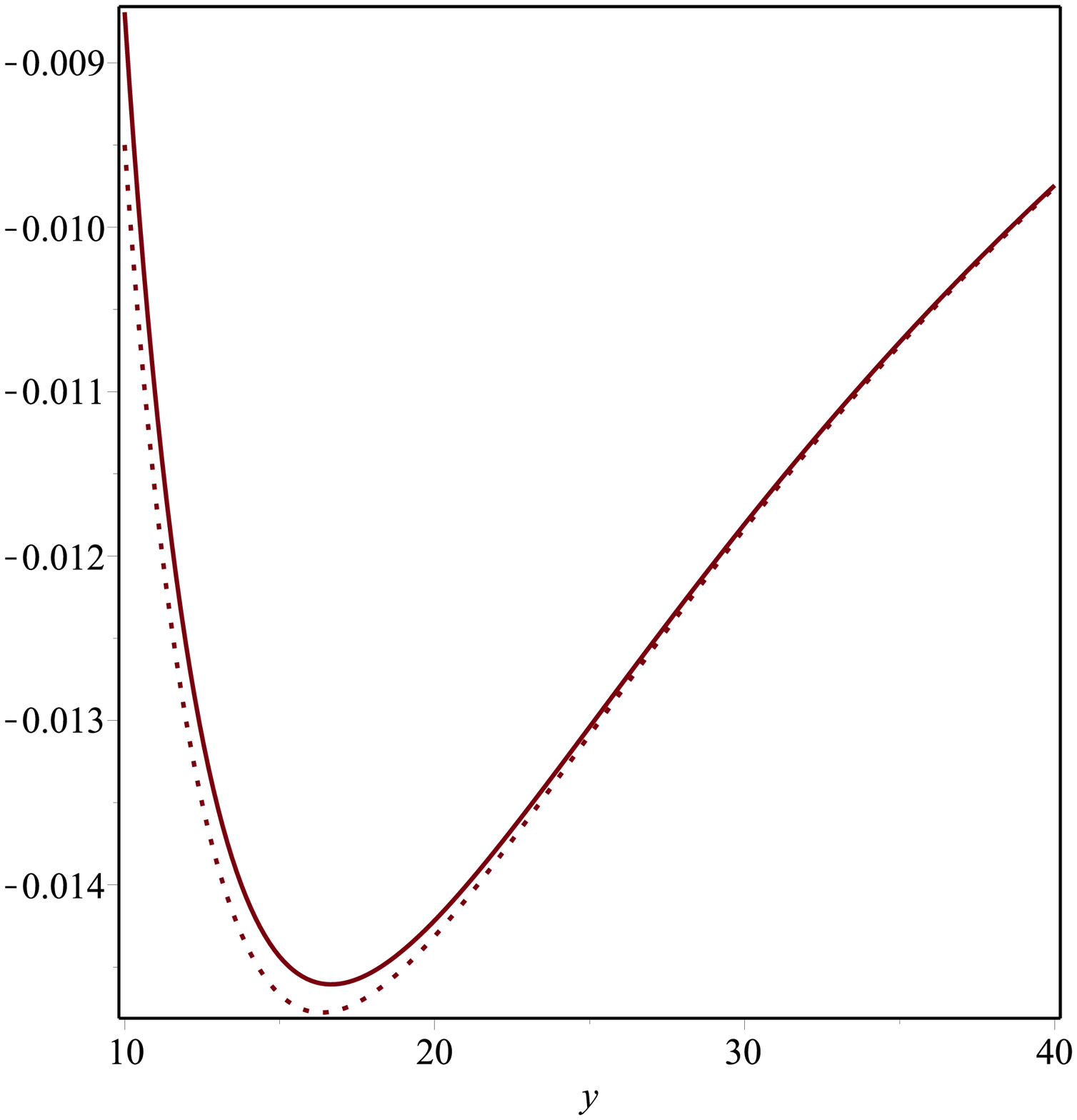}
    \includegraphics[width=0.3\textwidth]{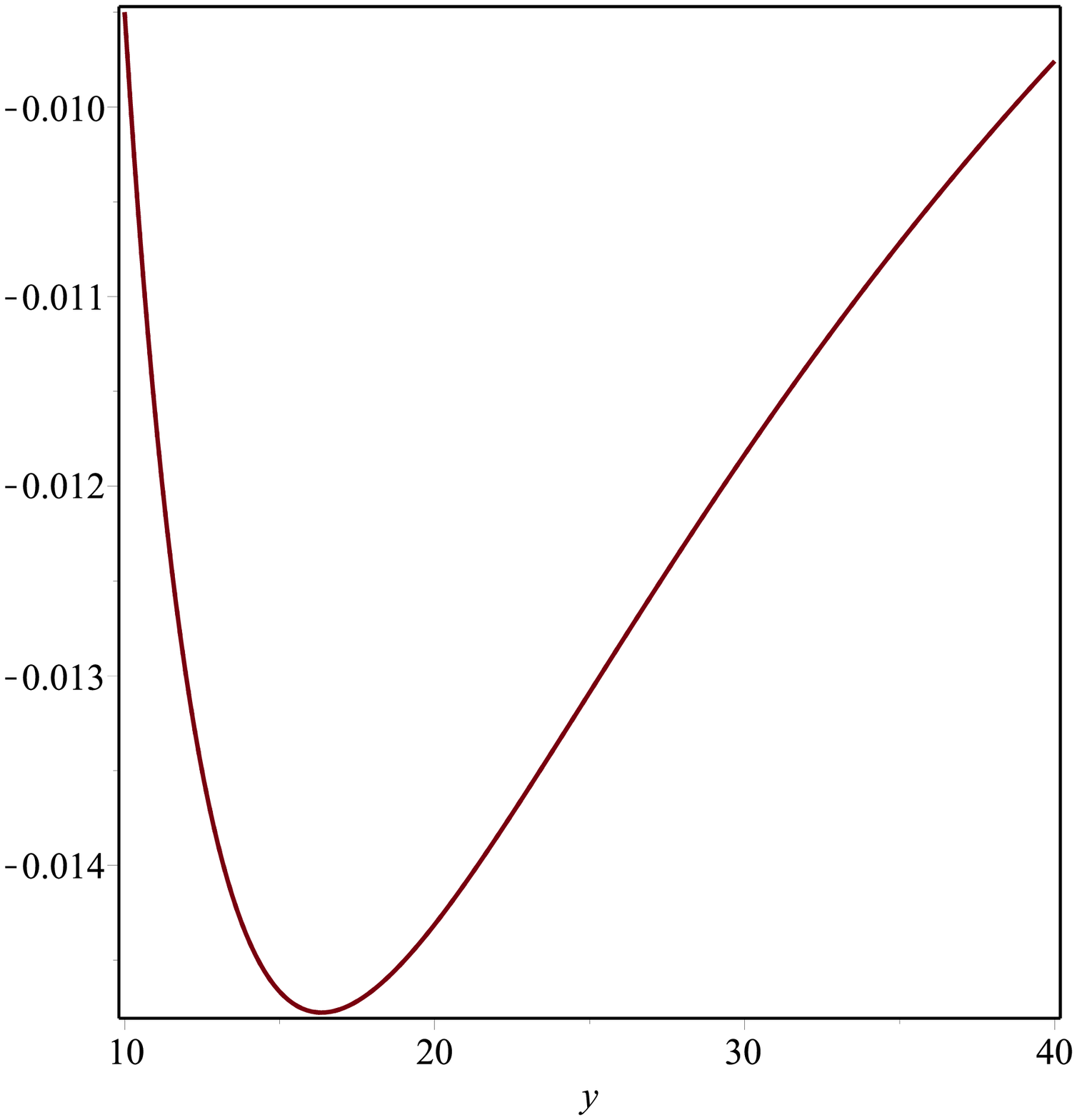}
\caption{\label{figure8}
Plots of the effective potential (\ref{VeffMIS}) (solid line) and the Schwarzschild effective potential (\ref{VSmassive}) (dotted line) in the massive case for $L=3$. The figure on the left refers to the case $\sigma=1$ while the one on the right to $\sigma=0.01$ for which the two potentials agree remarkably well in a large neighbourhood of the minimum.}
\end{figure}

\begin{figure}[!ht]\label{DMISmodelphoton}
\centering
   \includegraphics[width=0.3\textwidth]{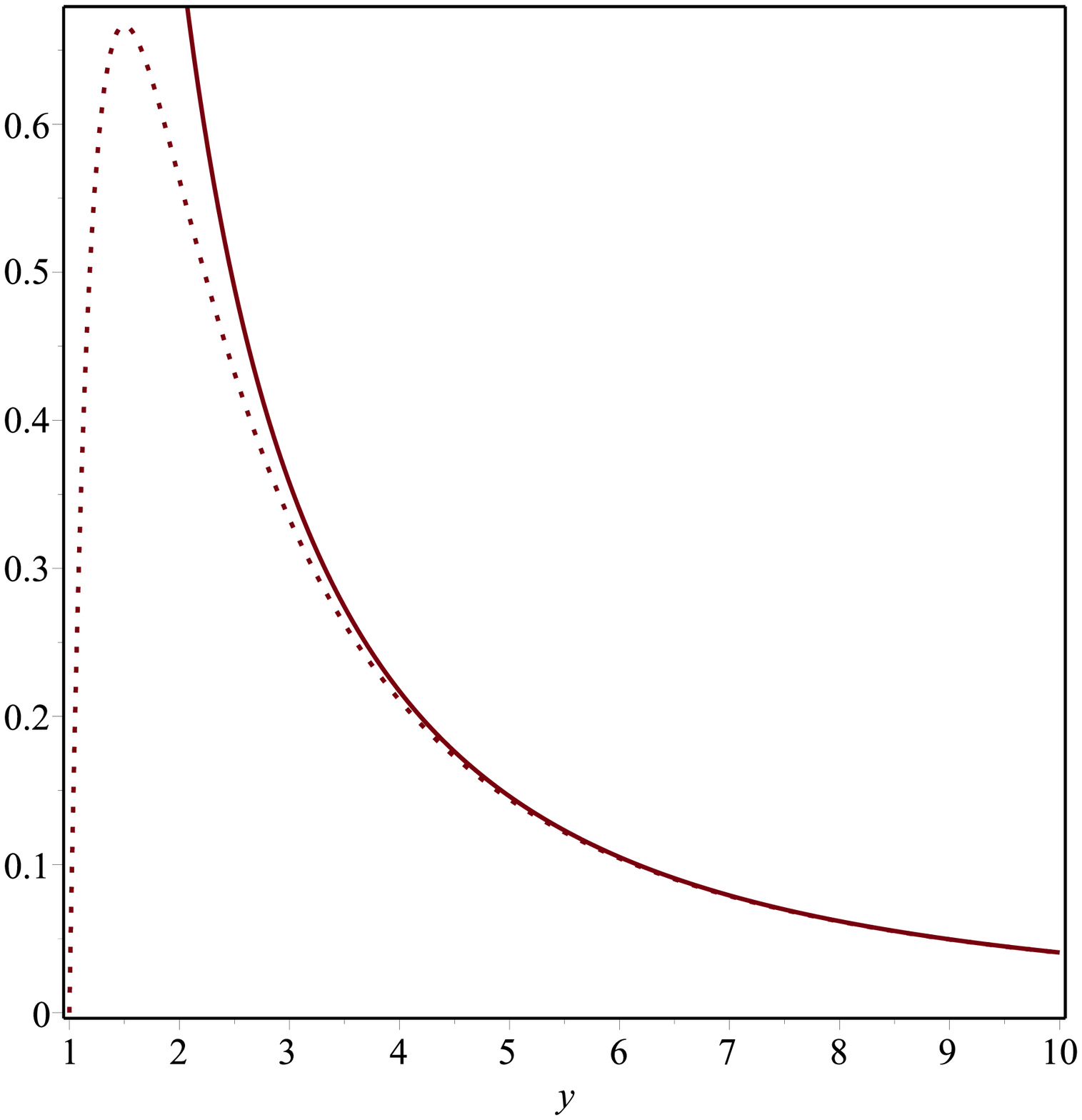}
    \includegraphics[width=0.3\textwidth]{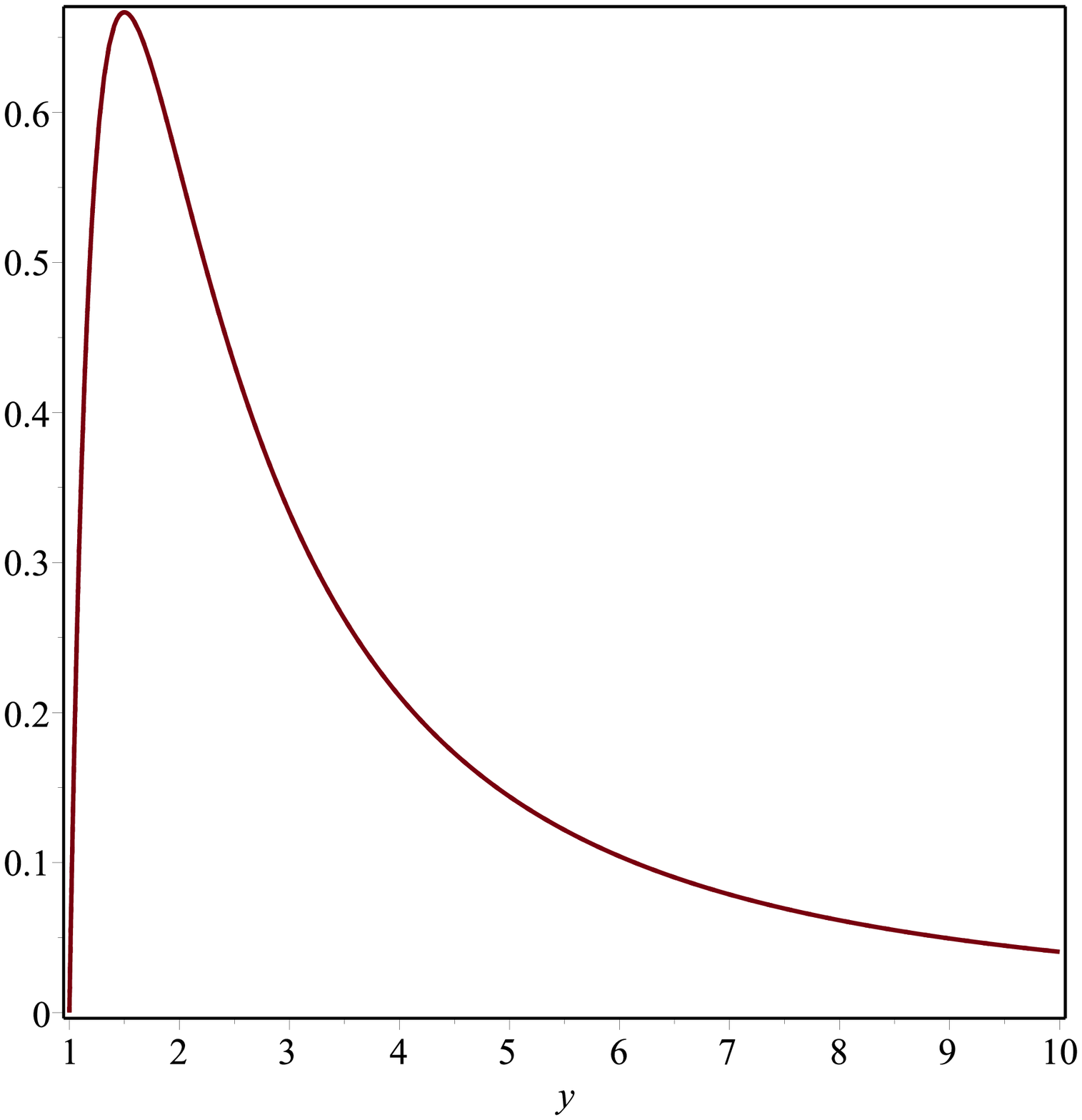}
\caption{\label{figure9}
Plots of the effective potential (\ref{VeffMIS}) (solid line) and the Schwarzschild effective potential (\ref{VSmassive}) (dotted line) in the massless case for $L=3$. The figure on the left refers to the case $\sigma=1$ while the one on the right to $\sigma=0.01$ for which both potentials shares the same photon sphere and both black holes have the same event horizon.}
\end{figure}

In this case, the density and mass function are 
\begin{equation}
\rho(r)=\frac{3M_{BH}}{4\pi r_0^3\left(1+\frac{r^2}{r_0^2}\right)^{5/2}},\quad
m(r)=M_{BH}\frac{r^3}{r_0^3\left(1+\frac{r^2}{r_0^2}\right)^{3/2}}
\end{equation}
As before we set $y=r/r_s$, $L=\ell/r_s$ and $r_0/r_s=\sigma$. We find that the effective potential is
\begin{equation}\label{VeffMIS}
V_{eff}(y)=\frac{L^2}{2y^2}-\left(\frac{\epsilon}{2y}+\frac{L^2}{2y^3}\right)\frac{y^3}{\sigma^3\left(1+\frac{y^2}{\sigma^2}\right)^{3/2}}.
\end{equation}
According to Table~\ref{tableEins}, there will be a black hole with two distinct horizons if $\sigma<2\sqrt{3}/9\approx 0.3849$. Moreover, (\ref{cond2}) requires that $y_{min}/\sigma>12$. Also in the present model, these constraints are easily met. From Fig.~\ref{figure8} we observe that in the massive case with $L=3$ the choice $\sigma=0.01$ already ensures that both potentials match well both at the minimum and in a large interval containing it. Moreover, Table~\ref{tableMIS} indicates that the choice of $\sigma$ is not sensitive to the angular momentum $L$ of the test particle. Finally, we observe in Fig.~\ref{figure9} that in the massless case with $L=3$ and for the choice $\sigma=0.001$ both potentials practically shares the same photon sphere and both black holes have almost the same event horizon. 
\begin{table}
\caption{MIS model: numerical values of the minima $y_{min}$ and $y_{min,s}$ in the effective potentials (\ref{VeffMIS}) and (\ref{VSmassive}) for $\sigma=0.01$ and different values of $L$.}
\begin{center}
\begin{tabular}{ | l | l | l | l|}
\hline
$L$             & $y_{min}$        &$y_{min,s}$          \\ \hline
2             & 6.000133327      &6.000000000                    \\ \hline
3             & 16.34850157      &16.34846923               \\ \hline
4             & 30.42222115      &30.42220510       \\ \hline
5             & 48.45208856      &48.45207880       \\ \hline
10            & 198.4885803      &198.4885780      \\ \hline
50            & 4998.499550      &4998.499550       \\ \hline
100           & 19998.49989      &19998.49989     \\ \hline
\end{tabular}
\label{tableMIS}
\end{center}
\end{table}

\subsection{The $(k,n)$-model}
\begin{figure}[!ht]\label{KN21model}
\centering
    \includegraphics[width=0.3\textwidth]{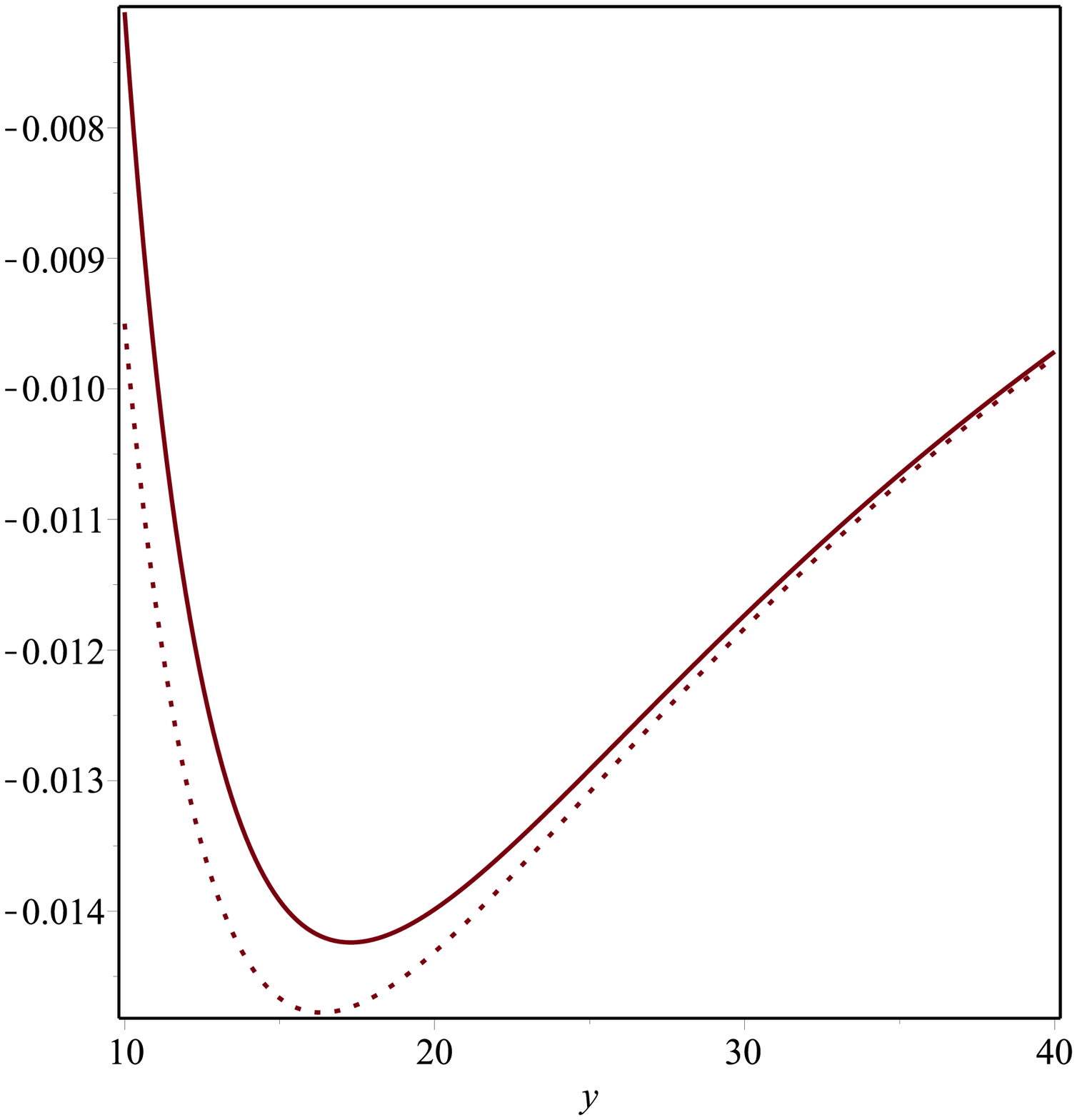}
    \includegraphics[width=0.3\textwidth]{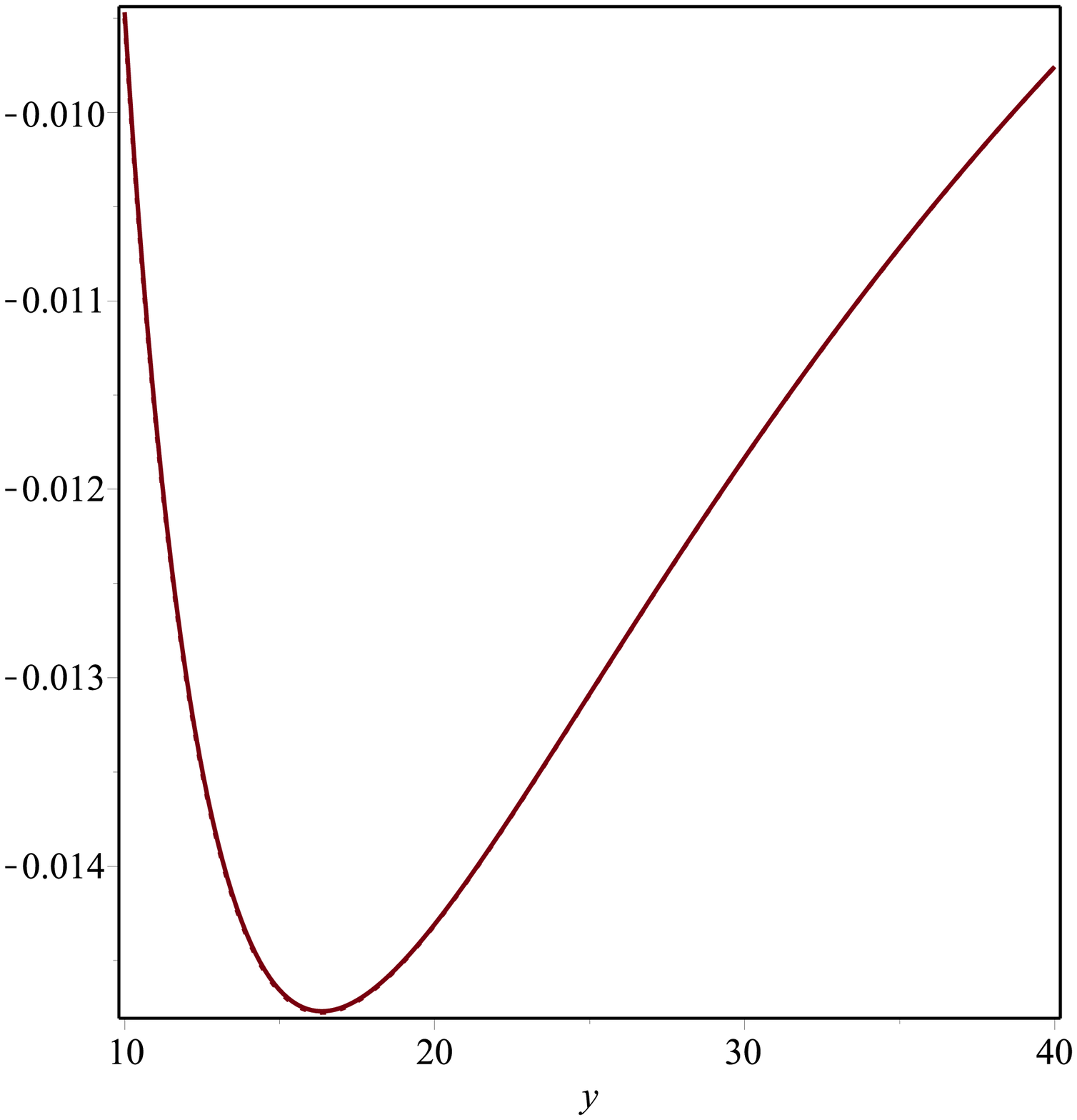}
\caption{\label{figure10}
Plots of the effective potential (\ref{Veffkn}) (solid line) with $(k,n)=(2,1)$ and the Schwarzschild effective potential (\ref{VSmassive}) (dotted line) in the massive case for $L=3$. The figure on the left refers to the case $\sigma=1$ while the one on the right to $\sigma=0.1$ for which the two potentials exhibit an excellent fit in a large neighbourhood of the minimum.}
\end{figure}

\begin{figure}[!ht]\label{KN21modelmassless}
\centering
    \includegraphics[width=0.3\textwidth]{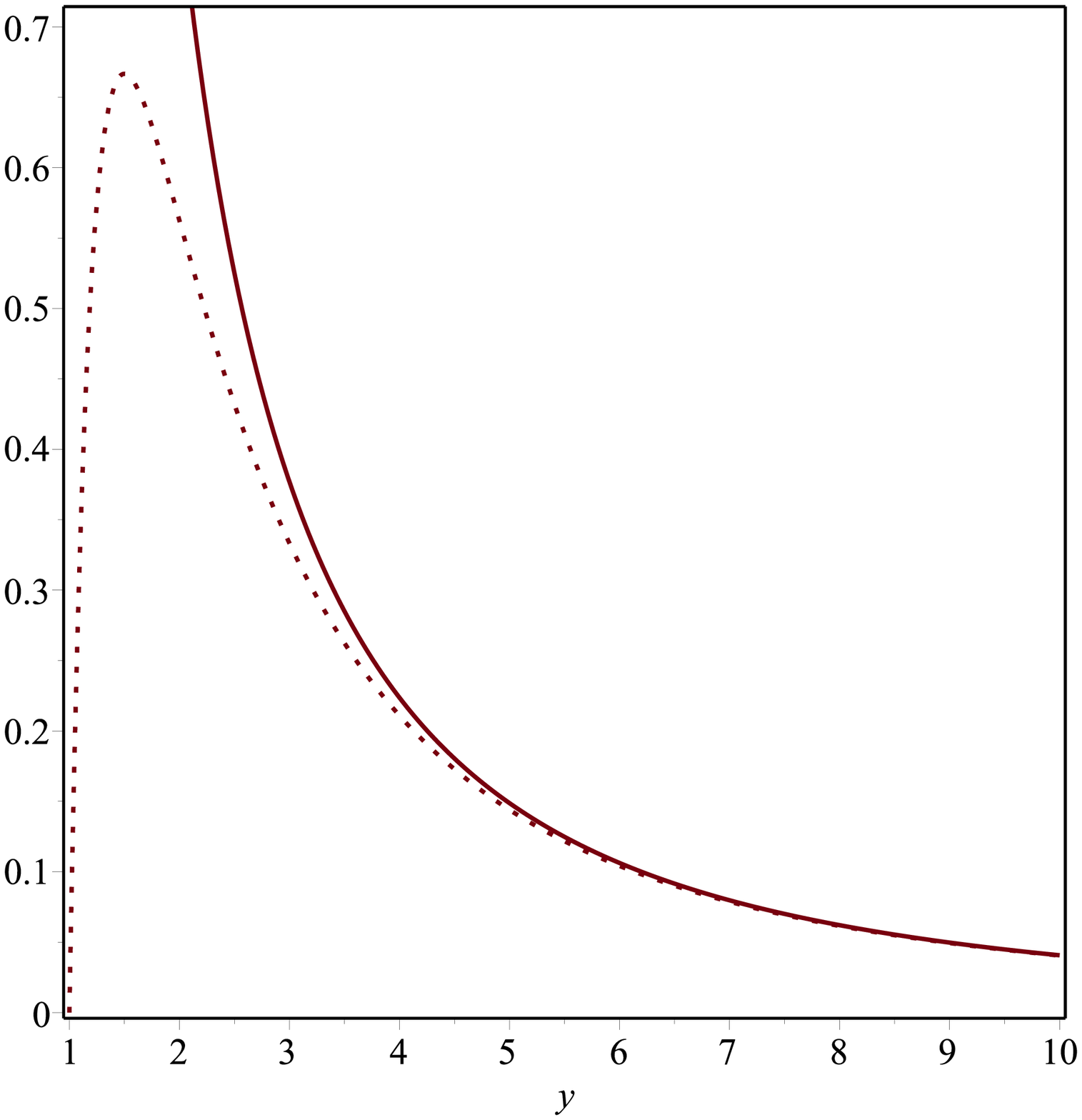}
    \includegraphics[width=0.3\textwidth]{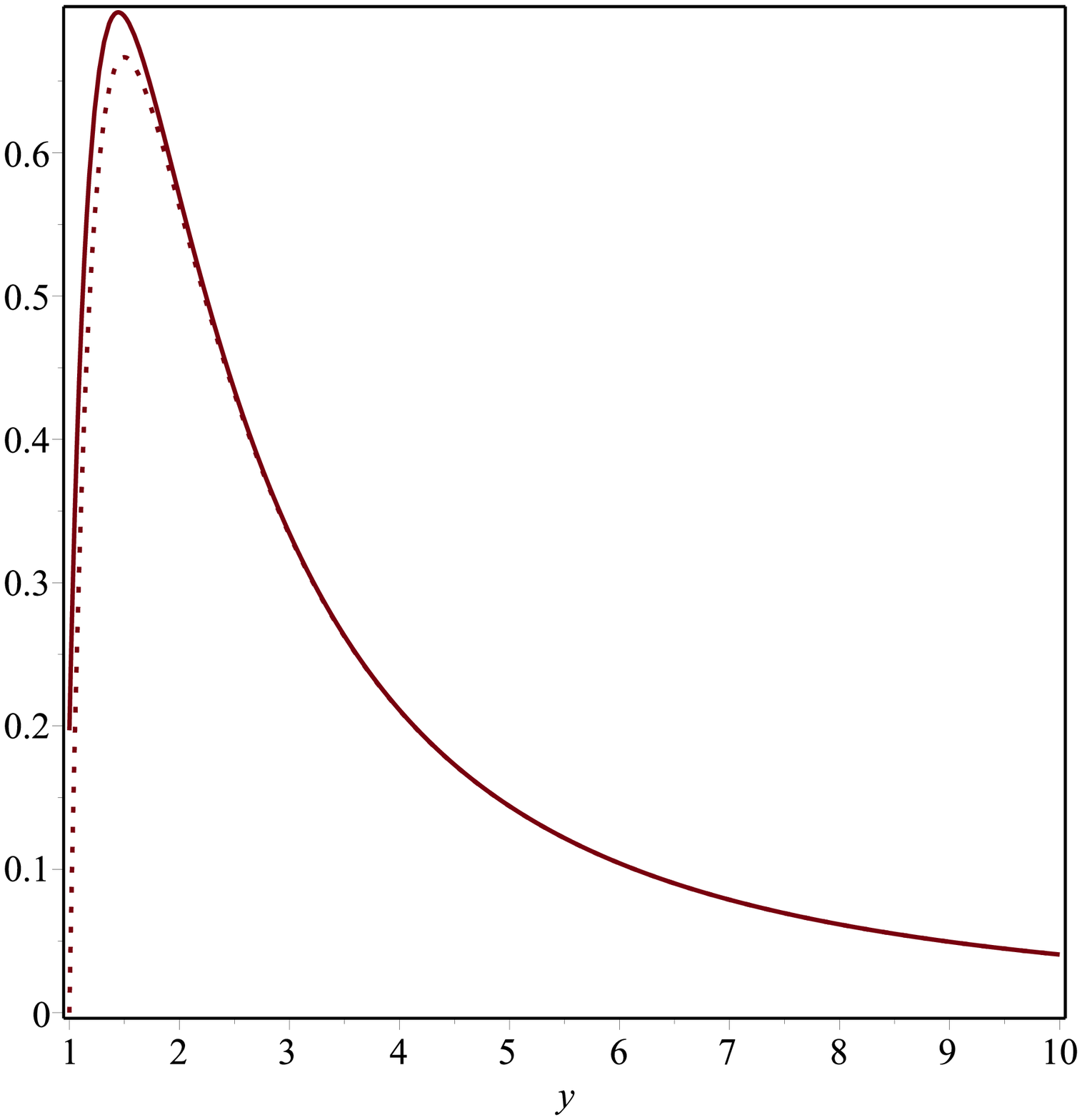}
    \includegraphics[width=0.3\textwidth]{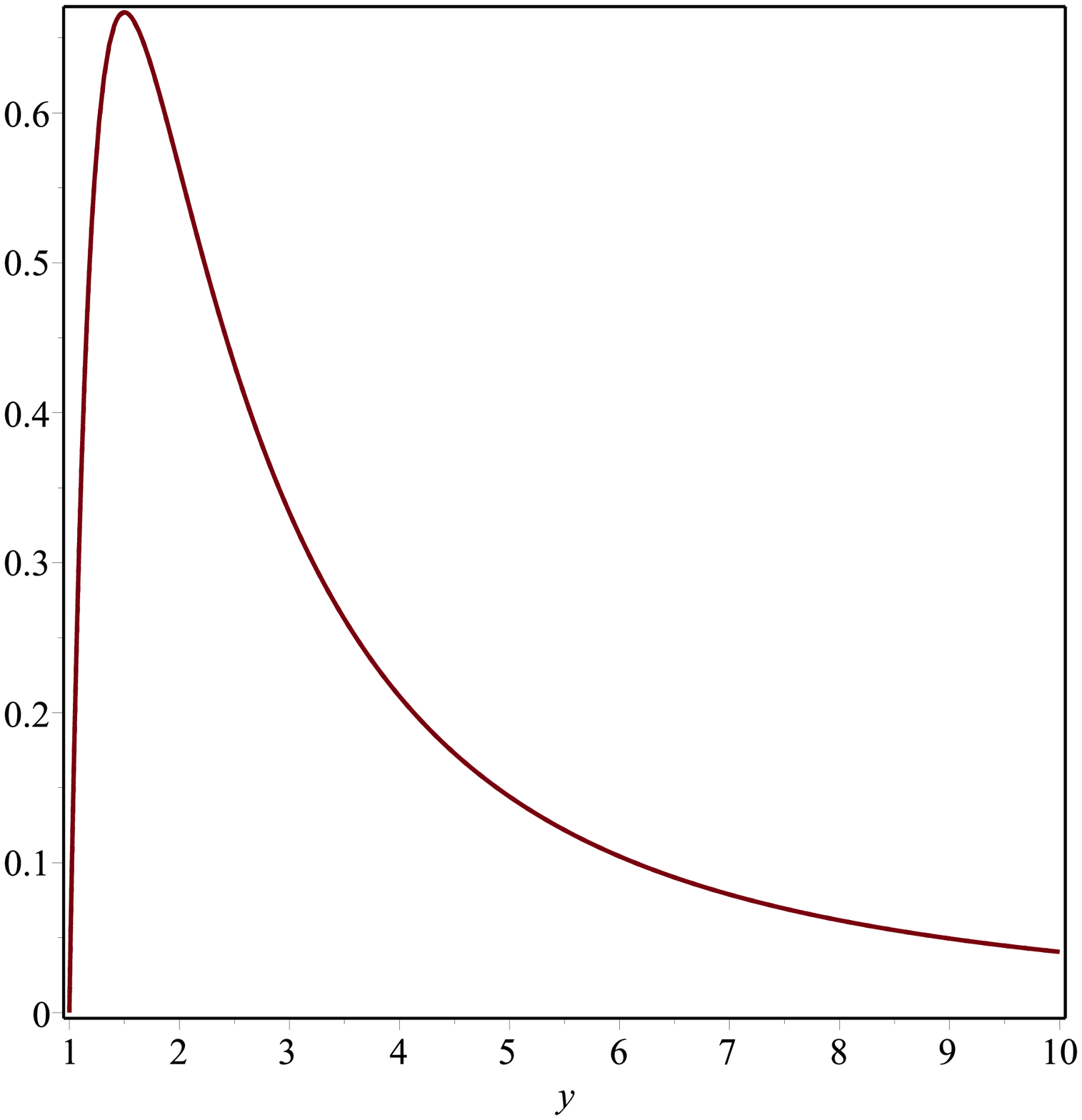}
\caption{\label{figure11}
Plots of the effective potential (\ref{Veffkn}) (solid line) and the Schwarzschild effective potential (\ref{VSmassive}) (dotted line) in the massless case for $L=3$ and $\sigma=1$ (far left), $\sigma=0.1$ (middle), $\sigma=0.01$ (far right).}
\end{figure}

For $k,n\in\mathbb{N}$ we find that the density and mass function are
\begin{equation}
\rho(r)=\frac{M_{BH}}{4\pi r_0^3 nB(k,3n)}\frac{1}{\left[1+\left(\frac{r}{r_0}\right)^{1/n}\right]^{3n+k}},\quad
m(r)=\frac{M_{BH}r^3}{3r_0^3 nB(k,3n)}{}_2 F_1(3n,3n+k;1+3n;-(r/r_0)^{1/n}).
\end{equation}
Let $y=r/r_s$, $L=\ell/r_s$ and $r_0/r_s=\sigma$. Then, the effective potential is found to be
\begin{equation}\label{Veffkn}
V_{eff}(y)=\frac{L^2}{2y^2}-\left(\frac{\epsilon}{2y}+\frac{L^2}{2y^3}\right)\frac{y^3}{3\sigma^3 nB(k,3n)}{}_2 F_1(3n,3n+k;1+3n;-(y/\sigma)^{1/n}).
\end{equation}
If we go back to Table~\ref{tableEins}, we realize that in order to have a black hole with two distinct horizons we need to impose that $\sigma<1/(2\mu_c)$. Furthermore, (\ref{cond2}) requires that $y_{min}/\sigma>\lambda$. Once the parameters $k$ and $n$ are fixed, the corresponding values of $\mu_c$ and $\lambda$ can be obtained from Table~\ref{tableEins} and Table~\ref{tableCond}. Also in the present model, these constraints can be easily fulfilled. As an example of the matching procedure at the minimum between the Schwarzschild effective potential and (\ref{Veffkn}), we consider the case $(k,n)=(2,1)$. Other choices of the parameters $k$ and $n$ can be treated similarly. If $(k,n)=(2,1)$, the constraints on $\sigma$ are represented by the inequalities $y_{min}/\sigma>23$ and $\sigma<0.3201$. From Fig.~\ref{figure10} we see that in the massive case with $L=3$ the choice $\sigma=0.1$ already ensures that both potentials match well both at the minimum and in a large interval containing it.
Furthermore, Table~\ref{tableKN} indicates that the choice of $\sigma$ is not sensitive to the angular momentum $L$ of the test particle.
\begin{table}
\caption{$(k,n)=(2,1)$-model: numerical values of the minima $y_{min}$ and $y_{min,s}$ in the effective potentials (\ref{Veffkn}) and (\ref{VSmassive}) for $\sigma=0.1$ and different values of $L$.}
\begin{center}
\begin{tabular}{ | l | l | l | l|}
\hline
$L$             & $y_{min}$        &$y_{min,s}$          \\ \hline
2             & 6.048923096      &6.000000000                    \\ \hline
3             & 16.36104864      &16.34846923               \\ \hline
4             & 30.42853117      &30.42220510       \\ \hline
5             & 48.45594864      &48.45207880       \\ \hline
10            & 198.4894936      &198.4885780      \\ \hline
50            & 4998.499586      &4998.499550       \\ \hline
100           & 19998.49990      &19998.49989     \\ \hline
\end{tabular}
\label{tableKN}
\end{center}
\end{table}
Finally, we observe in Fig.~\ref{figure11} that in the massless case with $L=3$ the choice $\sigma=0.1$ does not match well the radius of the Schwarzschild photon sphere. As it can be seen there, fixing $\sigma=0.001$ provides a good fit for both the photon sphere and the event horizon.

\section{Fuzzy self-gravitating Dark Matter droplets from a nonlocal EOS}
\begin{figure}[!ht]\label{radpress}
\centering
    \includegraphics[width=0.3\textwidth]{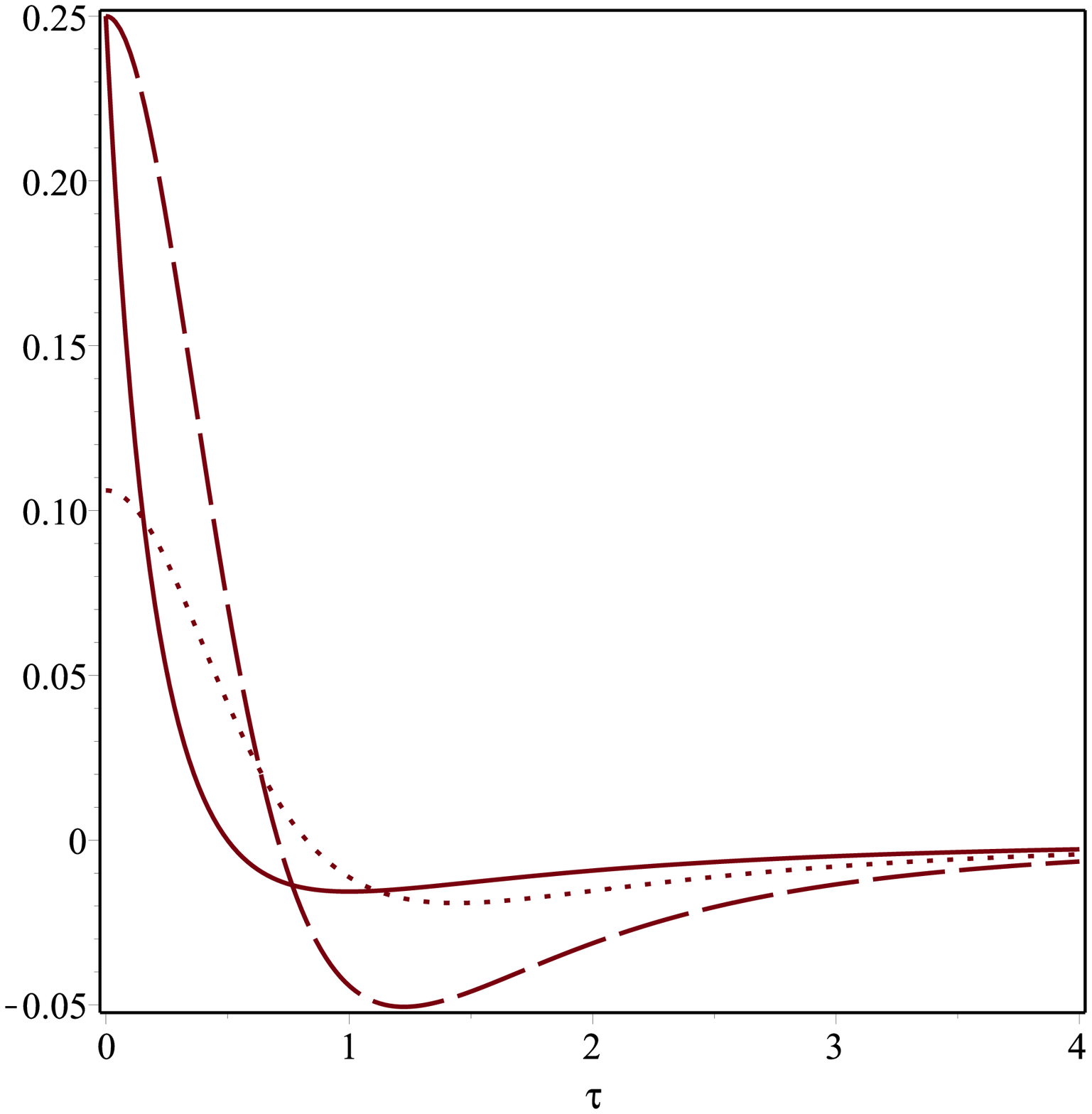}
    \includegraphics[width=0.3\textwidth]{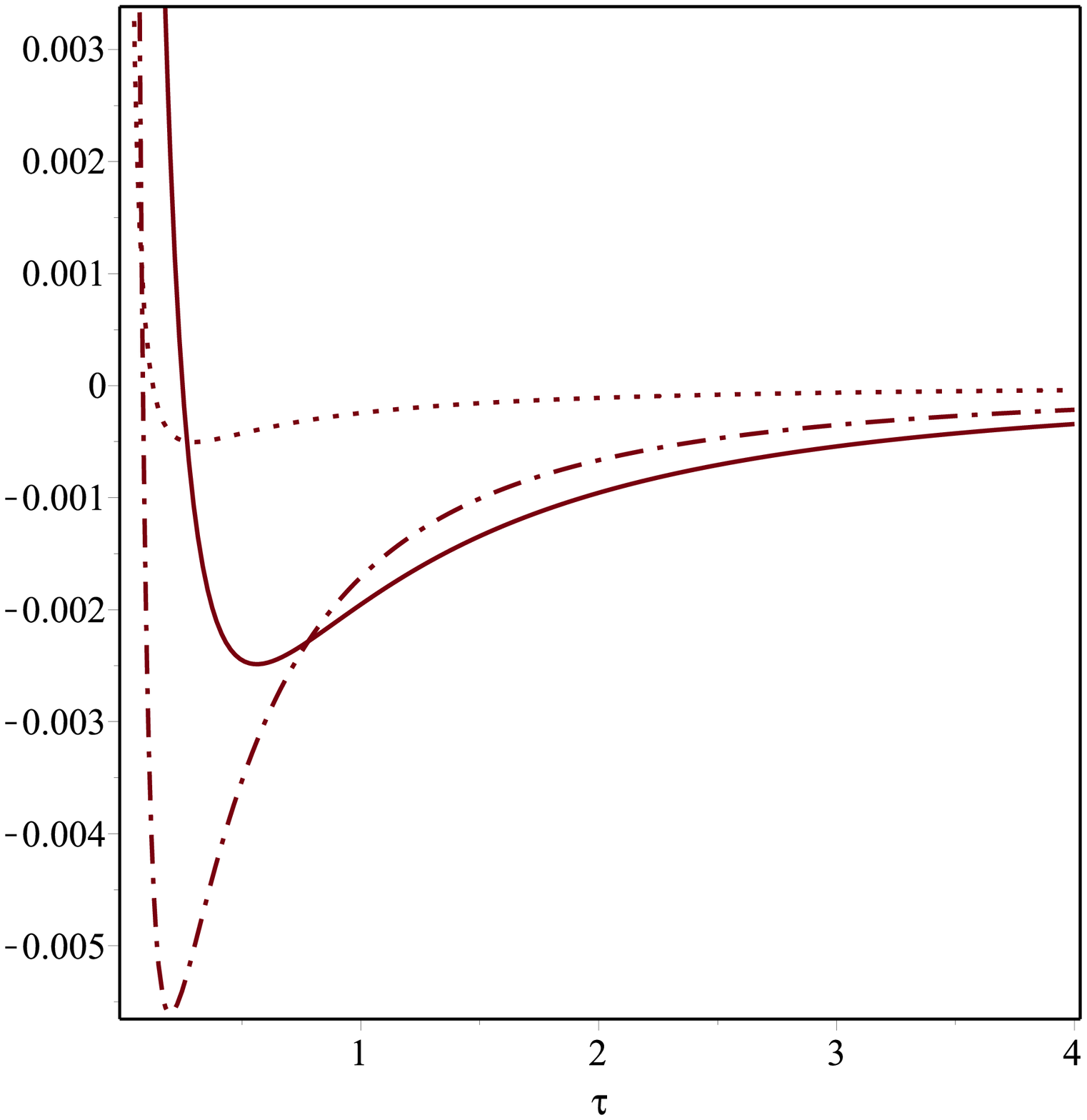}
    \includegraphics[width=0.3\textwidth]{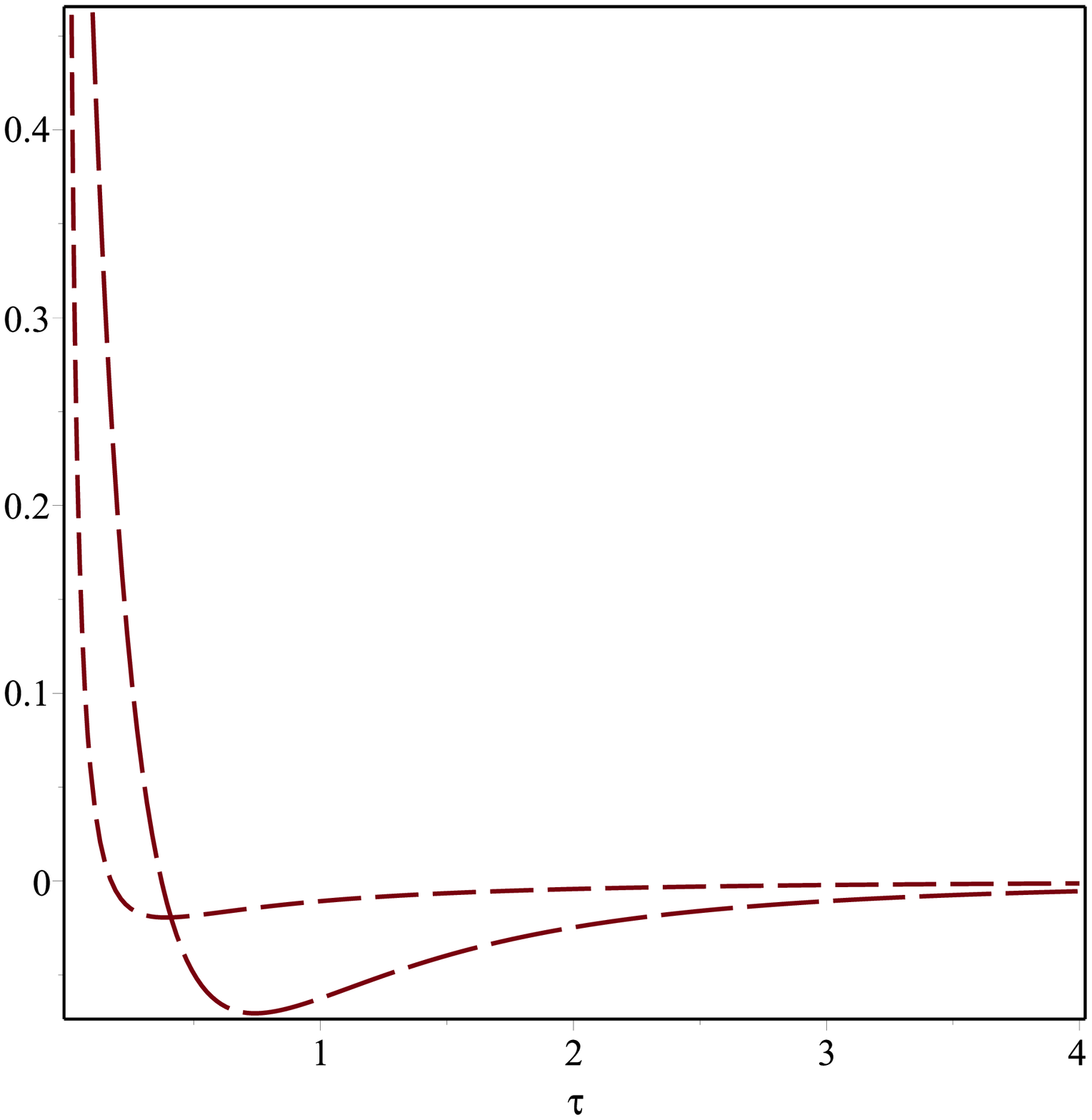}
\caption{\label{figure12}
Plot of the rescaled radial pressure $\pi r_0^3 p_r/M$ versus $\tau=r/r_0$ with $p_r$ as given in Table~\ref{radialp}. The radial pressure is positive in the inner region and it vanishes at some typical value of $\tau$ which depends on the particular model considered. On the right of such a value of $\tau$, the pressure becomes negative and it
exhibits a minimum. The far left picture showcases the following models: Dehnen $(1,4,0)$ (solid line), PS (dotted line) and MIS (longdashed line). The middle and far right pictures portrait several examples of the $(k,n)$-model. More precisely, the central panel displays the cases $(1,2)$ (solid line), $(1,3)$ (dotted line) and $(2,3)$ (dashdotted line) while the panel on the right contains the cases $(2,1)$ (longdashed line) and $(2,2)$ (dashed line).}
\end{figure}

\begin{figure}[!ht]\label{tanbotpress}
\centering
    \includegraphics[width=0.3\textwidth]{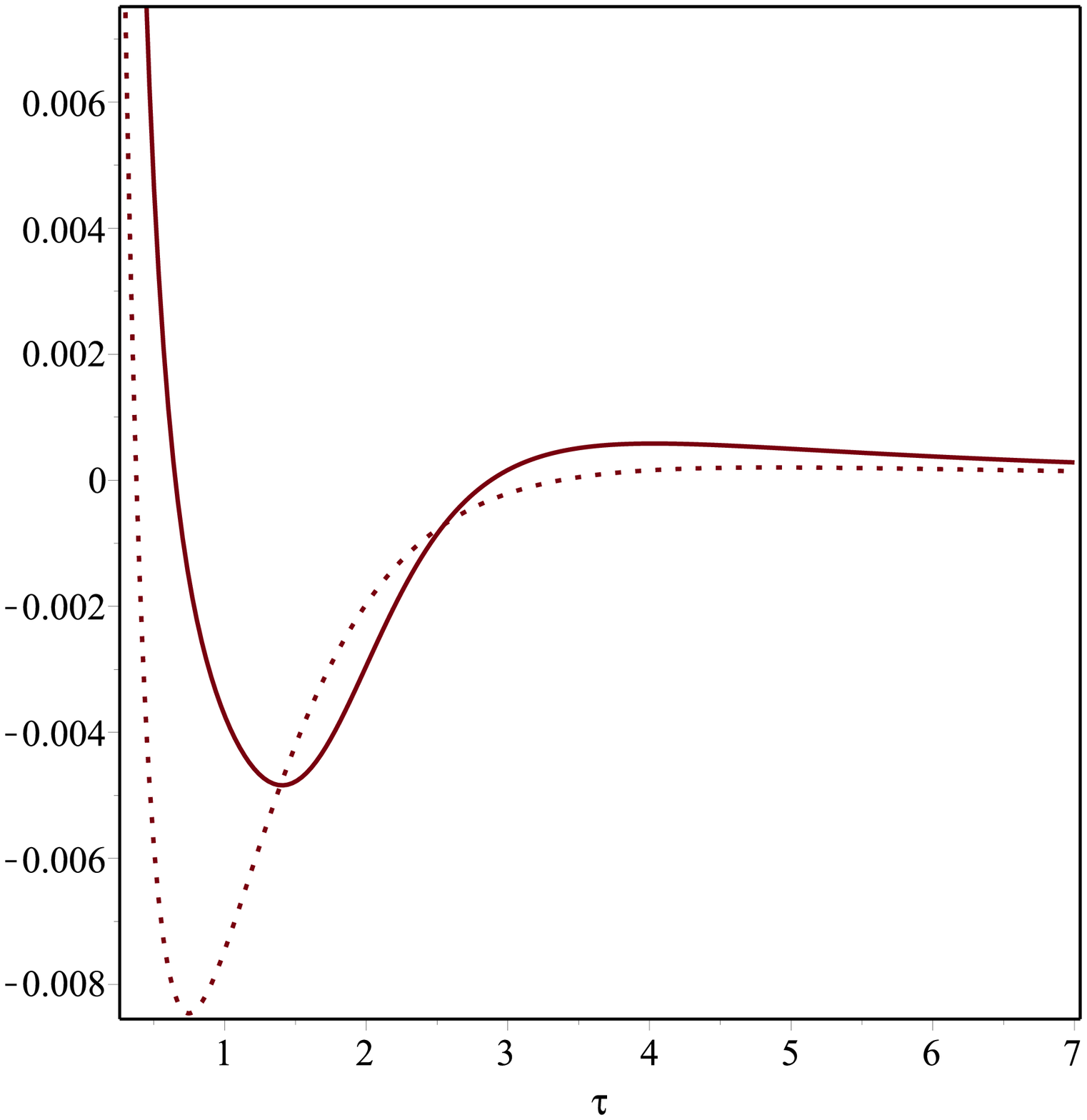}
    \includegraphics[width=0.3\textwidth]{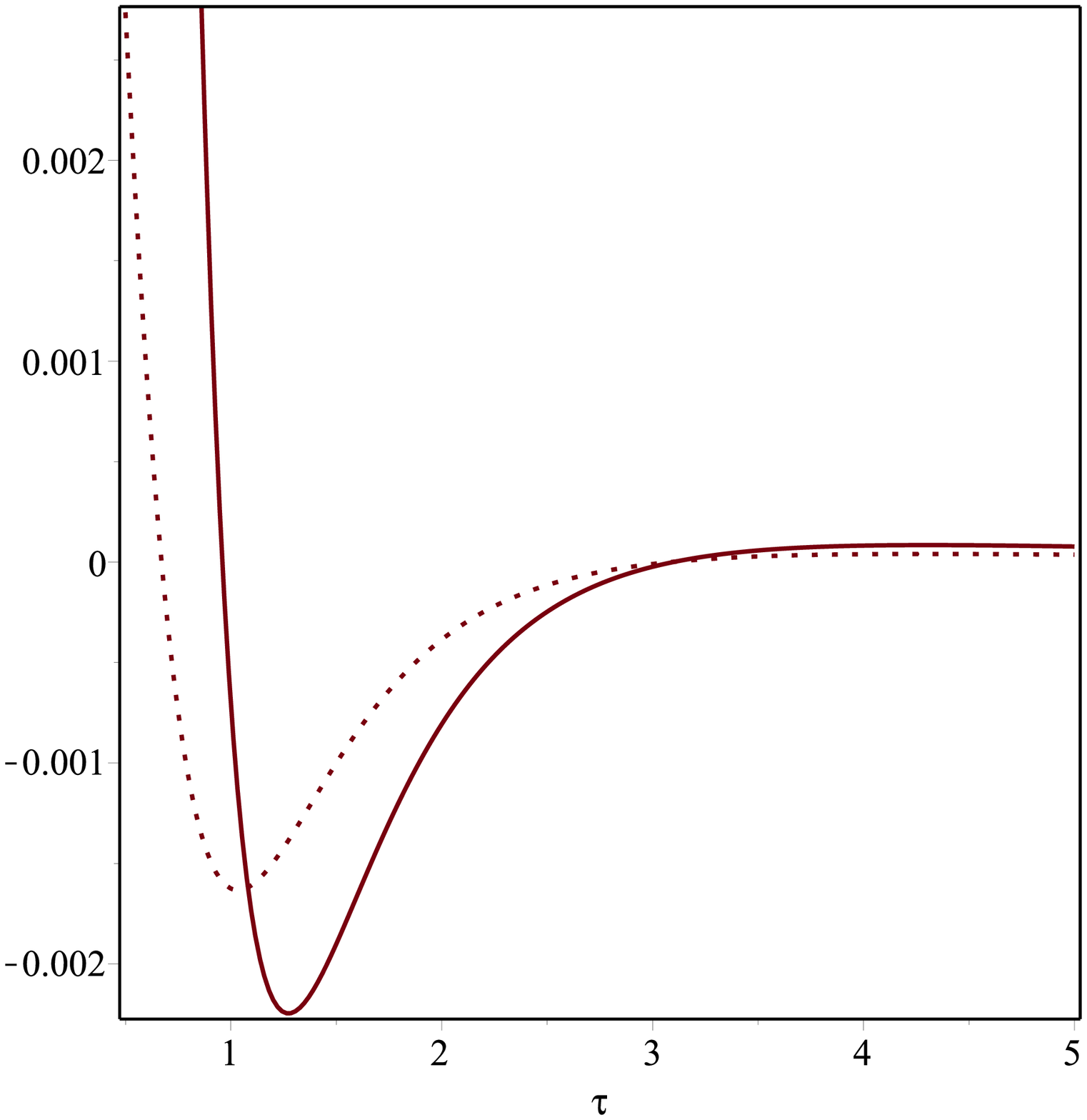}
    \includegraphics[width=0.3\textwidth]{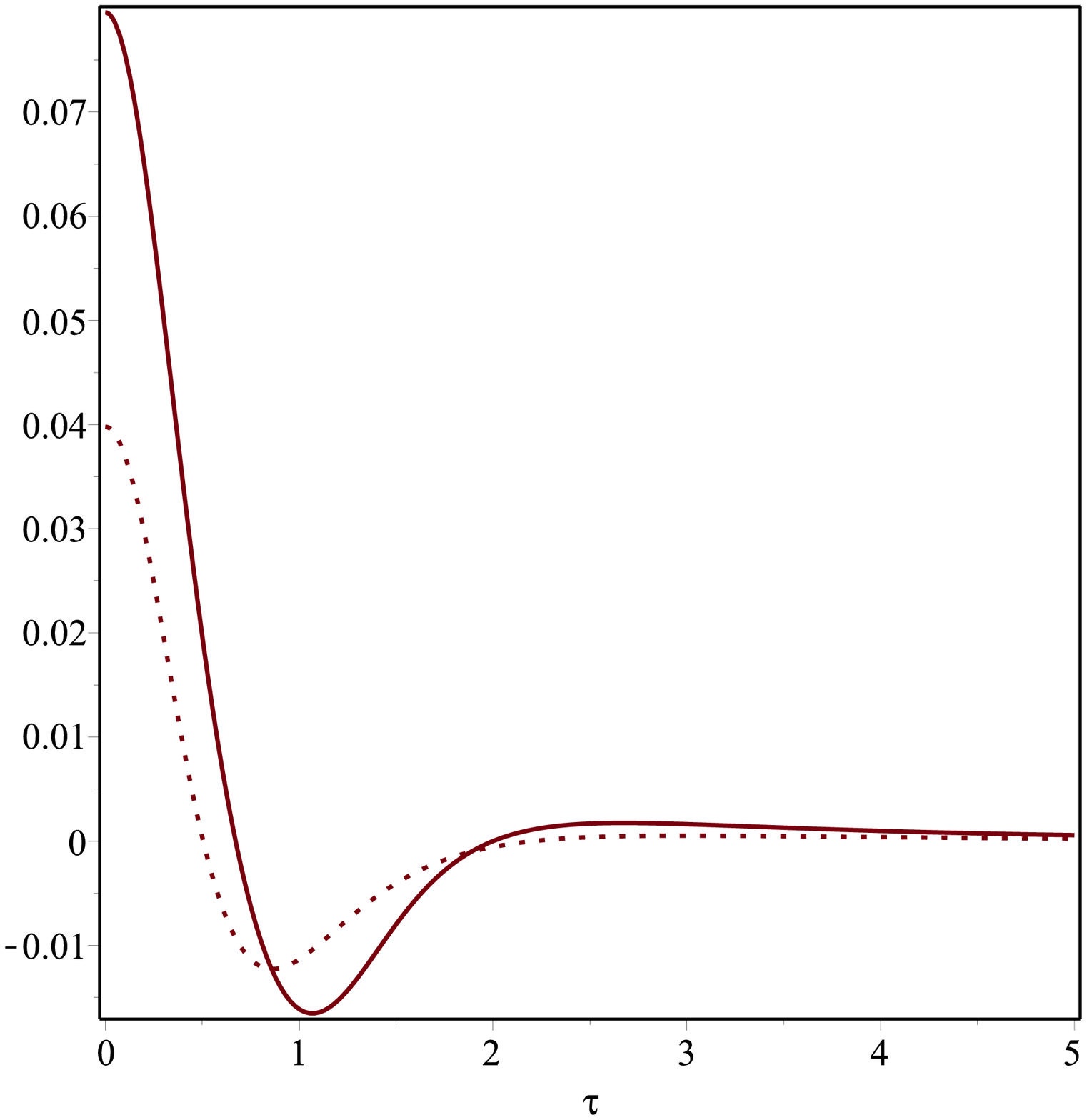}
\caption{\label{figure13}
Plot of the tangential pressure $p_\bot$ as a function of $\tau=r/r_0$ for different values of $\mu=M/r_0$. The left picture displays $p_\bot$ in the Dehnen $(1,4,0)$ model for $\mu=3$ (solid line) and $\mu=2$ (dotted line) while the right plot represents $p_\bot$ in the MIS model for $\mu=1$ (solid line) and $\mu=0.5$ (dotted line). In both models, the tangential pressure is finite at the origin and takes there the value $p_\bot(0)=\mu/4\pi r_0^2$. The central graph is related to the PS model where $p_\bot$ has been plotted for $\mu=0.39$ (solid line) and $\mu=0.19$ (dotted line). Also in this case the pressure stays finite at the centre.}
\end{figure}

\begin{figure}[!ht]\label{tanbotbotpress}
\centering
    \includegraphics[width=0.3\textwidth]{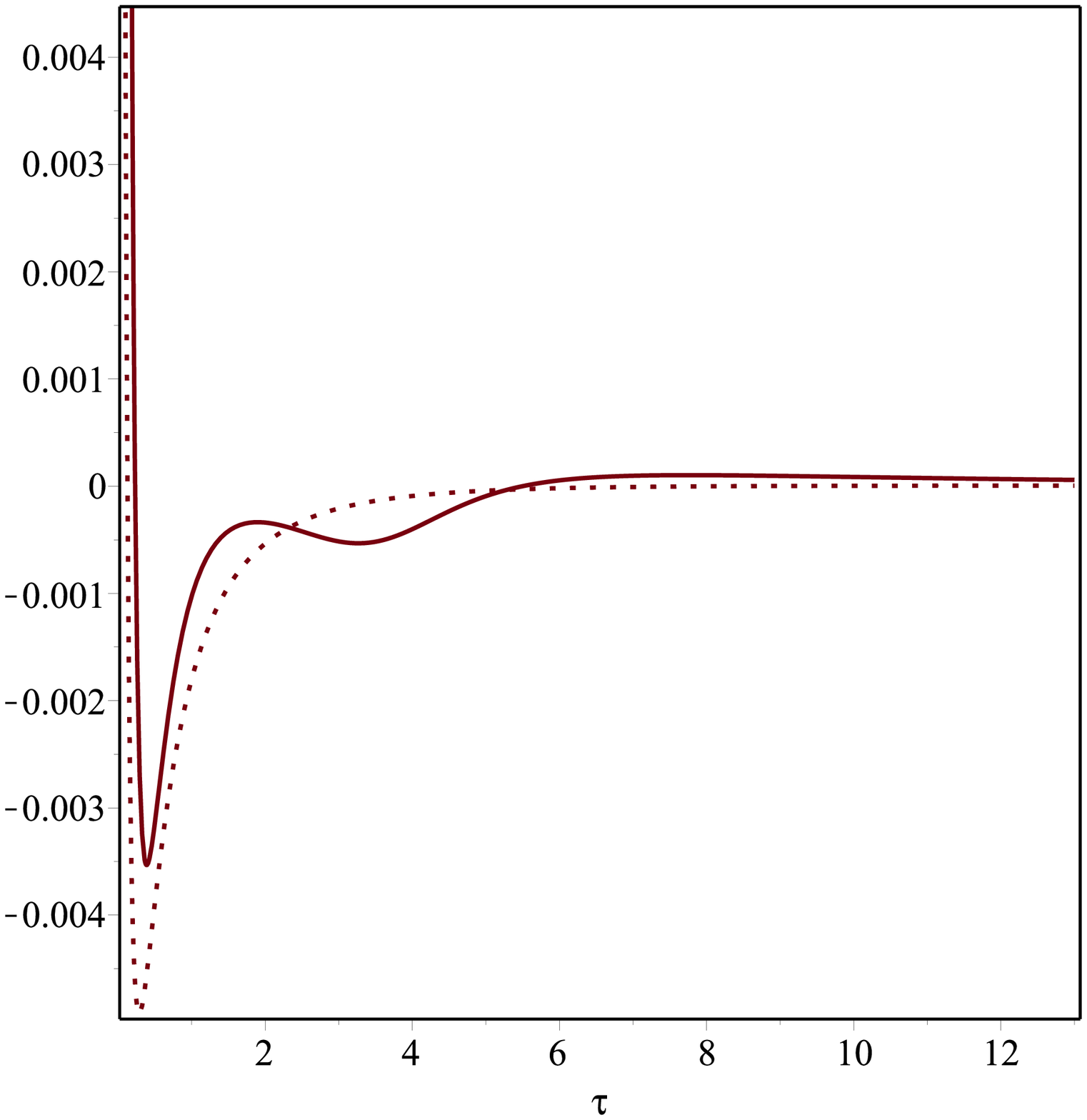}
     \includegraphics[width=0.3\textwidth]{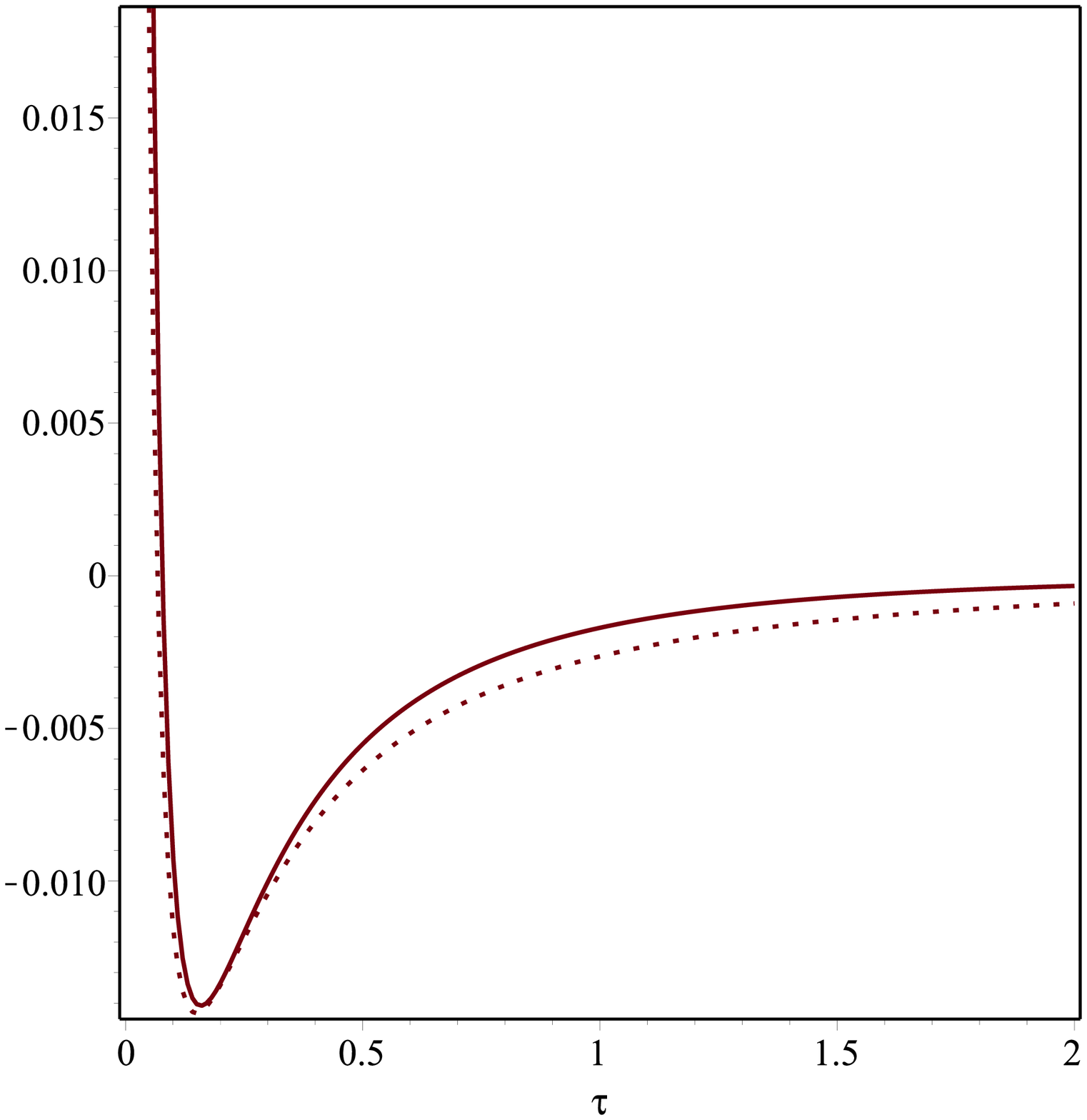}
    \includegraphics[width=0.3\textwidth]{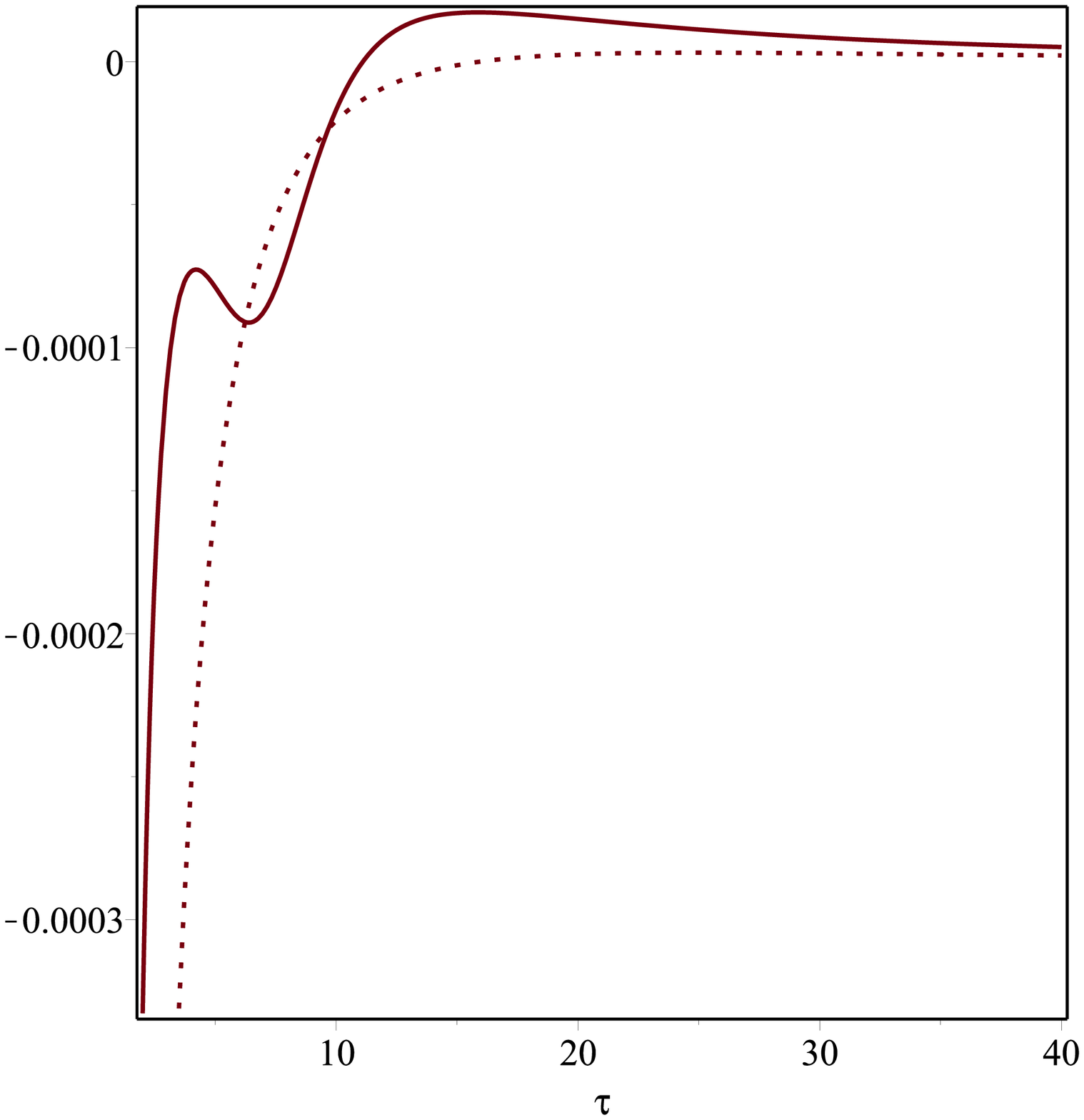}
\caption{\label{figure14}
Plot of the tangential pressure $p_\bot$ as a function of $\tau=r/r_0$ for different values of $\mu=M/r_0$ in the $(k,n)$-model. The left picture displays $p_\bot$ in the $(1,2)$-model for $\mu=22$ (solid line) and $\mu=5$ (dotted line).  In the central and right panels we considered the $(1,3)$-model over different ranges of the variable $\tau$. More precisely, the solid and dotted lines corresponds to the choices $\mu=150$ and $\mu=100$, respectively. The pressure remains always finite at the centre where $p_\bot(0)=\mu/4\pi r_0^2$. Interestingly $p_\bot$ can develop more than one minimum if $\mu$ lies close to $\mu_c$.}
\end{figure}

In the previous section, we considered an anisotropic fluid subject to a de Sitter-like   EOS for the radial pressure of the form, $p_r = -\rho$. Under these assumptions the Zhao energy density profile may give rise to regular black hole or  self-gravitating droplet configurations controlled by the rescaled mass parameter $\mu$. However, the diffusive nature of such profiles suggests that nonlocality may also play an important role because one would reasonably expect that variations in the radial pressure result from variations of the energy density within the entire volume. One possibility of implementing nonlocality is to follow the prescription described in \cite{Hern1,Hern2,Ab1} where it is assumed that the energy-momentum tensor components not only depend on the spacetime event but also on certain averages of the the energy density profile over the enclosed configuration. More precisely, we consider an anisotropic fluid described by a nonlocal EOS of the form \cite{Hern1,Hern2,Ab1}
\begin{equation}\label{prrr}
p_r(r)=\rho(r)-\frac{2}{r^3}\int_0^r u^2\rho(u)~du.
\end{equation}
In Table~\ref{radialp} we summarized the analytic formulae for the radial pressure emerging from the models considered in the present work. It is interesting to observe that $p_r$ displays the following behaviour: it is finite and positive at $r=0$ where it attains a maximum, then it decreases and becomes negative and finally, it vanishes asymptotically away from the gravitational object (see Figure~\ref{figure12}). 
\begin{table}[!ht]
\caption{Analytic results for the radial pressure. For the abbreviations we refer to Table~\ref{overview}.}
\begin{center}
\begin{tabular}{ | l | l | l | l|l|}
\hline
Model          &  $p_r(r)$  & $p_r(0)$    \\ \hline
Dehnen (1,4,0) &  $\frac{M}{4\pi r_0^3}\frac{1-\frac{2r}{r_0}}{\left(1+\frac{r}{r_0}\right)^4}$     & $\frac{M}{4\pi r_0^3}$            \\ \hline
PS             &  $\frac{M}{\pi^2 r_0^3}\left\{\frac{1}{\left(1+\frac{r^2}{r_0^2}\right)^2}+\left(\frac{r_0}{r}\right)^{-3}\left[\frac{r/r_0}{1+\frac{r^2}{r_0^2}}-\arctan{\left(\frac{r}{r_0}\right)}\right]\right\}$     & $\frac{M}{3\pi^2 r_0^3}$          \\ \hline
MIS            &  $\frac{M}{4\pi r_0^3}\frac{1-\frac{2r^2}{r_0^2}}{\left(1+\frac{r^2}{r_0^2}\right)^{5/2}}$     & $\frac{M}{4\pi r_0^3}$                        \\ \hline
$(k,n)$-model $\gamma=0$  & $\frac{M}{12\pi r_0^3 nB(k,3n)}\left\{3\left[1+\left(\frac{r}{r_0}\right)^{1/n}\right]^{-3n-k}-2{}_2 F_1(3n,3n+k;1+3n;-(r/r_0)^{1/n})\right\}$ & $\frac{M}{12\pi r_0^3 nB(k,3n)}$\\ \hline
\end{tabular}
\label{radialp}
\end{center}
\end{table}
Since we are interested in matter configurations at hydrostatic equilibrium, the fact that the radial pressure is positive in a region of finite extent but negative outside, allows us to introduce an effective size $R$ for the gravitational object by means of the condition $p_r(R)=0$. We warn the reader in advance that such a gravitational object will not have a finite radius because $\rho$ is not zero in the region $r>R$. For a list of numerical values of the quantity $\widehat{R}=R/r_0$ we refer to Table~\ref{tablerx}.
\begin{table}
\caption{For different choices of the triple $(\alpha,\beta,\gamma)$ in the Zaho model we present some typical values of the radial distance $\widehat{R}=R/r_0$ at which the radial pressure vanishes.}
\begin{center}
\begin{tabular}{ | l | l | l | l|}
\hline
Model             & $\widehat{R}=R/r_0$          \\ \hline
Dehnen $(1,4,0)$  & 0.5000000000               \\ \hline
PS                & 0.8242659494                \\ \hline
MIS               & 0.7071067812                 \\ \hline
$(k,n)=(1,2)$     & 0.2500000000           \\ \hline
$(k,n)=(1,3)$     & 0.1250000000          \\ \hline
$(k,n)=(2,1)$     & 0.3722813233                 \\ \hline
$(k,n)=(2,2)$     & 0.1729565347       \\ \hline
$(k,n)=(2,3)$     & 0.0835514834        \\ \hline
\end{tabular}
\label{tablerx}
\end{center}
\end{table}
If we consider a static spherically symmetric matter distribution given by the Zhao profile and we insist that the energy-momentum tensor is that of an anisotropic fluid as in (\ref{EMT}), the Einstein field equations $G_{\mu\nu}=-8\pi T_{\mu\nu}$ together with the conservation equation $T^{\mu\nu}{}_{;\nu}=0$ and the following ansatz for the line element
\begin{equation}\label{guappo}
ds^2=A^2(r)dt^2-\frac{dr^2}{B(r)}-r^2\left(d\vartheta^2+\sin^2{\vartheta}d\varphi^2\right)
\end{equation}
lead to the following solution
\begin{eqnarray}
B(r)&=&1-\frac{2m(r)}{r},\label{Bfun}\\
A^2(r)&=&e^{\phi(r)},\quad\phi(r)=\int_r^\infty\psi(u)~du,\quad\psi(r)=\frac{1}{B(r)}\left[8\pi rp_r(r)+\frac{2m(r)}{r^2}\right],\label{Afun}\\
p_\bot(r)&=&p_r(r)+\frac{r}{2}\left[\frac{dp_r}{dr}+\frac{p_r(r)+\rho(r)}{B(r)}\left(4\pi rp_r(r)+\frac{m(r)}{r^2}\right)\right].\label{ttp}
\end{eqnarray}
For more details in the derivation of the above solution we refer the reader to \cite{EPJCus}. At this point a remark is in order. Since the metric coefficient $B$ coincides with the $g_{rr}$ determined in the Section~\ref{DMG}, it immediately follows that $B$ and $g_{rr}$ will share the same roots provided that $\mu\geq\mu_c$. Furthermore, the tangential pressure $p_\bot$ blows up at the zeroes of $B$, since the latter appears in (\ref{ttp}) in the denominator. This rules out the possibility of interpreting the present solution as a dirty black hole metric because if this were the case, $p_\bot$ should remain finite at the horizons \cite{PIEROBOSS}. If we impose that $\mu<\mu_c$, we do not only circumvent the aforementioned problem but we also ensure the regularity of the function $\phi$ because $B$ entering in the denominator in the last expression in (\ref{Afun}) will never vanish. As a result of this analysis, we draw the conclusion that the line element (\ref{guappo}) represents a fuzzy self-gravitating DM droplet. In Figure~\ref{figure13} and \ref{figure14} we plotted the tangential pressure for the models considered in our work to show that it is indeed well-behaved for any value of $r$ provided that $\mu<\mu_c$.
Also in the case of a nonlocal equation of state, it turns out that the self-gravitating droplet does not exhibit any curvature singularity at $r=0$ made exception to the PS model. This can be easily verified by means of the following formula for the Kretschmann scalar adapted to the line element (\ref{guappo}) \cite{EPJCus}
\begin{equation}\label{kre}
\mathcal{K}=R^{\alpha\beta\gamma\delta}R_{\alpha\beta\gamma\delta}=\frac{2}{r^2}\left[\left(\frac{dB}{dr}\right)^2+B^2(r)\psi^2(r)\right]+\frac{1}{4}\left[B(r)\psi^2(r)+2B(r)\frac{d\psi}{dr}+\psi(r)\frac{dB}{dr}\right]^2.
\end{equation}
Note that because of the presence of the term $1/r^2$ in (\ref{kre}), it is not clear a priori whether the Kretschmann scalar is singularity free at $r=0$ for the models considered here. With the help of Maple we verified that all models except the PS model do not possess a central curvature singularity while in the PS model the self-gravitating droplet exhibits a naked singularity as it can be seen in Table~\ref{tablekre}. 
\begin{table}
\caption{For different choices of the triple $(\alpha,\beta,\gamma)$ in the Zaho model we present the corresponding expressions of the Kretschmann scalar at $r=0$ where the PS model has a curvature singularity. Here $\mu=M/r_0$}
\begin{center}
\begin{tabular}{ | l | l | l | l|}
\hline
Model             & $\mathcal{K}$          \\ \hline
Dehnen $(1,4,0)$  & $80\mu^2/r_0^2$               \\ \hline
PS                & $\frac{64\mu^2(3\pi^2-56\pi\mu+272\mu^2)}{\pi^2(\pi-8\mu)^2}\frac{1}{r^4}+\mathcal{O}\left(\frac{1}{r^2}\right)$                \\ \hline
MIS               & $80\mu^2/r_0^2$                 \\ \hline
$(k,n)=(1,n)$     & $80\mu^2/r_0^2$           \\ \hline
$(k,n)=(2,1)$     & $1280\mu^2/r_0^2$                \\ \hline
$(k,n)=(2,2)$     & $3920\mu^2/r_0^2$       \\ \hline
$(k,n)=(2,3)$     & $8000\mu^2/r_0^2$        \\ \hline
\end{tabular}
\label{tablekre}
\end{center}
\end{table}
In order to complete the analysis of the geometry relative to the line element (\ref{guappo}), we observe that $B\to 1$ at space-like infinity. Moreover,the fact that  $\psi(r)=-2M/r^2+\mathcal{O}(1/r^3)$ ensures that $e^{\phi(r)}\to 1$ as $r\to\infty$. Hence, the manifold described by (\ref{guappo}) goes over into the Minkowski metric asymptotically away. We conclude this section by showing that our droplet allows for bound states of massive and massless particles.

A few remarks are in order here. With the new nonlocal EOS we will treat the emerging astrophysical objects in their own rights as they display new physical features worth to focus upon. This is to say, we will not concentrate the discussion on the agreement of the respective effective potentials with the corresponding quantity of a Schwarzschild BH. However, we will come back to this point later at an appropriate point.

With the help of ($25.16$) in \cite{Fliessbach} we immediately find that the effective potential for the droplet is given by
\begin{equation}\label{effp}
U_{eff}(r)=\frac{e^{\phi(r)}}{2}\left(\epsilon+\frac{\ell^2}{r^2}\right), 
\end{equation}
where $\epsilon$ and $\ell$ have been already defined in the previous section. If we consider for instance the Dehnen $(1,4,0)$ model, a simple computation shows that the metric coefficients are
\begin{equation}
A^2(r)=B(r)=1-\frac{2Mr^2}{(r+r_0)^3}
\end{equation}
and if in addition we let $y=r/r_s$, $L=\ell/r_s$ and $r_0/r_s=\sigma$, the effective potential becomes
\begin{equation}\label{effpdehnen}
U_{eff}(y)=\frac{1}{2}\left[1-\frac{y^2}{\sigma^3\left(1+\frac{y}{\sigma}\right)^3}\right]\left(\epsilon+\frac{L^2}{y^2}\right).
\end{equation}
Taking into account that $\sigma$ and the rescaled mass $\mu$ are linked through the relation $\sigma=1/2\mu$, the condition $\mu<\mu_c$ constrains the possible choices for $\sigma$ according to the inequality $\sigma>\sigma_c=1/2\mu_c$. Going back to Table~\ref{tableEins} we find that $\sigma>0.148148$.
\begin{figure}[!ht]\label{effdropletdehnen}
\centering
    \includegraphics[width=0.3\textwidth]{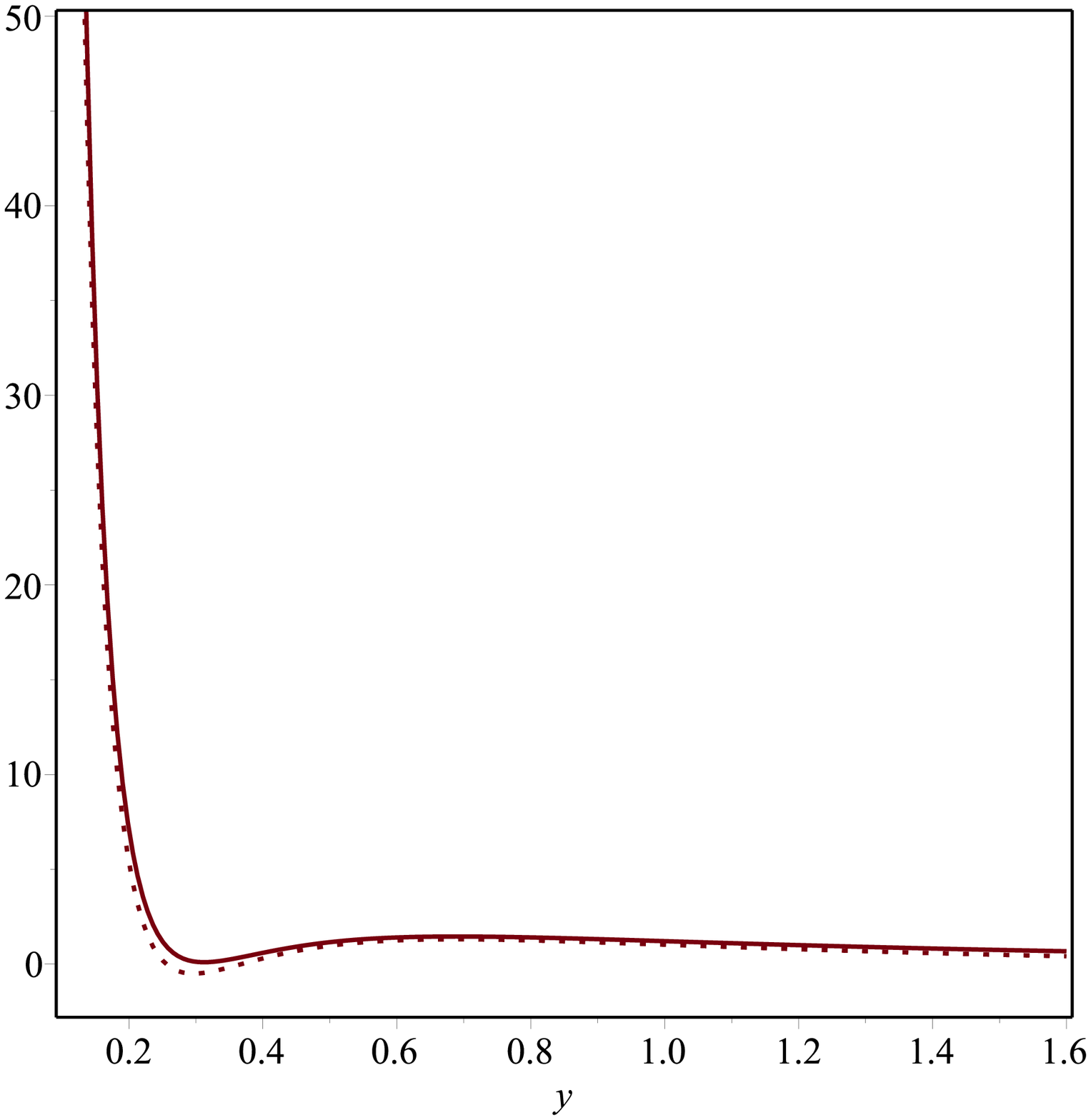}
     \includegraphics[width=0.3\textwidth]{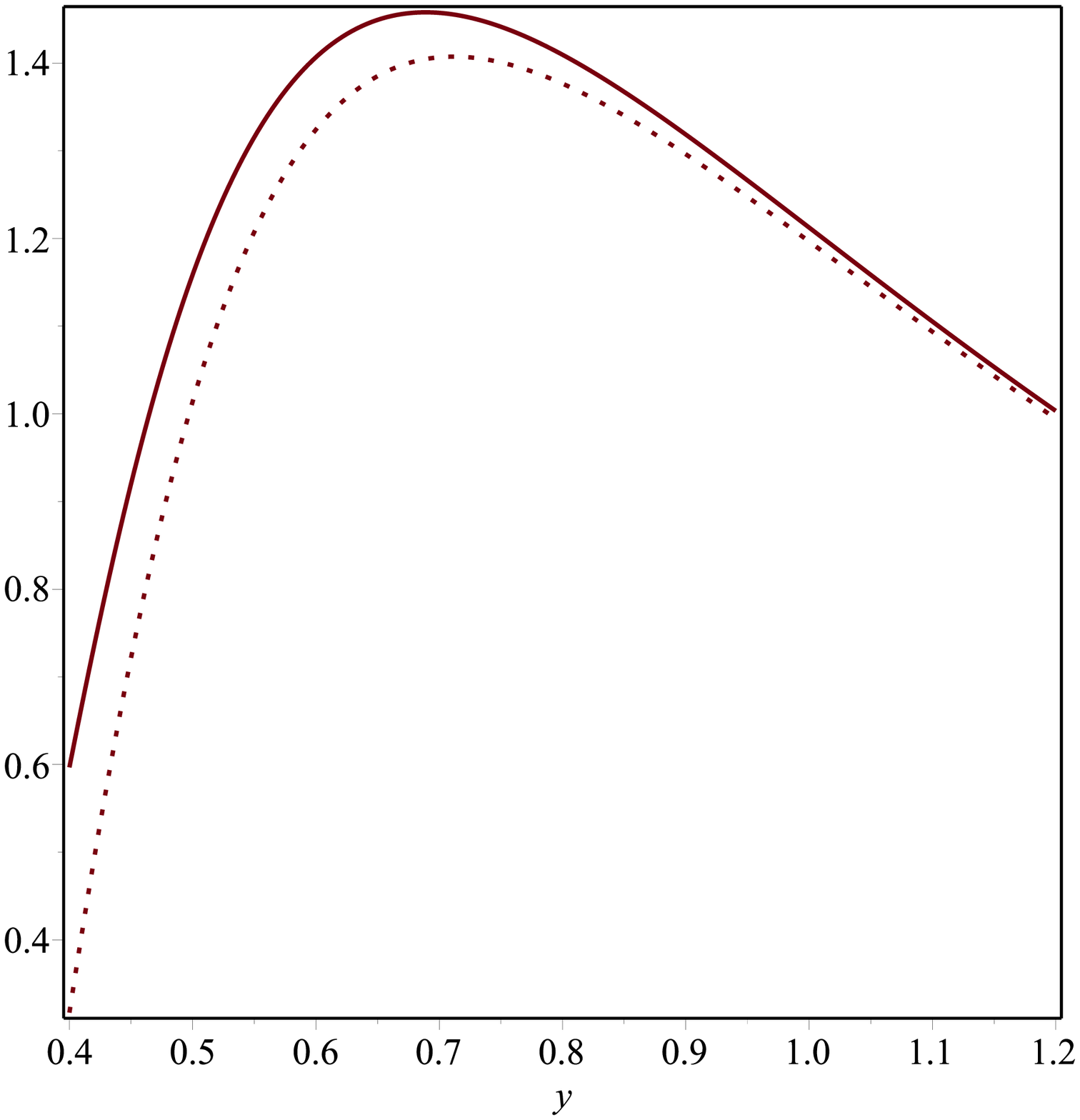}
     \includegraphics[width=0.3\textwidth]{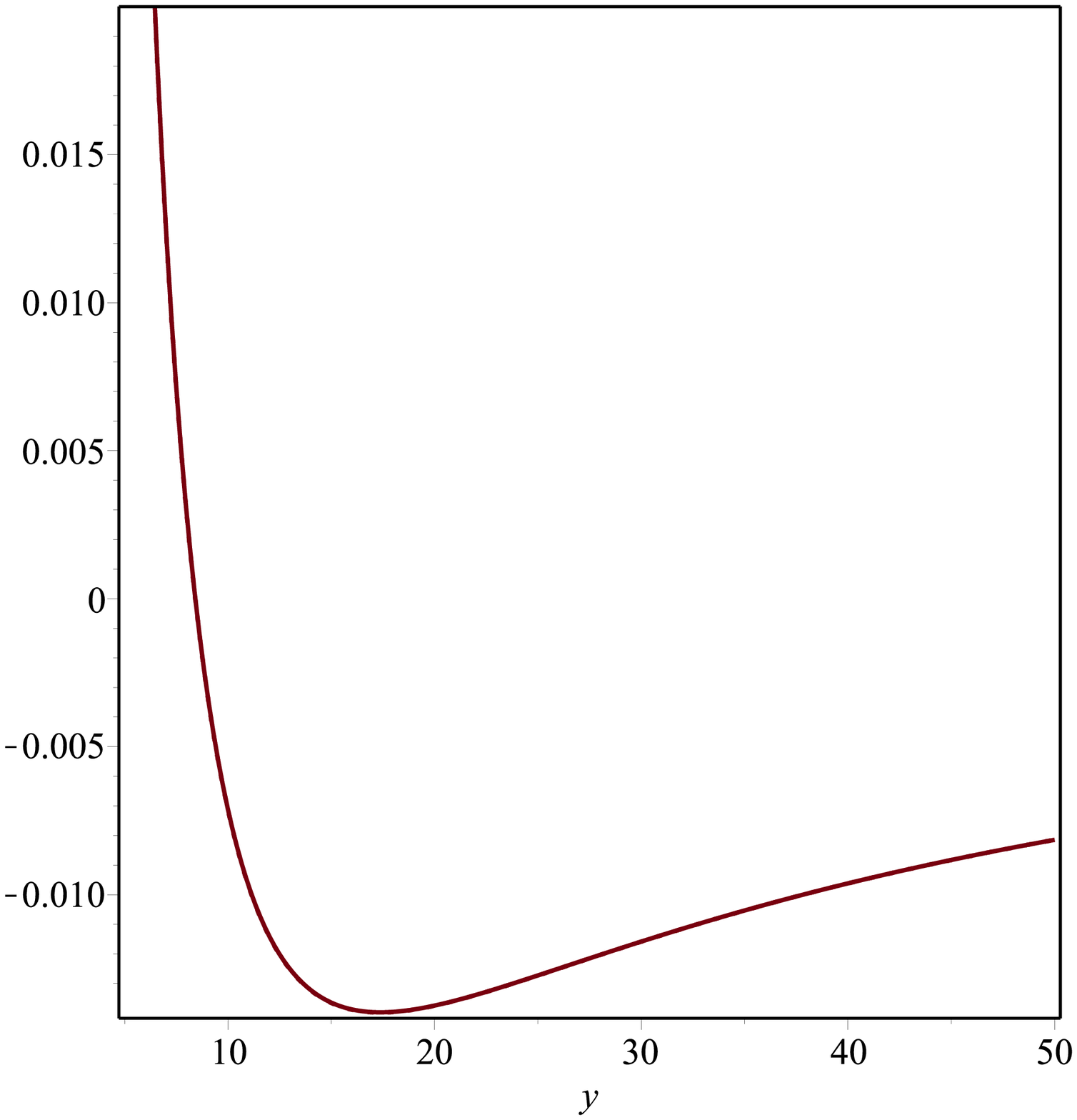}
\caption{\label{figure15}
Plot of the effective potential (\ref{effpdehnen}) for a massive particle with $L=3$ when $\sigma=0.15$ (solid line) and $\sigma=0.14815$ (dotted line). In plotting (\ref{effpdehnen}) we took advantage of the fact that the potential is defined up to an additive constant in order to subtract a factor one half ensuring that $U_{eff}\to 0$ as $y\to\infty$. In the case $\sigma=0.15$ the minima occur at $y=0.311$ and $y=17.376$ while the maximum is located at $y=0.689$. $U_{eff}$ is positive at the first minimum and negative at the second minimum. For $\sigma=0.14815$ the minima are at $y=0.296$ and $y=17.363$ while the maximum is at $y=0.712$. The effective potential is negative at both minima.}
\end{figure}
\begin{figure}[!ht]\label{effdropletdehnenmassless}
\centering
    \includegraphics[width=0.3\textwidth]{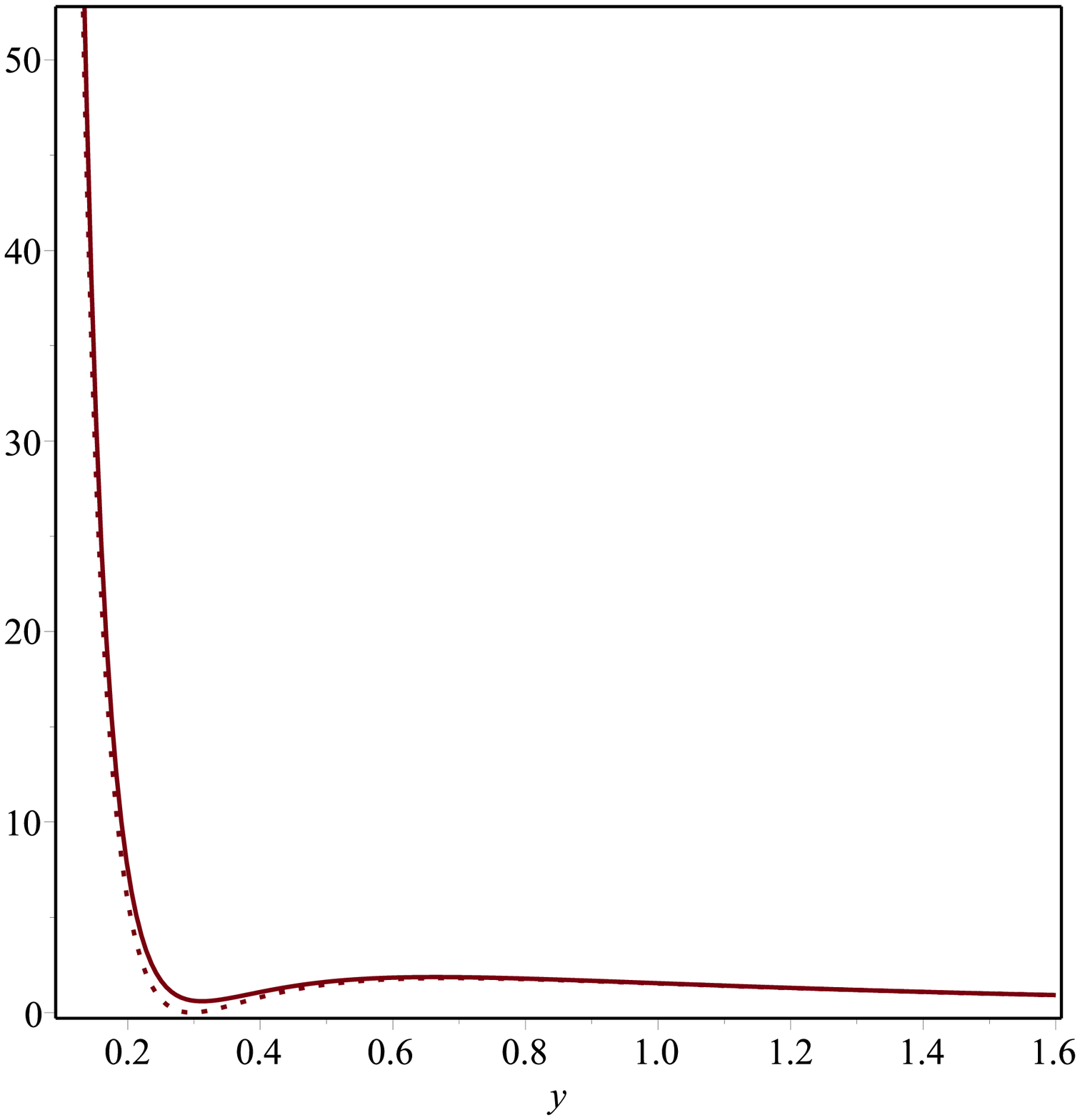}
     \includegraphics[width=0.3\textwidth]{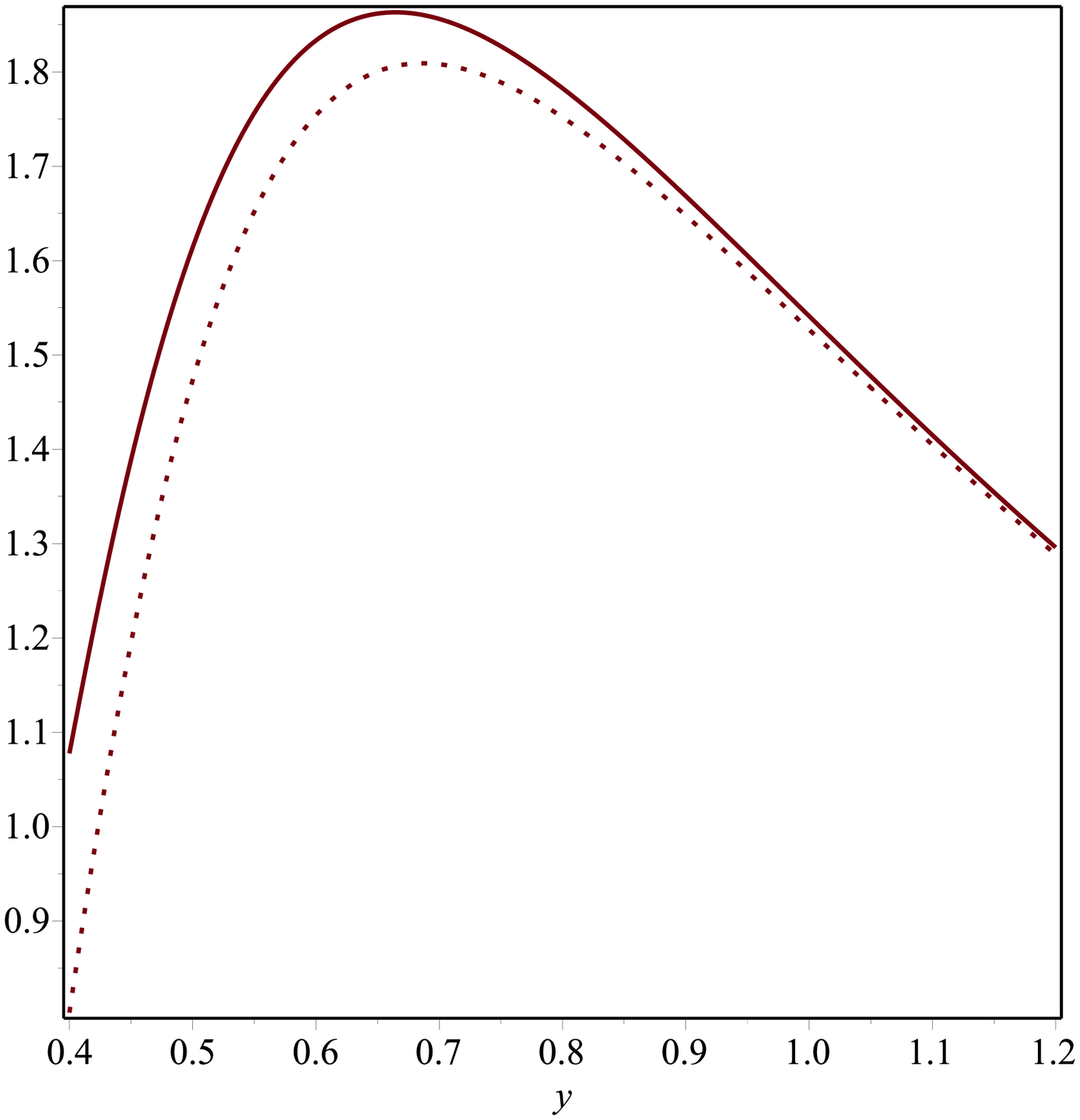}
     \includegraphics[width=0.3\textwidth]{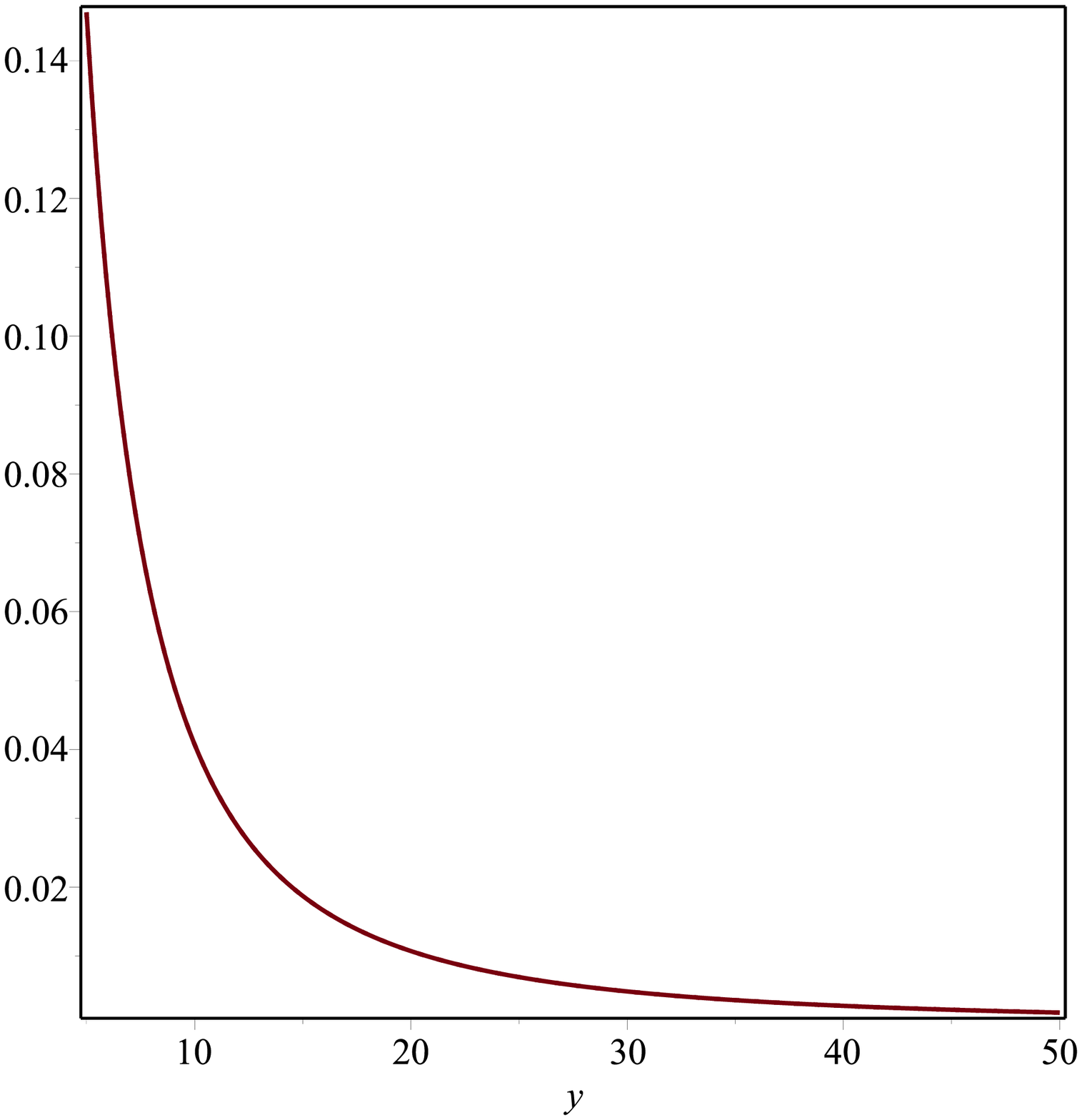}
\caption{\label{figure16}
Plot of the effective potential (\ref{effpdehnen}) for a massless particle with $L=3$ when $\sigma=0.15$ (solid line) and $\sigma=0.14815$ (dotted line). In the case $\sigma=0.15$ a minimum and a maximum occur at $y=0.312$ and $y=0.664$, respectively. For $\sigma=0.14815$ the minimum is at $y=0.296$ and the maximum at $y=0.686$. For both choices of $\sigma$ the effective potential is everywhere positive.}
\end{figure}
As it can be seen in Figure~\ref{figure15} and Figure~\ref{figure16} bound states corresponding to stable and unstable orbits of massive or massless particles are allowed. Moreover, all orbits take place outside the effective size $R$ of the droplet which according to Table~\ref{tablerx} and after the rescaling introduced above is given by $Y=R/r_s=r_0/2r_s=\sigma/2$. In the massive case for $L$ fixed and increasing $\sigma$, the effective potential admits only one global minimum. For instance, if $L=3$ and $\sigma=1$, the minimum is located at $y=22.248$ while for $\sigma=10$ the minimum is at $y=54.728$. Since the effective potential in the massive case is nonnegative, a matching procedure with $U_{eff}$ for a Schwarzschild BH is not suitable in this case. In the presence of light, the gravitational object may exhibit an outer unstable photon sphere and an inner stable photon sphere if $\sigma$ is chosen appropriately. Furthermore, for fixed $L$ and increasing $\sigma$ the droplet does not need to possess a photon sphere. For instance, if $L=3$ and $\sigma>0.159$ there is no photon sphere. Another model which can be analytically solved is for instance the $(k,n)$-model with $k=1$ and $n=2$. In this case, we find that the metric coefficients are
\begin{equation}
A^2(r)=B(r)=1-\frac{2Mr^2}{\left(\sqrt{r}+\sqrt{r_0}\right)^6}.
\end{equation}
Proceeding as above the effective potential reads
\begin{equation}\label{effpkn12}
\mathcal{U}_{eff}(y)=\frac{1}{2}\left[1-\frac{y^2}{\sigma^3\left(1+\sqrt{\frac{y}{\sigma}}\right)^6}\right]\left(\epsilon+\frac{L^2}{y^2}\right).
\end{equation}
As before we impose the condition $\mu<\mu_c$ which according to Table~\ref{tableEins} translates into the equivalent constraint $\sigma>0.02195$.
\begin{figure}[!ht]\label{effdropletKN12}
\centering
    \includegraphics[width=0.3\textwidth]{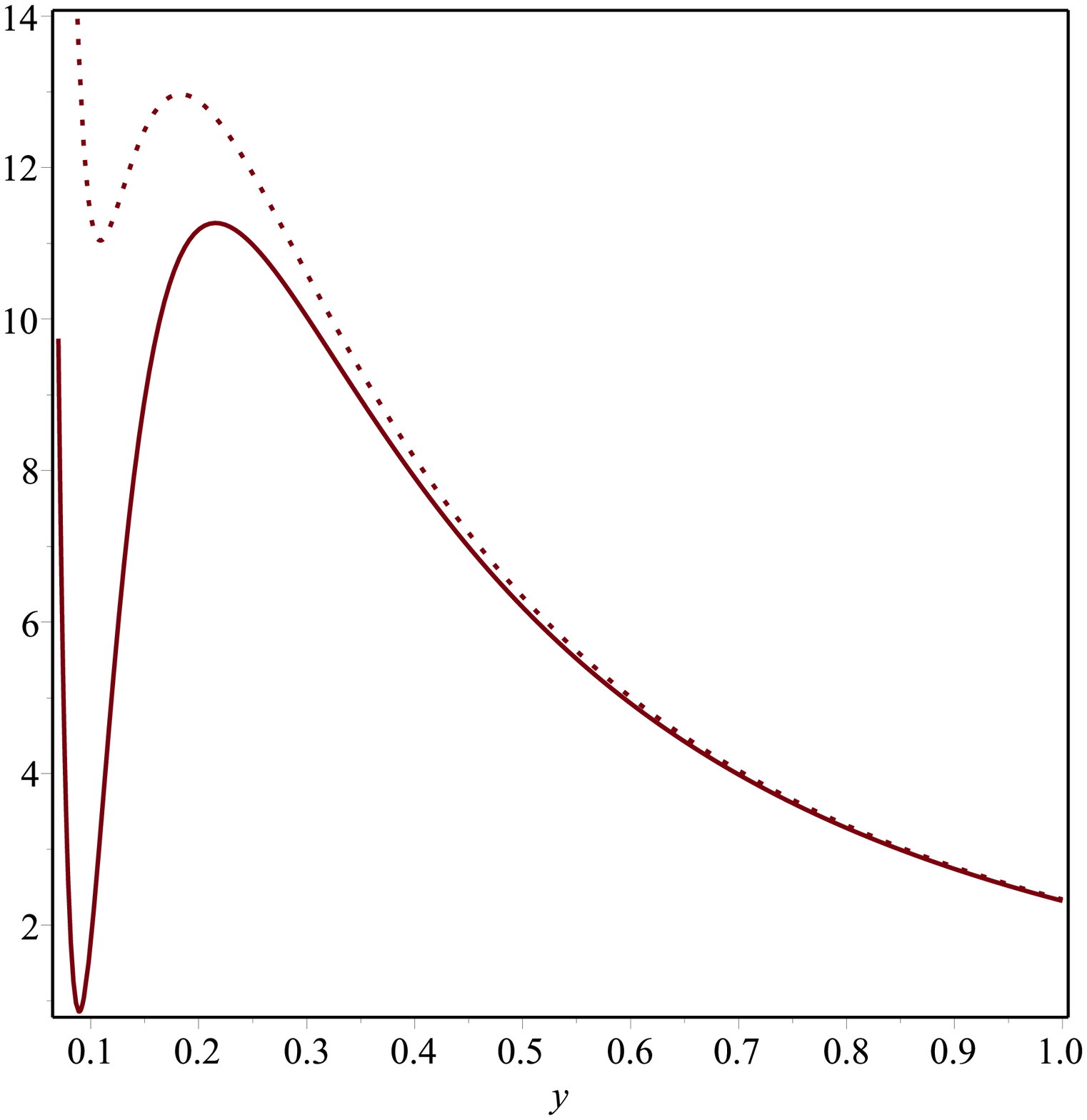}
     \includegraphics[width=0.3\textwidth]{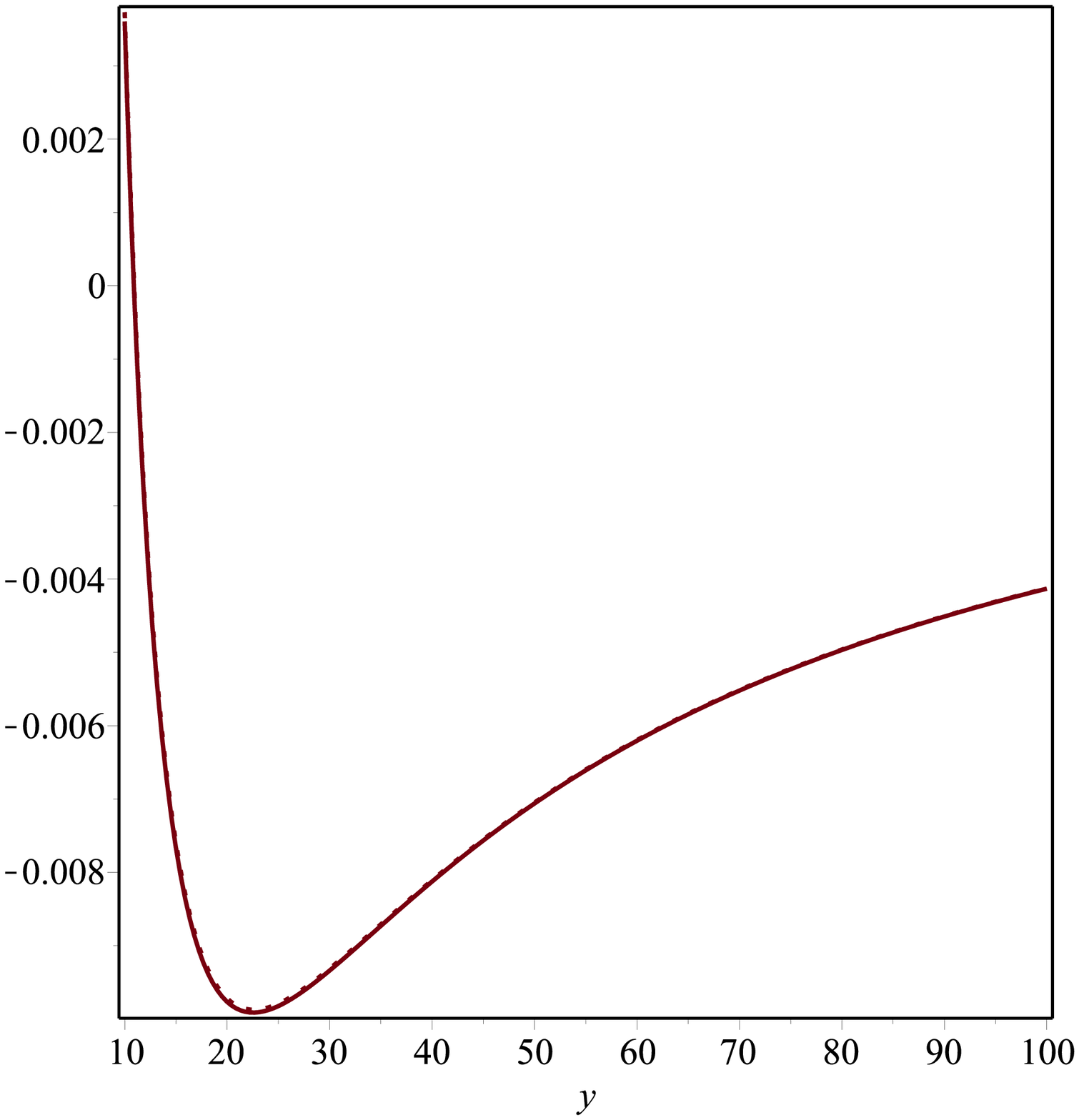}
\caption{\label{figure17}
Plot of the effective potential (\ref{effpkn12}) for a massive particle with $L=3$ when $\sigma=0.0220$ (solid line) and $\sigma=0.0225$ (dotted line). In the case $\sigma=0.0220$ the minima occur at $y=0.089$ and $y=22.536$ while the maximum is located at $y=0.216$.  For $\sigma=0.0225$ the minima are at $y=0.109$ and $y=22.604$ while the maximum is at $y=0.183$. The effective potential is negative at both minima. In both cases $\mathcal{U}_{eff}$ is positive at the first minimum and negative at the second minimum.}
\end{figure}
\begin{figure}[!ht]\label{effdropletKN12massless}
\centering
    \includegraphics[width=0.3\textwidth]{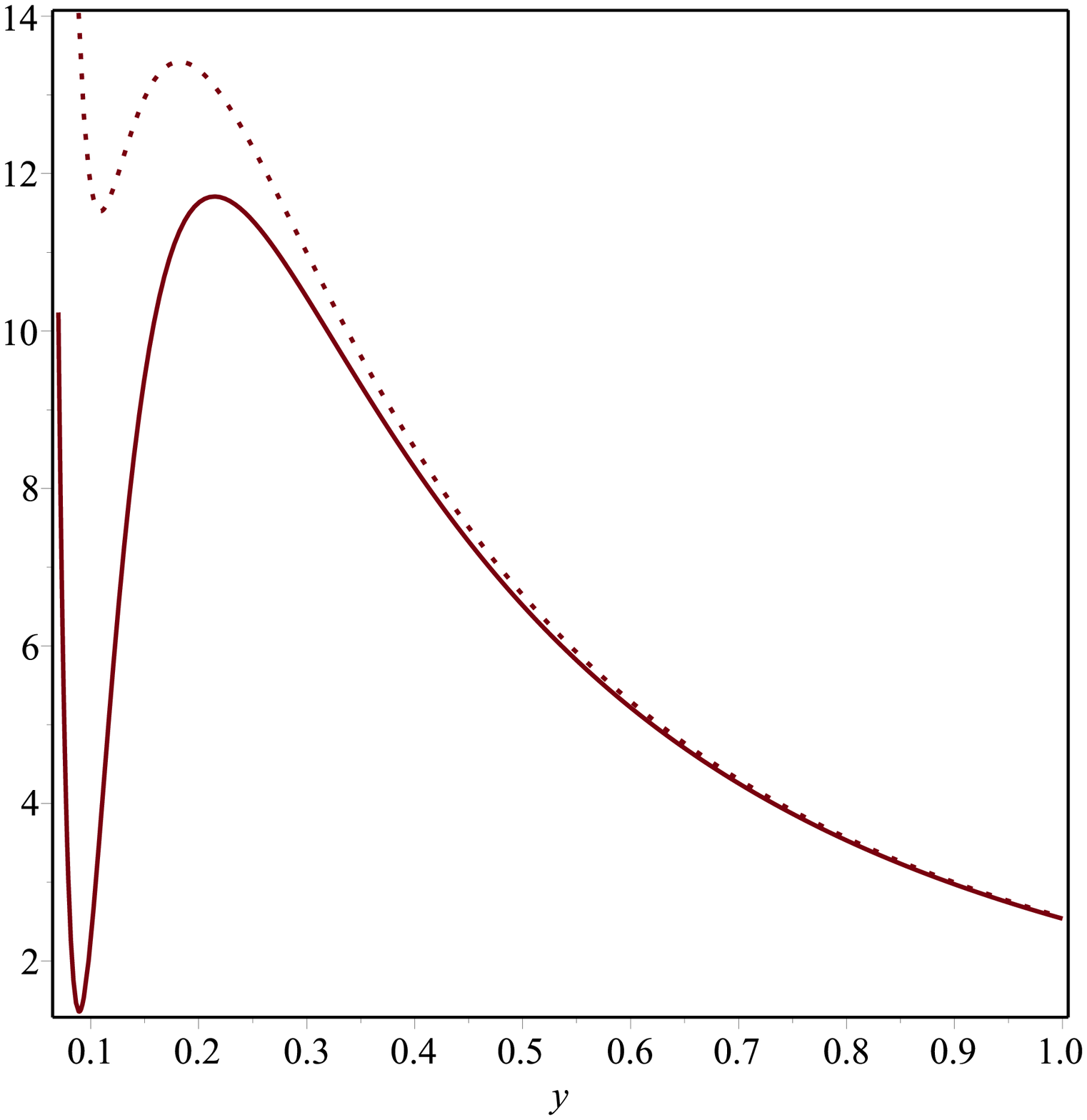}
     \includegraphics[width=0.3\textwidth]{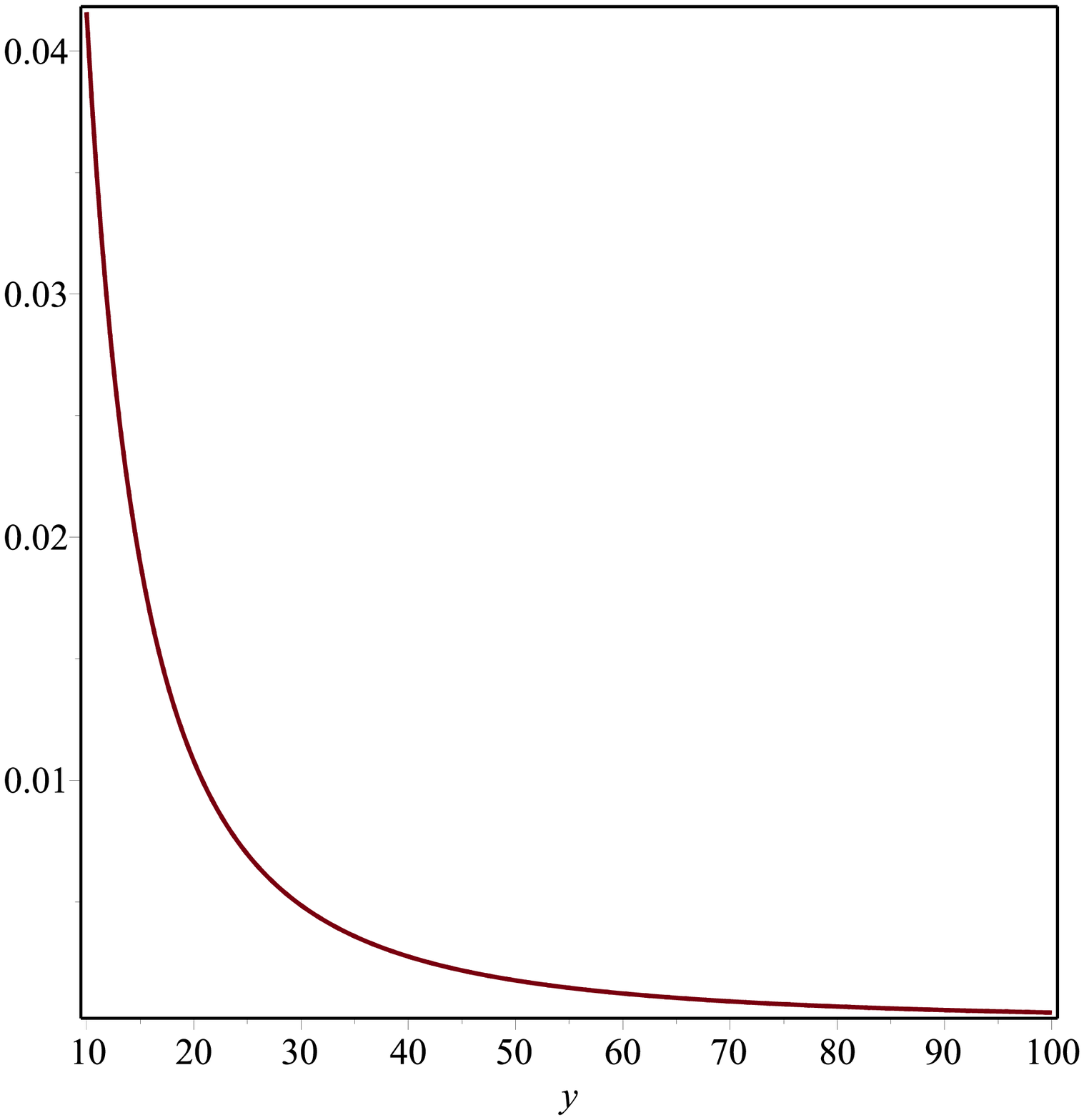}
\caption{\label{figure18}
Plot of the effective potential (\ref{effpkn12}) for a massless particle with $L=3$ when $\sigma=0.0220$ (solid line) and $\sigma=0.0225$ (dotted line). In the case $\sigma=0.0220$ a minimum and a maximum occur at $y=0.089$ and $y=0.215$, respectively. For $\sigma=0.0225$ the minimum is at $y=0.109$ and the maximum at $y=0.182$. For both choices of $\sigma$ the effective potential is everywhere positive.}
\end{figure}
\begin{figure}[!ht]\label{dropsch}
\centering
    \includegraphics[width=0.3\textwidth]{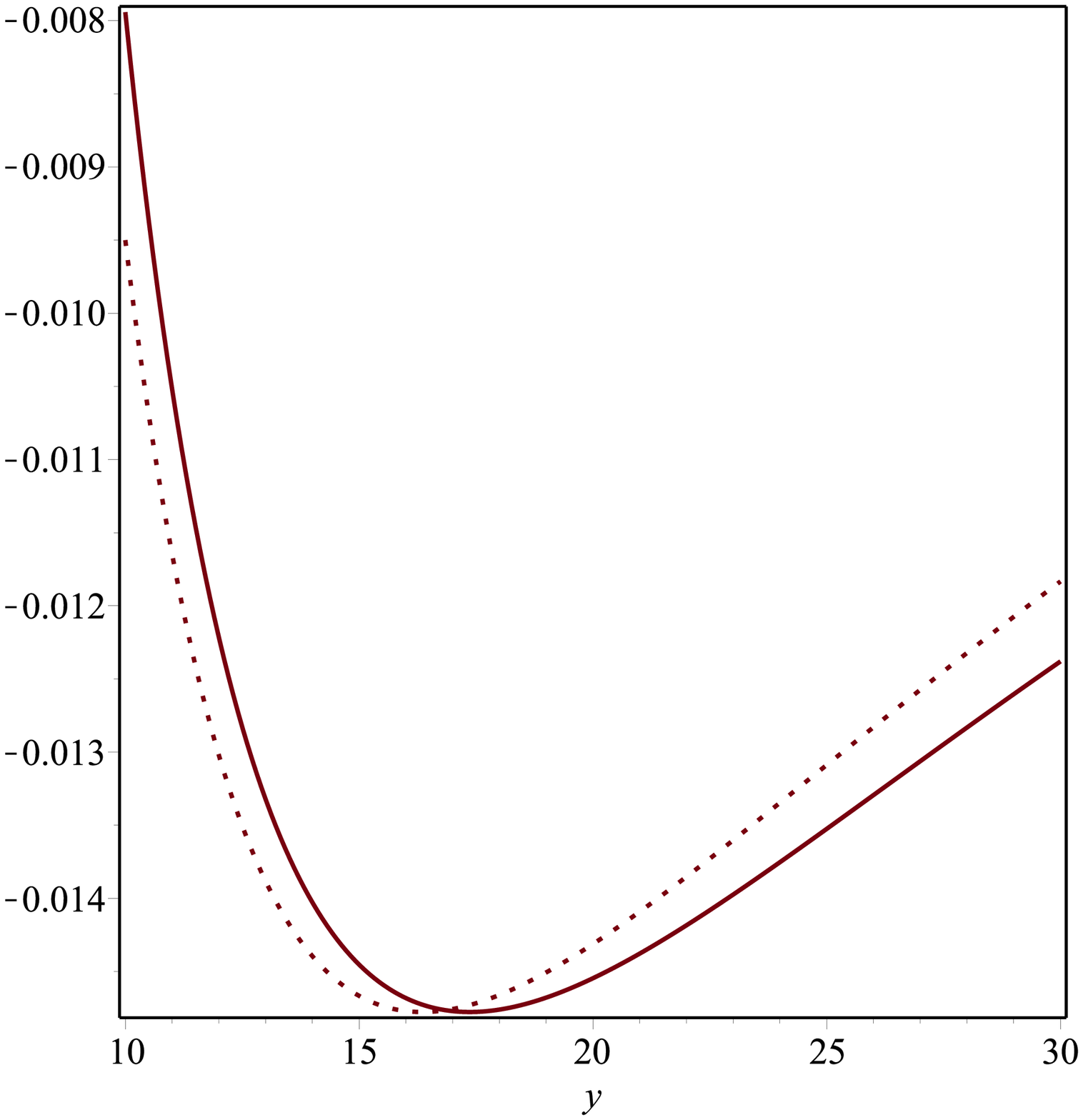}
     \includegraphics[width=0.3\textwidth]{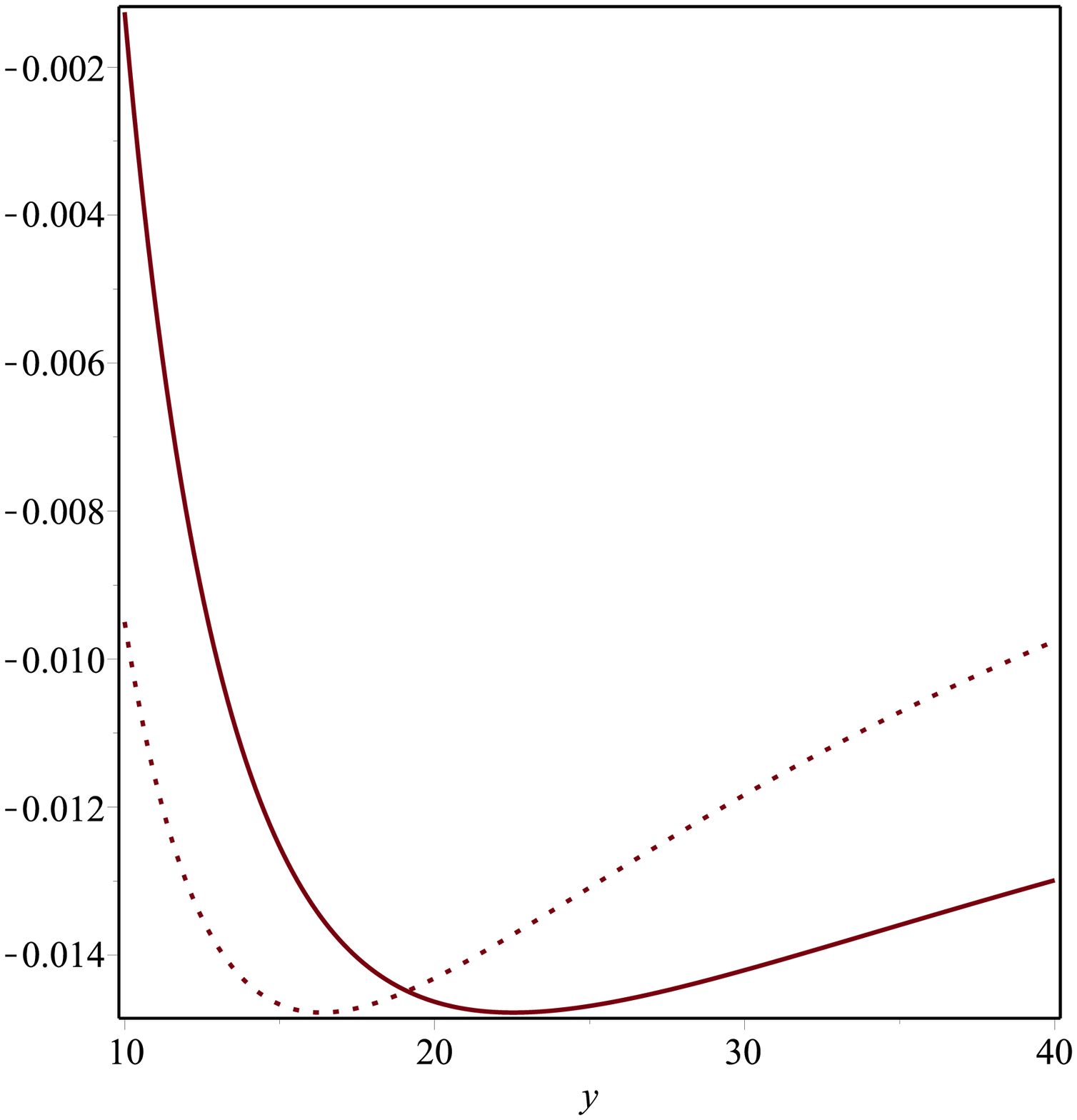}
\caption{\label{figure19}
Comparison between the minimum in the Schwarzschild effective potential for the massive case with $L=3$ and the outer minima of the effective potentials (\ref{effpdehnen}) (left panel, $\sigma=0.14815$, solid line) and (\ref{effpkn12}) (right panel, $\sigma=0.0220$, solid line). }
\end{figure}
Figure~\ref{figure17} and Figure~\ref{figure18} clearly show the presence of bound states for both massive and massless particles. Such orbits can be stable or unstable depending whether they occur in a neighbourhood of a minimum or a maximum in the effective potential. It is interesting to observe that also in this model all orbits are located in the region outside the effective size $R$ of the droplet which according to Table~\ref{tablerx} and after the rescaling introduced above is given by $Y=R/r_s=r_0/4r_s=\sigma/4$. In the massive scenario if we keep $L$ fixed and increase $\sigma$, we find that the potential may admit only one global minimum. For example, for $L=3$ it turns out that the effective potential exhibits only one minimum for $\sigma\geq 0.0228$.
In the context of massless particles, it is gratifying to see that also the present model indicates the formation possibility of an outer unstable photon sphere and an inner stable photon sphere. A further common feature between the two models treated here is that if we keep $L$ fixed and we slowly increase $\sigma$ the gravitational object will not possess a photon sphere. For example, for $L=3$ this happens whenever $\sigma\geq 0.0228$.

It appears that in the present case a matching procedure of $U_{eff}$ with  a Schwarzschild BH cannot be achieved due to the non-negativity of the emerging effective potential. However, due to the form of the equation of motion (\ref{C}) one can add to the effective potential
negative constant and absorb it on the right hand side in the constant $C$. This is shown in Figure~\ref{figure19} with the result that the matching procedure to the Schwarzschild case modeling the central BH, is admittedly not as good as in the previous cases with the de Sitter EOS. If this superficial agreement is good enough remains to be seen. At the same time, we do not see it as a drawback but rather as a new chance to probe into exotic astrophysical objects that are solutions to the Einstein field equations emerging from DM profiles. We draw the reader attention to the fact that the droplets described in this section may exhibit stable bound orbits for photons.

We conclude this section by studying the shadow of the self-gravitating droplets emerging from the Dehnen and the $(k,n)=(1,2)$ models coupled to a nonlocal EOS for those values of the parameter $\sigma$ generating a photon sphere. To this purpose, we recall that the shadow is defined to be the lensed image at infinity of the photon sphere \cite{EHT}.
Since the detection of a black hole photon sphere is within the capabilities of the Event Horizon Telescope, it is interesting to investigate how the shadow generated by our gravitational object compares with respect to the shadow of a Schwarzschild black hole. According to \cite{EHT} for a metric of the form (\ref{guappo}), the radius of the photon sphere $r_\gamma$ is defined as a positive real root of the equation
\begin{equation}
r=2A^2\left(\frac{dA^2}{dr}\right)^{-1},
\end{equation}
while the relation connecting the radius $r_{sh}$ of the black hole shadow with $r_\gamma$ is
\begin{equation}\label{ratio}
\frac{r_{sh}}{r_\gamma}=\frac{1}{\sqrt{A^2(r_\gamma)}}.
\end{equation}
We recall that in the case of a Schwarzschild black hole $A^2=1-2M/r$, the photon sphere is located at $r_\gamma=3M$ and the above ratio turns out to be $r_{sh}/r_{\gamma}=\sqrt{3}\approx 1.732$. In what follows it is convenient to introduce the rescaled quantities $y=r/r_s$ and $r_0/r_s=\sigma$. Moreover, we will focus our attention to the outer photon sphere oft the droplet. In the case of the Dehnen model $(1,4,0)$ the equation for the photon sphere turns out to be
\begin{equation}
\left(\frac{y}{\sigma}+1\right)^4-\frac{3y^3}{2\sigma^4}=0
\end{equation}
while (\ref{ratio}) becomes
\begin{equation}
\frac{y_{sh}}{y_\gamma}=\frac{1}{\sqrt{1-\frac{y^2_\gamma}{\sigma^3\left(1+\frac{y_\gamma}{\sigma}\right)^3}}}.
\end{equation}
Then, for $\sigma=0.14815\div 0.15$, we find $y_\gamma=0.664568\div 0.686474$ and therefore, $y_{sh}/y_\gamma=2.29\div 2.34$. If we consider instead the model $(k,n)=(1,2)$, the equation of the photon sphere reads
\begin{equation}
\left(\sqrt{\frac{y}{\sigma}}+1\right)^7-\frac{3y^{5/2}}{2\sigma^{7/2}}=0
\end{equation}
and in this case
\begin{equation}
\frac{y_{sh}}{y_\gamma}=\frac{1}{\sqrt{1-\frac{y^2_\gamma}{\sigma^3\left(1+\sqrt{\frac{y_\gamma}{\sigma}}\right)^6}}}.
\end{equation}
A computation similar to the one done above shows that if we take $\sigma=0.0220\div 0.0225$, then $y_\gamma=0.182328\div0.214684$ and hence, $y_{sh}/y_\gamma=2.89\div 3.18$. We found that in the Dehnen and the model $(k,n)=(1,2)$ the ratio $r_{sh}/r_\gamma$ is $1.3$ and $1.7$ times larger than the corresponding ratio for a Schwarzschild black. Hence, we hope that in the next future the existence of these gravitational objects may be confirmed or disproved with the help of EHT.

\section{Conclusions and outlook}
We found new astrophysical objects based on known DM profiles and de Sitter or nonlocal equations of state. We emphasize that these models are solutions of Einstein field equations and they describe stable regular fuzzy objects with no horizons (droplets) or with one or two horizons (BHs). In many cases of the DM profiles, choosing suitable parameters one can show that the effective potential describing the motion of a test particle is close to the corresponding motion in the BH picture. We concentrated here on the local minimum in the effective potential. In such a case the motion of the S-stars around the central object gives the same qualitative result. In the case of a nonlocal EOS, we also found exotic properties of the effective potential in addition to the usual unstable maximum, namely a minimum corresponding to stable orbits of photons. This implies that such objects may admit two photon spheres. Furthermore, the small differences in the effective potential of different models can be used to discriminate between them. In addition, as a tool to distinguish between the different models we suggested to use the shadow of the droplet. This seems to us appropriate and timely because the detection of such a shadow is well within the capabilities of the Event Horizon Telescope. Finally, while finishing this manuscript, we found some recent papers \cite{V1,V2} which are relevant to our future work. In the aforementioned literature, the author shows that just the observation of relativistic images  (no information about the masses and distances are required) provides an incredibly accurate value for the upper bound to the compactness of massive dark objects. It would be a worthwhile undertaking to apply the new method developed in \cite{V1,V2} in order to study the gravitational lensing signature of the new gravitational objects derived in the present work.

\end{document}